\documentclass{article}
\usepackage[utf8]{inputenc}
\usepackage{float}
\usepackage{graphicx}
\usepackage{xcolor}
\newcommand{\comment}[1]{}
\usepackage{amsmath}
\usepackage{authblk}
\usepackage{xr}
\usepackage[papersize={8.5in,11in}]{geometry}

\title{Experimental Demonstration of Attosecond Pump-Probe Spectroscopy with an X-ray Free-Electron Laser}

\author[1, 2]{Zhaoheng Guo*}
\author[1, 4]{Taran Driver
\thanks{These two authors contributed equally.}}
\author[4, 7]{Sandra Beauvarlet}
\author[1]{David Cesar}
\author[1]{Joseph Duris}
\author[1,2]{Paris L. Franz}
\author[9]{Oliver Alexander}
\author[1]{Dorian Bohler}
\author[17, 18]{Christoph Bostedt}
\author[9]{Vitali Averbukh}
\author[1,5]{Xinxin Cheng}
\author[10]{Louis F. DiMauro}
\author[15]{Gilles Doumy}
\author[1,4,5]{Ruaridh Forbes}
\author[13]{Oliver Gessner}
\author[1,5]{James M. Glownia}
\author[1,2,4]{Erik Isele}
\author[1,5]{Andrei Kamalov}
\author[1,4]{Kirk A. Larsen}
\author[1]{Siqi Li}
\author[1,5]{Xiang Li}
\author[1,5]{Ming-Fu Lin}
\author[10]{Gregory A. McCracken}
\author[1,5]{Razib Obaid}
\author[1,3,4]{Jordan T. O'Neal}
\author[1,2]{River R. Robles}
\author[14]{Daniel Rolles}
\author[9]{Marco Ruberti}
\author[14]{Artem Rudenko}
\author[13]{Daniel S. Slaughter}
\author[1]{Nicholas S. Sudar}
\author[1,2,4]{Emily Thierstein}
\author[10]{Daniel Tuthill}
\author[11]{Kiyoshi Ueda}
\author[14]{Enliang Wang}
\author[1,2,4]{Anna L. Wang}
\author[1,2,4]{Jun Wang}
\author[13]{Thorsten Weber}
\author[1,4]{Thomas J. A. Wolf}
\author[15, 16]{Linda Young}
\author[1]{Zhen Zhang}
\author[1,2,3,4] {Philip H. Bucksbaum}
\author[9]{Jon P. Marangos}
\author[1,2,6]{Matthias F. Kling}
\author[1,2,6]{Zhirong Huang}
\author[1,5]{Peter Walter}
\author[8]{Ludger Inhester}
\author[7]{Nora Berrah}
\author[1,4]{James P. Cryan\thanks{jcryan@slac.stanford.edu}}
\author[1,6]{Agostino Marinelli\thanks{marinelli@slac.stanford.edu}}

\affil[1]{SLAC National Accelerator Laboratory, Menlo Park, CA 94025, USA}
\affil[2]{Department of Applied Physics, Stanford University, Stanford, CA 94025, USA }
\affil[3]{Department of Physics, Stanford University, Stanford, CA 94025, USA}
\affil[4]{Stanford PULSE Institute, SLAC National Accelerator Laboratory, Menlo Park, CA 94025, USA}
\affil[5]{Linac Coherent Light Source, SLAC National Accelerator Laboratory, Menlo Park, CA 94025, USA}
\affil[6]{Department of Photon Science, Stanford University, SLAC National Accelerator Laboratory, Menlo Park, CA 94025, USA}
\affil[7]{Physics Department, University of Connecticut, Storrs, CT 06269, USA.}
\affil[8]{Center for Free-Electron Laser Science CFEL, Deutsches Elektronen-Synchrotron DESY, Notkestr. 85, 22607 Hamburg, Germany}
\affil[9]{Quantum Optics and Laser Science Group, Blackett Laboratory, Imperial College London,
London, SW7 2BW, United Kingdom}
\affil[10]{Department of Physics, The Ohio State University, Columbus, Ohio 43210, USA}
\affil[11]{Department of Chemistry, Tohoku University, Sendai 980-8578, Japan}
\affil[12]{Department of Physics and James Franck Institute, The University of Chicago, Chicago, IL, USA}
\affil[13]{Chemical Sciences Division, Lawrence Berkeley National Laboratory, Berkeley, California 94720, USA}
\affil[14]{J.R. Macdonald Laboratory, Department of Physics, Kansas State University, Manhattan, Kansas 66506, USA}
\affil[15]{Chemical Sciences and Engineering Division, Argonne National Laboratory, Argonne, IL, 60439, USA}
\affil[16]{Department of Physics and James Franck Institute, The University of Chicago, Chicago, IL, USA}
\affil[17]{Paul-Scherrer Institute, CH-5232, Villigen PSI, Switzerland}
\affil[18]{LUXS Laboratory for Ultrafast X-ray Sciences, Institute of Chemical Sciences and Engineering, \'{E}cole Polytechnique F\'{e}d\'{e}rale de Lausanne (EPFL), CH-1015, Lausanne, Switzerland}

\date{\today}

\begin{document}
\maketitle

\begin{abstract}
Pump-probe experiments with sub-femtosecond resolution are the key to understanding electronic dynamics in quantum systems. 
Here we demonstrate the generation and control of sub-femtosecond pulse pairs from a two-colour X-ray free-electron laser~(XFEL). 
By measuring the delay between the two pulses with an angular streaking diagnostic, we characterise the group velocity of the XFEL and demonstrate control of the pulse delay down to $270~$as. 
We demonstrate the application of this technique to a pump-probe measurement in core-excited para-aminophenol. 
These results demonstrate the ability to perform pump-probe experiments with sub-femtosecond resolution and atomic site specificity. 
\end{abstract}

\section{Introduction}
Pump-probe spectroscopy is the primary workhorse for ultra-fast experiments~\cite{zewail1988laser}.
This form of stroboscopic measurement requires an initial light pulse which excites, or ``pumps", the system into a non-equilibrium state. 
A subsequent light pulse is then used to probe the time evolution of the system.
When either the pump or probe pulse is tuned to X-ray wavelengths, the interaction of the pulse is strongest with highly localised core-level electrons, resulting in an atomic site-specific interrogation of electronic densities~\cite{siegbahn1982electron}.
This technique is routinely applied to follow photo-induced chemical and material transitions on femtosecond timescales~\cite{picon2016hetero, berrah2019femtosecond, ilchen2021site, barillot2021correlation, schnorr2013time, liekhus2015ultrafast, schwickert2022electronic}. 
In extending the pump-probe technique to sub-femtosecond timescales we can probe the motion of valence electrons in quantum systems.

The ultrafast dynamics of electrons are integral to many chemical and physical processes. 
For example, electron-light interactions, and the resulting electronic motion, are the fundamental mechanisms by which light is absorbed in matter, a basic building block of solar energy technology and photosynthesis~\cite{engel2007evidence, brixner2005two}. 
Furthermore, electron dynamics mediate chemical change and can influence biological function~\cite{scholes2017using}. 
The timescale for this electronic motion is set by the eV-scale binding energy of the most chemically relevant, valence electrons to be in the range of few-femtosecond to sub-femtosecond. 
Therefore, pump/probe experiments capable of accessing coherent electronic motion require the use of sub-femtosecond pulse pairs with sufficient intensity to enable multi-photon interactions.

Isolated light pulses with sub-femtosecond pulse duration were initially demonstrated as early as 2001~\cite{hentschel2001attosecond}, by exploiting the process of strong-field driven high harmonic generation~(HHG)~\cite{li1989multiple}. 
Recently, it has been demonstrated that X-ray free-electron lasers (XFELs), which can produce X-ray pulses with sufficient intensity to drive nonlinear interactions, can also generate isolated attosecond pulses~\cite{duris2020tunable, macarthur2019phase, zhang2020experimental, malyzhenkov2020single} and pulse trains~\cite{duris2021controllable, maroju2020attosecond}.
Pump-probe techniques that combine an external laser pulse with attosecond XFELs present serious challenges in the attosecond regime. 
For XFELs based on self-amplified spontaneous emission~\cite{bonifacio1994spectrum}, the XFEL arrival-time jitter is dominated by the RF phase jitter, and is on the order of tens to hundreds of  femtoseconds~\cite{glownia2010time,prat2020compact,kang2017hard,decking2020mhz}. 
To mitigate this effect, the arrival time of the X-rays with respect to an external laser can be reliably measured on a shot-to-shot basis with femtosecond accuracy~\cite{harmand2013achieving, bionta2011spectral}; sub-fs time-sorting has been reported~\cite{hartmann2014sub, maroju2023attosecond} but with relatively long laser and X-ray pulses (tens of femtoseconds and few femtoseconds, respectively). 
Laser pump/X-ray probe experiments can exploit X-ray observables to measure laser-driven dynamics, but in all such reported experiments the resolution is limited to tens of femtoseconds by either the pulse duration or the arrival time uncertainty. 
Timing stability can be greatly improved using multi-pulse FEL techniques. 
In this case both pump and probe pulses are emitted by the same electron beam, and the pump-probe delay is independent of the arrival time jitter of the electrons~\cite{lutman2013experimental, hara2013two, lutman2016fresh, ferrari2016widely, marinelli2013multicolor, marinelli2015high}.

In this work, we demonstrate the generation, diagnostic, and control of two-colour pulse pairs with attosecond-scale duration and delays at the Linac Coherent Light Source~\cite{emma2010first}. 
The delay between the two pulses can be controlled with hundreds of attosecond accuracy and their relative timing jitter can be estimated to be less than $270~$as.
We also report the application of this novel technique to a time-dependent pump-probe measurement of core-ionised dynamics in the para-aminophenol molecule.
These results extend the X-ray-pump/X-ray-probe techniques to attosecond timescales.
 
\section{Demonstration of Two-colour Attosecond Pulse Pairs}

\begin{figure}[H]
\centering
\includegraphics[width=0.85\linewidth]{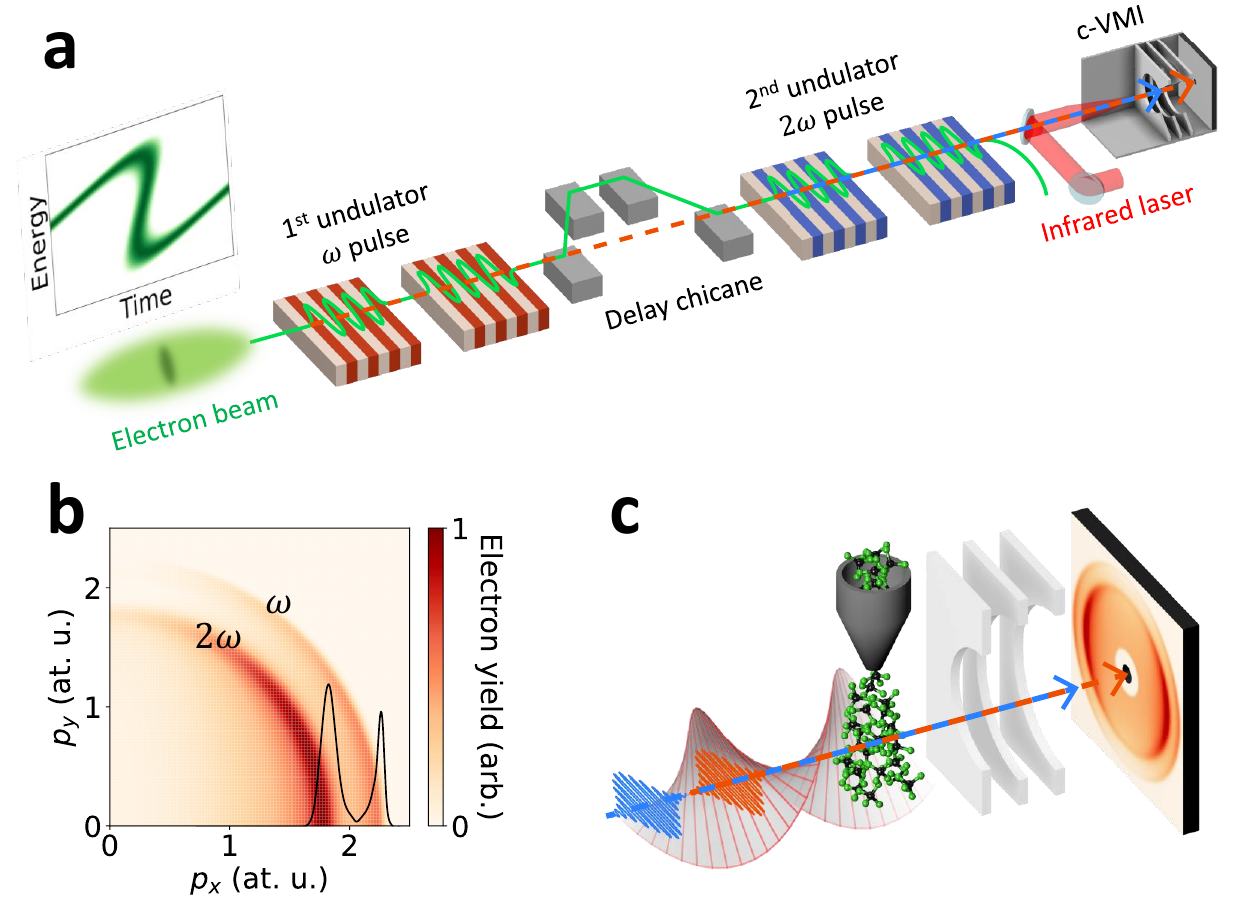}
\caption{Experimental configuration for generating~(a) and diagnosing~(c) sub-femtosecond pulse pairs with an XFEL. 
In panel~\textbf{a}, a modulated electron bunch with a high-current spike lases in a first set of magnetic undulators with resonant wavelength~$370$~eV~(orange in the figure). 
A magnetic chicane can be used to delay the electron bunch with respect to this first pulse before lasing in the second undulator section, with a resonant wavelength that is half that of the initial undulator, $740$~eV~(blue in the figure). 
In panel~\textbf{c}, X-ray pulses ionise the target sample~(CF$_4$) and the photoelectrons are collected with a co-axial velocity map imaging spectrometer~(see text). 
Panel~\textbf{b} shows the measured two-dimensional projection of the photoelectron momentum distribution recorded by the c-VMI in the absence of the streaking field.
The $p_x$ is chosen to lie along the X-ray polarisation direction. 
The black line shows the electron momentum distribution retrieved from the inverse Abel transform of the data.
}
\label{fig:Split_Undulator_and_Streaking_Measurment_Setup}
\end{figure}

Figure~\ref{fig:Split_Undulator_and_Streaking_Measurment_Setup}~(a) shows a schematic representation of the experiment. 
An enhanced SASE~(ESASE)\cite{zholents2005method} method is used to produce a short current spike by shaping the photocathode laser pulse~\cite{zhang2020experimental}. 
This high current spike is capable of generating sub-femtosecond X-ray pulses. 
The LCLS-II soft X-ray undulator beamline is divided into two halves, each tuned to a different resonant wavelength. 
Both undulators are divided in individually tunable modules (see Methods). 
The shaped electron bunch emits a single-spike soft X-ray pulse from each half of the undulator, and the relative delay is controlled with a magnetic chicane. 
The minimum delay achievable with a split undulator is limited to a few femtoseconds by the relative slippage of the electrons with respect to the first pulse~\cite{lutman2013experimental}. 
In order to access sub-femtosecond delays, the second pulse must be tuned to a harmonic of the first. 
In this case, the first pulse introduces strong harmonic microbunching in the electron beam, which seeds the emission of the second pulse. 
Therefore, the second pulse can saturate in a short undulator length, minimising the relative slippage of the two pulses. 
In our demonstration, we up-convert to the second harmonic, but, in principle, this scheme applies to higher harmonics as well.

The delay between the two pulses is measured with attosecond angular streaking~\cite{eckle2008attosecond,hartmann2018attosecond, duris2020tunable,li2018co}. 
Figure~\ref{fig:Split_Undulator_and_Streaking_Measurment_Setup}~(c) illustrates our experimental setup.
The co-propagating $\omega/2\omega$ pulse pairs are incident on a molecular target (tetrafluoromethane, CF$_4$) and drive x-ray ionisation in the presence of a circularly polarised, 1.3~$\mu$m laser field. 
We record the transverse momentum distribution~(with respect to the propagation axis of the X-rays) of the ionised electrons using a co-axial velocity map imaging spectrometer~(c-VMI)~\cite{li2018co, walter2022time}.
The effect of the laser field is to shift~(or streak) the measured electron momentum distribution in the opposite direction of the laser vector potential at the time of X-ray ionisation~\cite{kitzler2002quantum,itatani2002attosecond,hartmann2018attosecond}:
\begin{equation}
    \vec{p}(t\to\infty) = \vec{p}_0+e\vec{A}(t_0), \label{eq:semi_classical_streaking_model}
\end{equation}
where $\vec{p}(t\to\infty)$ is the momentum of the electron~(in atomic units) measured at the detector, 
$\vec{p}_0$ is the momentum of the electron in the absence of the laser field, $\vec{A}(t_0)=-\int_{-\infty}^{t_0}\vec{\mathcal{E}}_L(t^{\prime}) dt^{\prime}$ is the vector potential of the laser field $\vec{\mathcal{E}}_L$ at the time of ionisation $t_0$, and $e$ is the electron charge, taken as $e=-1$ in atomic units.
The streaking interaction therefore establishes a mapping from angle to time: the direction of the transverse momentum shift determines the time of arrival of the x-ray pulses within a laser period. 

We measure the delay by recording the difference in the angular shift of the photoelectrons produced by each pulse.
The angle, $\Delta \phi$, between the momentum shift of the two photoelectron features relates to the delay $\Delta \tau$ between the two pulses according to the relationship:
\begin{equation}
    \Delta \tau = \frac{\Delta \phi}{2\pi}\times T_L, \label{eq:streaking_angle_difference_and_delay}
\end{equation}
where $T_L$ is the period of the streaking laser~($T_L = 4.3~$fs in our experiment).
The two photoelectron features are shown in Fig.~\ref{fig:Split_Undulator_and_Streaking_Measurment_Setup}~(b): $68$~eV electrons from ionization from the carbon~$K$-shell by the $\omega$~($370$~eV) pulse and $45$~eV electrons from ionization from the fluorine~$K$-shell by the $2\omega$~($740$~eV) pulse.

\begin{figure}[H]
\includegraphics[width=1\linewidth]{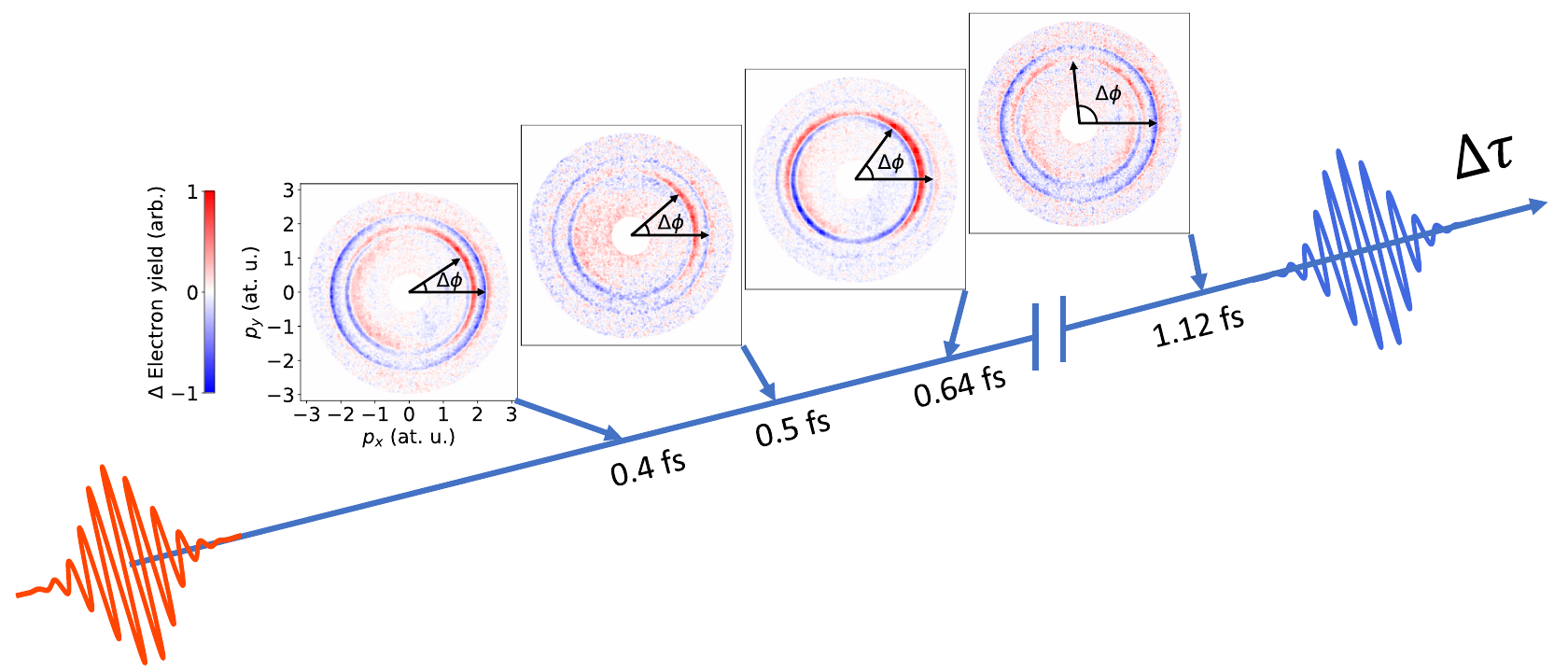}
\caption{
Differential measurements of two-dimensional projections of the photoelectron momentum distributions as a function of average time-delay $\Delta \tau$ between $\omega/2\omega$ pulses. 
Each delay is obtained using different undulator and chicane configurations. 
From left to right: Chicane off with 2/3/4 undulator modules for $2\omega$ pulses, and chicane on with 4 undulator modules for $2\omega$ pulses. 
We have subtracted backgrounds recorded in the absence of the streaking laser in all insets. 
The difference $\Delta \phi$ between two streaking directions is related to the average time-delay $\Delta \tau$ between $\omega/2\omega$ pulse pairs in each undulator beamline configuration.}
\label{fig:Delay_Analysis}
\end{figure}

Figure~\ref{fig:Delay_Analysis} shows the differential transverse momentum distribution (with and without the IR laser) for different $\omega/2\omega$~delays.
To extract the average delay between the pulses we employ a correlation-based method that exploits the large shot-to-shot variation in the laser/X-ray arrival time in our experiment~($\sim500$~fs~\cite{glownia_time-resolved_2010}). 
This method was benchmarked by comparison with a single-shot analysis for a subset of the delays~(see Sec.~S2.2 in Supplementary Information).

\begin{figure}[H]
\centering
\includegraphics[width=0.85\linewidth]{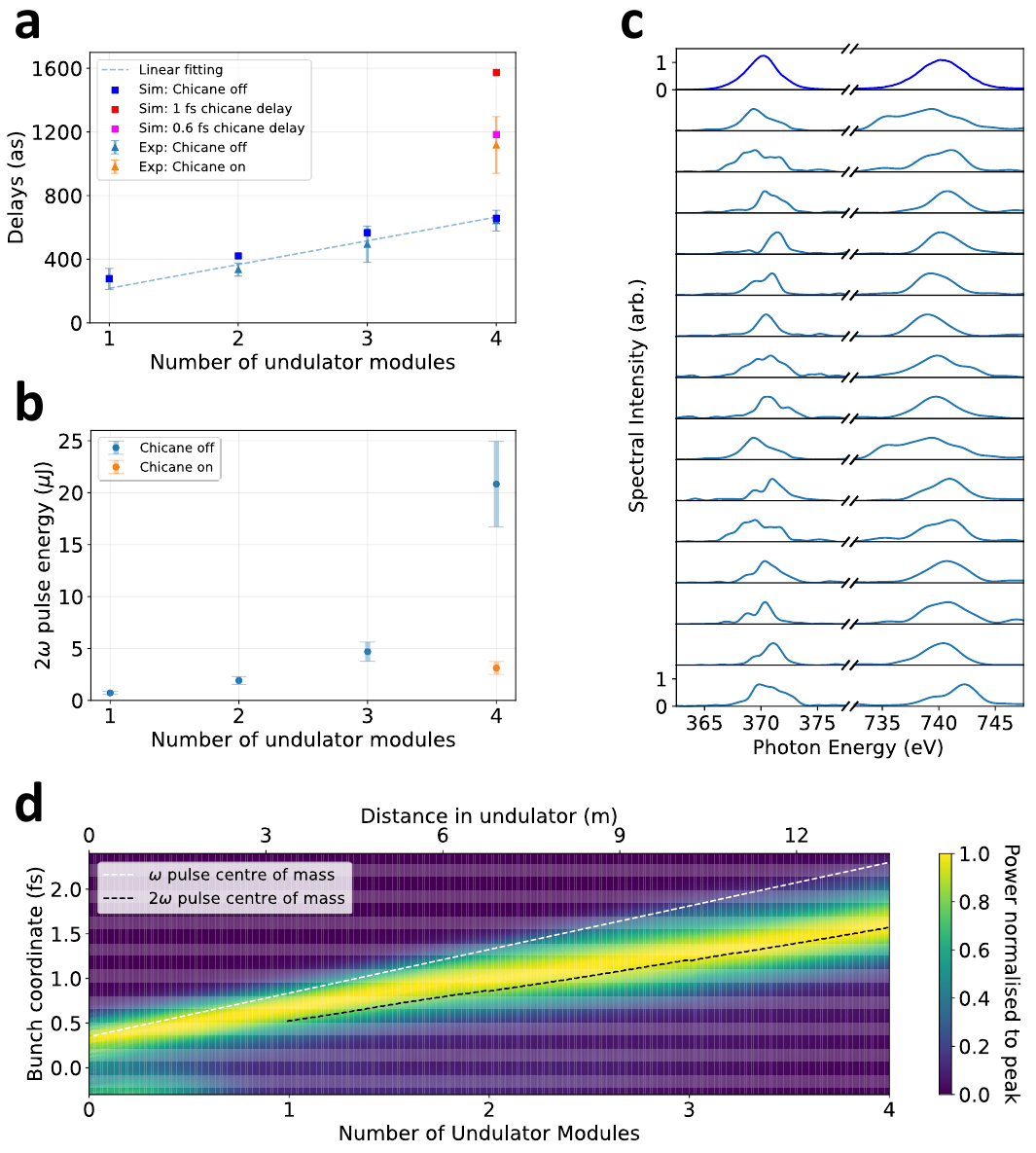}
\caption{\textbf{a}, Measured and calculated delays between $\omega/2\omega$ pulse pairs at different beamline configurations. 
\textbf{b}, Measured average $2\omega$ pulse energy at different beamline configurations. 
The error bar shows the uncertainty in calibrating the average $2\omega$ pulse energy from the average spectrum. 
\textbf{c}, Measured spectra of $\omega/2\omega$ pulse pairs with no chicane delay and 3 undulator modules for $2\omega$ pulses. 
The top panel in \textbf{c} shows the average two-colour spectrum without smoothing. 
The bottom 15 panels show 15 single-shot spectra smoothed with a Gaussian kernel $\sigma$ of 5 spectral pixels (0.35~eV for $\omega$ spectra, 0.85~eV for $2\omega$ spectra). 
\textbf{d}, The start-to-end simulation of the average $\omega$/$2\omega$ pulse pair (averaged over 100 shots) in the 2nd undulator section with no chicane delay. 
The power of the $2\omega$ pulse at each location in $z$ is normalised by the maximum at that location. 
The white (black) dashed line shows the centre of mass of the $\omega$~($2\omega$) pulse as a function of the electron beam's travel distance in undulator. 
The forward propagation of XFEL pulses in quadrupole magnets and free-space drifts with respect to the bunch coordinate has been removed in panel \textbf{d}.
}
\label{fig:Meaured_Delay_and_Pulse_Energy}
\end{figure}

Figure~\ref{fig:Meaured_Delay_and_Pulse_Energy}~(a) shows the measured average delay between the $\omega/2\omega$ pulse pairs for different undulator configurations.
From the correlation method it is also possible to infer an upper limit for the arrival time-jitter between the two pulses, which we estimate to be $270~$as (see Sec.~S2.1.2 in Supplementary Information).

To control the delay in the sub-fs regime we exploit the relative slippage of the two pulses and vary the number of undulator modules used in the second stage. 
The delay between $\omega/2\omega$ pulses increases roughly linearly in the second stage at a rate of $130 \pm 19$~as per undulator module~(87~periods/module).
This approach offers control of the delay up to roughly $0.7~$fs. 
The delay can be further increased using a magnetic chicane to increase the path length of the electron bunch, and thus increase the delay between the pulses, as shown in the data point labeled ``Chicane on".

\section{Discussion and Interpretation of Angular Streaking Measurements}

The dependence of the temporal separation of the pulses on the number of undulator modules is due to a mismatch in the group velocity between the resonant $2\omega$ pulse and the non-resonant $\omega$ pulse. 
Figure~\ref{fig:Meaured_Delay_and_Pulse_Energy}~(d) shows the results of a start-to-end simulation of the experiment. 
The radiation power of the probe is plotted as a function of the undulator length and the bunch coordinate.  
The mismatch in the group velocities between the $\omega$ and $2\omega$ pulses can be clearly seen in the divergence between the centre of mass of the $\omega$ pulse~(dotted white line) and the $2\omega$ pulse (dotted black line).

To interpret our experimental results, we parameterise the FEL group velocity as follows:
\begin{equation}
    v_g = v_b \left(1+ \alpha\frac{\lambda_r}{\lambda_u}\right),
    \label{eqn:vg}
\end{equation}
where $v_g$ and $v_b$ are the group velocity and the beam velocity (both close to the speed of light), $\lambda_r$ is the radiation wavelength, $\lambda_u$ is the period of the magnetic undulator and $\alpha$ is a positive number, typically between 0 and 1, which accounts for the slippage effects during the lasing process~(a value of $\alpha = 1$ corresponds to propagation at the speed of light, or $v_g = c$). 
In the one-dimensional limit, a free-electron laser pulse in the exponential gain regime has a group velocity given by the electron beam velocity plus one third of the slippage rate~\cite{kim1997temporal, pellegrini2016physics, huang2007review}, corresponding to $\alpha = 1/3$.
For a pulse generated by a high-current ESASE spike, the radiation remains close to the current spike because the gain in the high-current region is larger than in the rest of the bunch, an effective temporal gating effect. 
This limits the group velocity to the average velocity of the electrons~\cite{baxevanis2018time}, thus $\alpha = 0$.
In the nonlinear regime, the group velocity is typically higher than in the exponential gain regime and can exhibit exotic behaviours such as superluminal propagation~($\alpha > 1$)~\cite{yang2020postsaturation, watanabe2007experimental, hajima2001analyses}.

Our experiment is the first measurement of the group velocity of an XFEL pulse. 
By comparing the change of delay between $\omega/2\omega$ pulse pairs with the slippage per undulator module, we determine the group velocity of the $2\omega$ pulse as,
\begin{equation}
    v_{g,2\omega} = v_b\left( 1+(0.73\pm0.04)
    \frac{\lambda_r}{\lambda_u}\right) \label{eq:group_velocity_2w}.
\end{equation}
We measure the group velocity of the 2$\omega$ pulse to be faster than in the exponential gain regime, because the pulse is generated close to the saturation point. 
By comparing the experimental data with the simulation, we find that the $2\omega$ pulse is quickly amplified to above gigawatt-level and propagates ahead of the ESASE current spike, exhibiting behaviour consistent with the early onset of superradiant propagation~\cite{bonifacio1990superradiant, mirian2021generation}. 
Our observations are consistent with the start-to-end simulations plotted in Fig.~\ref{fig:Meaured_Delay_and_Pulse_Energy}(a).

To further increase the delay, one can use the magnetic chicane up to a delay value of tens of femtoseconds. 
We measured the time-delay for a nominal magnetic chicane value of $1$~fs (``Chicane On" data point).  
The discrepancy between the simulated chicane delay and the nominal experimental chicane delay is likely due to the magnetic hysteresis of the chicane dipoles operating far from saturation. 
In future experiments, one could minimise this effect by using one or more off-resonance magnetic undulators to introduce such short delays.
For this small value of the chicane delay, the harmonic microbunching is not entirely suppressed and the delay remains consistent with a $2\omega$ group velocity larger than the beam velocity.

Figure~\ref{fig:Meaured_Delay_and_Pulse_Energy}~(b) also shows the average pulse energy of the $2\omega$ pulse as a function of the undulator length. 
The pulse energy increases with the number of undulator modules and decreases as the chicane delay is turned on, due to the suppression of microbunching induced by chicane dispersion~\cite{pellegrini2016physics}.
This delay-dependent pulse energy can be easily measured on a shot-to-shot basis using, for example, a grating spectrometer. 
Figure~\ref{fig:Meaured_Delay_and_Pulse_Energy}~(c) shows an example of such single-shot characterisation of these pulses, showing simultaneous $\omega/2\omega$ spectra measured with a Variable Line Spacing grating spectrometer~\cite{obaid2018lcls, hettrick1988resolving,chuang2017modular}. 
The FWHM bandwidths are $2.6 \pm 0.8~$eV and $3.4 \pm 0.6~$eV for the $\omega$ and $2\omega$ pulses, respectively.

\section{Demonstration of a Time-Resolved Experiment}

To demonstrate the performance of these two-colour pulse pairs, we carry out a simple pump-probe experiment to measure the dynamics of core-ionisation in para-aminophenol molecules.
We tune the lower photon energy pulse to $\hbar\omega=295.5$~eV, to prepare the system in a non-equilibrium state or ``pump'' the system. 
At this photon energy, the pump pulse is most likely to remove electrons from the carbon $K$-shell of the para-aminophenol molecule, thus creating single core hole states~(SCH). 
Based on the measured fluence of the pump pulse we estimate that nearly every molecule in the focal volume is ionised. 
The binding energy~(BE) of the carbon $K$-shell electrons in para-aminophenol is between $289.71 -291.41$~eV, depending on the relative position of the core-level vacancy to the nitrogen or oxygen atomic-sites~\cite{zhaunerchyk2015disentangling}. 
Ionisation by the pump pulse produces slow photoelectrons with $\sim5$~eV of kinetic energy. 
We probe the dynamics induced by the pump pulse with subsequent ionisation by the probe pulse tuned to $2\hbar\omega=591$~eV.

\begin{figure}[H]
    \centering
    \includegraphics[width=0.7\textwidth]{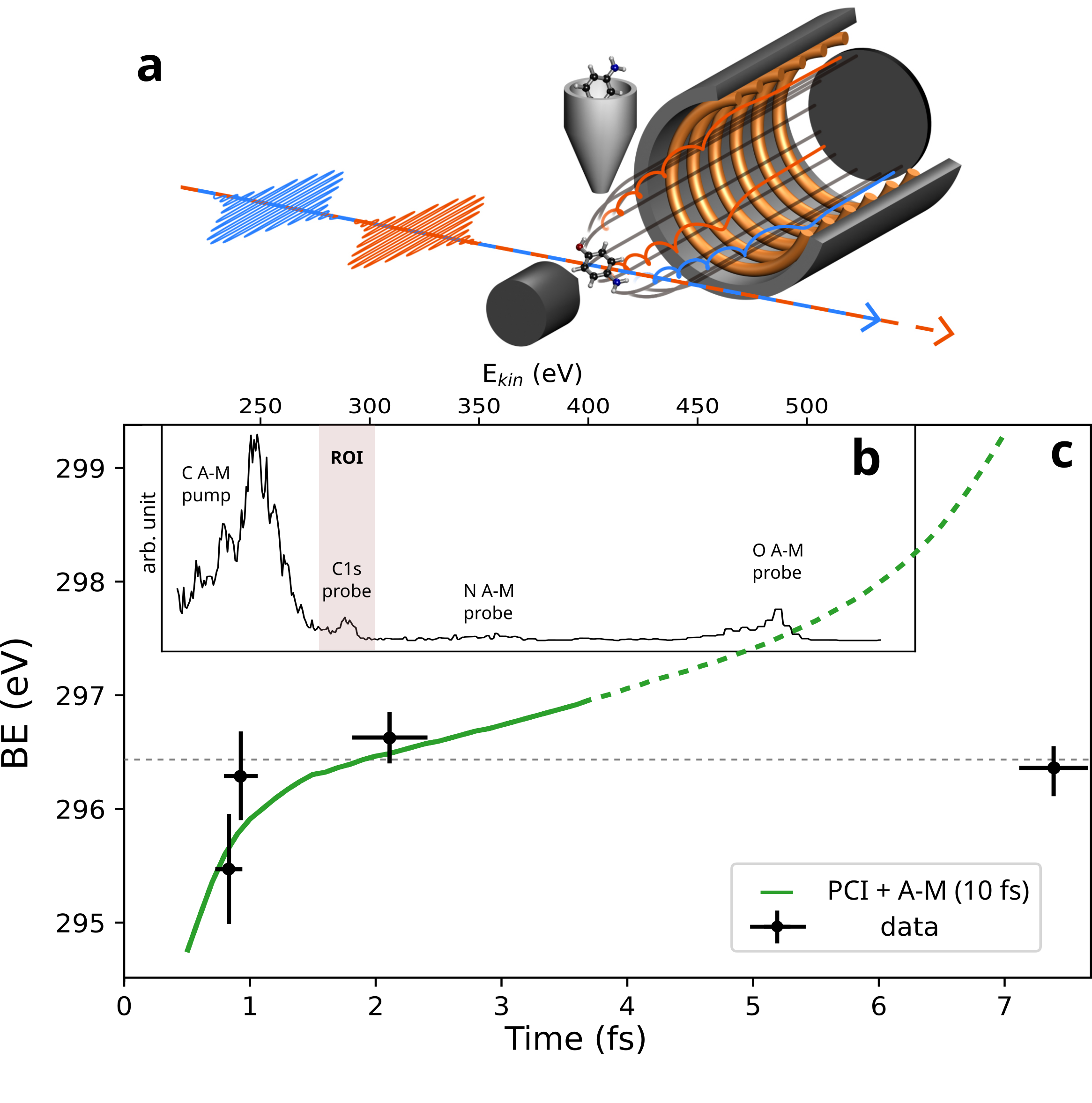}
    \caption{\textbf{(a)} Schematic of the experimental setup, electrons ionised by the probe pulse are collected with a magnetic bottle electron time-of-flight spectrometer.
    \textbf{(b)}~Raw photoelectron spectrum showing Auger-Meitner (A-M) features for carbon, nitrogen and oxygen sites in aminophenol molecule. 
    Here the pump photon energy is $295.5$~eV (responsible for the carbon A-M) and the probe is $591$~eV. 
    The Region of Interest (ROI) is depicted in red and showing the carbon $1s$ peak produced by the probe. 
    \textbf{(c)} Black dots show the binding energies (BE) obtained by means of covariance analysis vs. pump-probe delay. 
    The grey dashed line shows the static BE of a 1s electron in the core-ionised cation (referred to as BE$_{DCH}$ in the Supplementary Information). 
    The green curve shows a model that includes (i) classical simulation for PCI~\cite{russek_mehlhorn_1986} considering a slow pump electron of $10~$eV of kinetic energy and a fast probe electron with $300~$eV  kinetic energy and (ii) the A-M decay of a single core hole state with a $10~$fs timescale.  
    }
    \label{fig:mbes_ana}
\end{figure}

Figure~\ref{fig:mbes_ana}~(a) shows a schematic view of the experimental setup. 
Photoelectrons ionised by pump and probe pulses are collected with a magnetic bottle electron time-of-flight spectrometer~(MBES). 
Figure~\ref{fig:mbes_ana}~(b) shows the electron kinetic energy spectrum recorded with a delay of $2.1$~fs between pump and probe pulses. 
We analyse the photoelectron feature corresponding to the ionisation of the carbon $K$-shell ($C1s$) by the probe pulse in the kinetic energy window between $285-300$~eV~(region of interest, ROI). 
We extract the binding energy of this feature, defined as $BE= 2\hbar\omega - E_{kin}$, and plot the result as a function of the delay between the two X-ray pulses, in Figure~\ref{fig:mbes_ana}~(c). 
We find that the measured binding energy of this photoelectron feature increases by $\sim1.5$~eV in the first few femtoseconds.

The fast photoelectrons~($E_{kin}\simeq300$~eV) produced by the probe pulse will rapidly overtake the slower photoelectrons produced by the pump pulse~($E_{kin} \simeq 5$~eV).
As a consequence of this interaction, the fast electrons gain energy because they experience additional screening from the molecular ion. 
This kinetic energy shift decreases as the spatial position where the fast electron overtakes the slow electron moves further away from the ion, and thus the shift vanishes for long pump-probe delays~\cite{artemyev2019controlling}.
Such shifts have been referred to as a post collision interaction~(PCI) effect, and can be modelled with classical propagation in a Coulomb potential. 
In Figure~\ref{fig:mbes_ana}~(c) we compare our data to this model~(see Sec.~\ref{Subsection:SI_PCI} of Supplementary Information for more details of the model). 
The green curve shows the result of a calculation that includes both PCI and the effect of a finite core-hole lifetime, which is consistent with the short-time behaviour of the data. 
For the longer time delay~($7$~fs) other effects should be taken into account, for instance dissociation of a proton~(H$^+$) could happen on this time scale. 
This would lead to a shift of the binding energy to lower values, closer to the core-ionized cation binding energy~(dashed line in Fig.~\ref{fig:mbes_ana}~(c)).
A detailed understanding of the dynamics in core-ionised pAp is beyond the scope of this work. 
However, the measured time-dependent shift demonstrates the ability to perform pump-probe measurement with sub-femtosecond resolution using an X-ray free-electron laser.

\section{Conclusion}

We have demonstrated the generation of gigawatt-level, two-colour~($\omega/2\omega$), attosecond soft X-ray pulse pairs with controllable synchronised delays.
This is achieved by employing the split undulator method in a harmonic configuration, where the harmonic microbunching induced by the first pulse can seed the second one. 
We directly measure the average delay between the $\omega/2\omega$ pulse pair using an angular streaking technique for each beamline configuration.
The angular streaking measurement shows that sub-femtosecond delays can be controlled in steps of $(27 \pm4)\%$ of the slippage rate in the second undulator section ($130\pm 19$~as at $2\hbar\omega = 740$~eV) by changing the number of undulator modules in the second stage. 
This is a result of the small deviation of the group velocity of the second pulse from the speed of light.
Longer delays can be accessed using a magnetic chicane to delay the electron bunch with respect to the fundamental pulse.

The FEL group velocity has so far been a point of theoretical investigation \cite{yang2020postsaturation,curry2016thz,bartolini1991theoretical}, relevant to long-wavelength experiments where the radiation slippage strongly affects the FEL dynamics~\cite{fisher2022single, hajima2001analyses}. 
However, the level of experimental sophistication required for attosecond X-ray science makes it a key element in understanding and executing pump-probe measurements at the sub-femtosecond scale.
This experiment has produced the first measurement of the group velocity of an XFEL and it provides a new benchmark of time-dependent and nonlinear FEL theory.

Our two-colour attosecond pump-probe setup can be scaled to the next generation of XFEL facilities with megahertz repetition rates, since it relies on passive beam-shaping methods.
This method can be applied to any wavelengths of $\omega/2\omega$ by simply adjusting the undulator parameter~$K$. 
This enables probing of different absorption edges, as long as the second pulse is a harmonic of the first.
For example, $\omega/2\omega$ pulse pairs can be exploited in non-resonant-pump/resonant-probe experiments of ionic charge motion~\cite{barillot2021correlation}, or resonant-pump/non-resonant probing of core-excited state dynamics~\cite{al2022observation}.
We have demonstrated the application of this two-colour technique to a pump-probe experiment with sub-femtosecond resolution. 
The pulse energies achieved with the $\omega$-pulse enable nonlinear excitation techniques, such as impulsive stimulated X-ray Raman scattering~\cite{o2020electronic,mukamel2013multidimensional, keefer2023ultrafast}, which can be probed resonantly with the $2\omega$-pulse.

\section{Acknowledgements}
Use of the Linac Coherent Light Source (LCLS), SLAC National Accelerator Laboratory, is supported by the U.S. Department of Energy, Office of Science, Office of Basic Energy Sciences under Contract No. DE-AC02-76SF00515.
A.M., D.C., P.F., and Z.G. acknowledge support from the Accelerator and Detector Research Program of the Department of Energy, Basic Energy Sciences division. 
Z.G. also acknowledges support from Robert Siemann Fellowship of Stanford University.
The effort from T.D.D., J.W., M.F.K, T.W., and J.P.C. is supported by DOE, BES, Chemical Sciences, Geosciences, and Biosciences Division (CSGB). 
L.F.D., D.T. and G.M. acknowledge support from U.S. Department of Energy, Office of Science, Basic Energy Sciences under Awards DE-FG02-04ER15614 and DE-SC0012462.
V.A. and M.R. acknowledge support from the UK’s Engineering and Physical Science Research Council (EPSRC) through the grant “Quantum entanglement in attosecond ionization”, grant number EP/V009192/1.
O.A and J.P.M. were supported by the U.K. Engineering and Physical Sciences Research Council Grants No. EP/R019509/1, EP/X026094/1 and No. EP/T006943/1.
Th.W., D.S.S. and O.G. are supported by the U.S. DOE Office of Basic Energy Sciences, Division of Chemical Sciences, Biosciences and Geosciences under the contract No. DE-AC02-05CH11231
L.Y. and G.D were supported by the US Department of Energy, Office of Science, Basic Energy Sciences, Chemical Sciences, Geosciences, and Biosciences Division under award DEAC02-06CH11357.
D.R., A.R., and E.W. are supported by the same funding agency under grant No. DE-FG02-86ER13491.
S.B. and N.B. are supported by the U.S. DOE Office of Basic Energy Sciences, Division of Chemical Sciences, Biosciences and Geosciences under the contract No. DE-SC0012376.
A.M. would like to acknowledge Luca Giannessi and Pietro Musumeci for useful discussions and suggestions. L.I. would like to acknowledge helpful discussion with Sang-Kil Son.
\newpage

\section{Methods}
\subsection{XFEL Setup}
The temporal profile of the photocathode laser was shaped to produce a high current spike in the electron bunch~\cite{zhang2020experimental}. 
Attosecond $\omega/2\omega$ pulse pairs with controllable sub-femtosecond delays were generated by lasing this high current spike~\cite{duris2020tunable} in the split-undulator~\cite{lutman2013experimental} configuration with the harmonic configuration. 
The main parameters used in the experiment are listed in Table~\ref{tab_s:Machine_Parameters}. 

\begin{table}[h]
\centering
\begin{tabular}{l cc } 
 \hline
 \hline
 Parameter & Value \\ 
 \hline
 Beam energy & $5~$GeV \\
 Bunch charge & $140~$pC \\ 
 BC1 current & $140~$A \\
 BC2 current & $1700~$A/$2700~$A \\
 XLEAP wiggler period & $35~$cm \\
 XLEAP wiggler gap & $25~$mm\\
 XLEAP wiggler $K_w$ & 24.7 \\
 \hline
 \hline
\end{tabular}
\caption{\label{tab_s:Machine_Parameters}Machine parameters for generating 370~eV/740~eV XFEL pulses using the split-undulator mode in the angular streaking experiment.}
\end{table}

\begin{figure}[!ht]
\includegraphics[width=\linewidth]{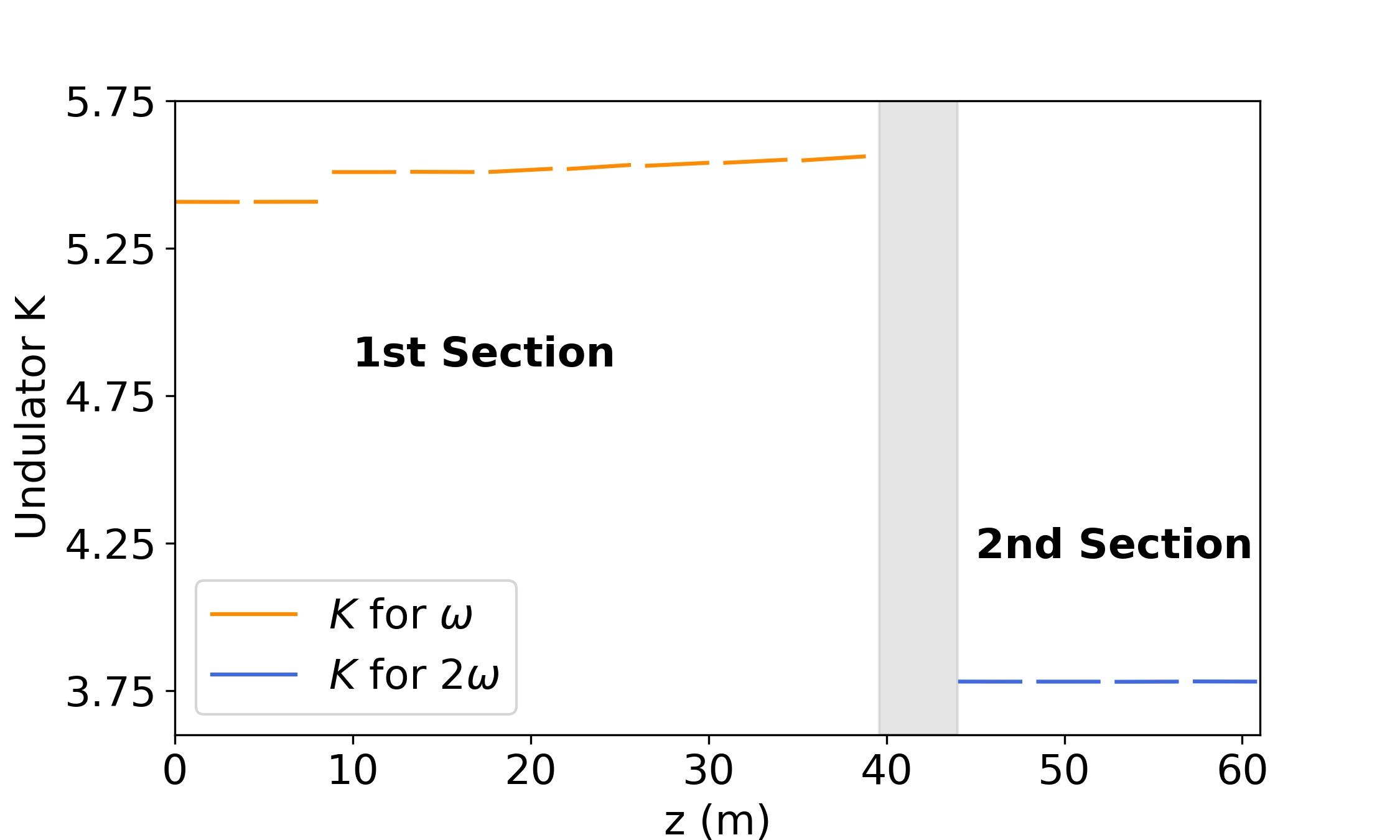}
\caption{ 
Values of undulator $K$ were set to be on resonance with $\omega$ and $2\omega$ pulses in the first and second undulator section, respectively. 
The transparent grey area shows the location of the magnetic chicane.}
\label{fig_s:omega_2omega_modes}
\end{figure}

The FEL resonance wavelength $\lambda_r$ is determined by~\cite{bonifacio1984collective},
\begin{equation}
    \lambda_r = \frac{\lambda_u}{2\gamma^2} \left(1+\frac{K^2}{2}\right),
\end{equation}
where $\lambda_u = 3.9~$cm is the length of one undulator period, $\gamma$ is the Lorentz factor of the electron beam, and $K$ is the normalised undulator strength parameter. 
The maximum possible value for the undulator parameter $K$ is $\sim5.77$. 
The two undulator sections are composed of undulator modules of 87 periods each. 
The values of $K$ for each module used to generate $370~$eV/$740~$eV attosecond XFEL pulses in the angular streaking experiment are shown in Fig.~\ref{fig_s:omega_2omega_modes}. 
In the split-undulator mode, the energy spread induced in the electron beam by the first pulse is harmful to the generation of the second pulse. 
Therefore, to avoid saturation, the parameter $K$ in the first two undulator modules was slightly detuned to reduce the $\omega$ pulse energy. 
Then, the undulator taper from the fifth to the ninth undulator modules matched the energy chirp to generate attosecond $370~$eV pulses and to suppress background radiation outside the high current spike. 
After the magnetic chicane, the undulator $K$ in the second section was decreased from $\sim 5.6$ to $\sim 3.8$ to generate attosecond $740~$eV pulses. 
The number of undulators in the second section can be controlled by varying the gap of individual undulator sections in the second stage, so that only the desired number of undulators is resonant with the second pulse.
A similar setup was used in the experiments in Fig.~\ref{fig:mbes_ana} to generate $295.5~$eV/$591~$eV attosecond XFEL pulses, with the electron beam energy at $4~$GeV and different undulator parameters $K$.

\subsection{Analysis of Angular Streaking Data}

We used two methods to extract the time-delay between $\omega/2\omega$ pulses. 
In the first method, correlation analysis was applied to extract the average delay and the delay jitter. 
Each single-shot 2D momentum distribution was polar-rebinned with respect to the centre of the unstreaked photoemission features with the IR laser intentionally mistimed. 
The electron yield was integrated over the momentum regions corresponding to the high-energy flank of each photoemission features to give two 1-D traces named $X(\theta_{\omega})$ and $Y(\theta_{2\omega})$. 
These two traces measure the change in electron yield due to streaking, as a function of the detector angle, for the carbon ($X(\theta_{\omega})$) and fluorine ($Y(\theta_{2\omega})$) photoemission features, respectively.
A 2D map was generated by calculating the correlation coefficient between anti-symmetric parts of two traces. 
The average differential angle $\Delta\phi$ can be measured as the deviation of the strongest correlation region (i.e. the shift of the brightest feature) from the diagonal line of the 2D correlation map. 
The average differential angle $\Delta\phi$ is determined by matching the experimental correlation maps with the simulations. 
The delay jitter was estimated from the amplitude of the 2D correlation map. 
In the second method, the momentum shift of the fluorine $K$-shell photoemission feature (the lower momentum photoelectron feature) was used as a single-shot measurement of the direction $\phi_{2\omega}$ of the streaking laser vector potential $\vec{A}$ at the time of arrival of the probe ($2\omega$) pulse. 
All measured 2D momentum distributions were divided into bins of $\phi_{2\omega}$ and averaged. 
We performed cosine fittings $A\cos(\phi_{2\omega}-\Phi)+B$ of the electron yield in two regions corresponding to high-energy flanks of two photoemission features, the same as in the correlation analysis. 
A global phase shift $\Delta \phi$ was fitted between two photoemission features and used as a measure of the average delay $\Delta \tau = T_{L}\times \Delta \phi/2\pi$ between $\omega/2\omega$ pulses. 
More details on two methods of delay analysis are given in the Sec.~S2 of Supplementary Information.

\subsection{Pump-Probe Measurements in aminophenol} 

The measurements were performed at the time-resolved molecular, and optical science (TMO) experimental hutch, where the X-ray pulses were focused using a pair of Kirkpatrick-Baez focusing mirrors~\cite{walter2022time}. 
The para-aminophenol~(pAp) was introduced using an in-vacuum oven heated to $\sim 160~^{\circ}$C.
The focused X-ray beam intercepted the molecular sample in the interaction
region of a two meter~($2$~m) magnetic bottle electron spectrometer (MBES). 
A static $140$~V retardation voltage was applied to the flight tube of the MBES to increase the kinetic energy resolution near the region of interest~(ROI). 
The spectrum of the probe pulse was measured shot-to-shot by a variable-line spaced grating spectrometer~\cite{obaid2018lcls, hettrick1988resolving}. 

To retrieve the binding energy shift analysed in Figure~\ref{fig:mbes_ana} we applied a covariance analysis procedure. 
The initial step was to calculate the covariance matrix between the shot-to-shot spectrum of the probe pulse and the measured photoelectron spectrum in the vicinity of the features ionised by the probe pulse. 
Covariance analysis is a common method employed in the analysis of FEL experiments, since FEL pulses are subject to shot-to-shot fluctuations of the pulse properties (pulse energy, photon energy) and covariance uses these fluctuations to gain resolution. 
At the same time, analysing the covariance of the photoelectron spectra with the probe photon spectra also allows us to isolate the contributions to the photoelectron spectrum from the probe pulse, removing most of the pump contributions.
 
Figure~\ref{fig_s:covariance_analysis}(a) in the Supplementary Information shows the covariance maps over the full photoelectron energy range recorded in the measurement. 
Each column of the figure shows a different delay (increasing from left to right), and for each delay we use between $20~$k to $30~$k shots to compute the covariance maps. 
The red curve shows the averaged probe photon spectra. 
The broad features around $250~$eV, $350~$eV, and $470~$eV of kinetic energy show the A-M electrons from the carbon, nitrogen, and oxygen sites of the pAp molecule, respectively. 
We note that for the two earliest delays, the intensities of the pump and the probe are strongly correlated, which explains why the covariance shows a stronger signal of the carbon A-M feature caused by the pump. 
Around $300~$eV, one can see the dispersive feature corresponding to the carbon $1$s photoemission associated with the probe pulse. 
We do not observe a feature corresponding to ground state un-pumped molecules, expected $5~$eV higher in the photoelectron kinetic energy spectrum. 
This is consistent with the fluences used for which we expect complete saturation of the $K$-shell absorption. 
Fig.~\ref{fig_s:covariance_analysis}(b) shows a zoom over this carbon $K$-shell photoelectron feature. 
To follow along the photoemission feature, we mark its maximum for each photon energy and extract the corresponding peak position in kinetic energy. 
Fig.~\ref{fig_s:covariance_analysis}(c) enforces the dispersion law $2\hbar\omega$ = $E_{kin}$ + BE. 
By fitting the offset to the curve, we retrieve the binding energy values per time-delay which we report in Fig.\ref{fig:mbes_ana}. 
The error on the measured binding energy is estimated using the bootstrapping method over 150 redistribution of $20~$k to $30~$k shots per time-delay.

\bibliographystyle{unsrt}
\bibliography{refs, referencesJPC}

\end{document}


\maketitle

\tableofcontents

\section{Accelerator Setup and Beam Shaping}\label{sec_s:accelerator_setup_and_beam_shaping}

\begin{figure}[!ht]
\includegraphics[width=\linewidth]{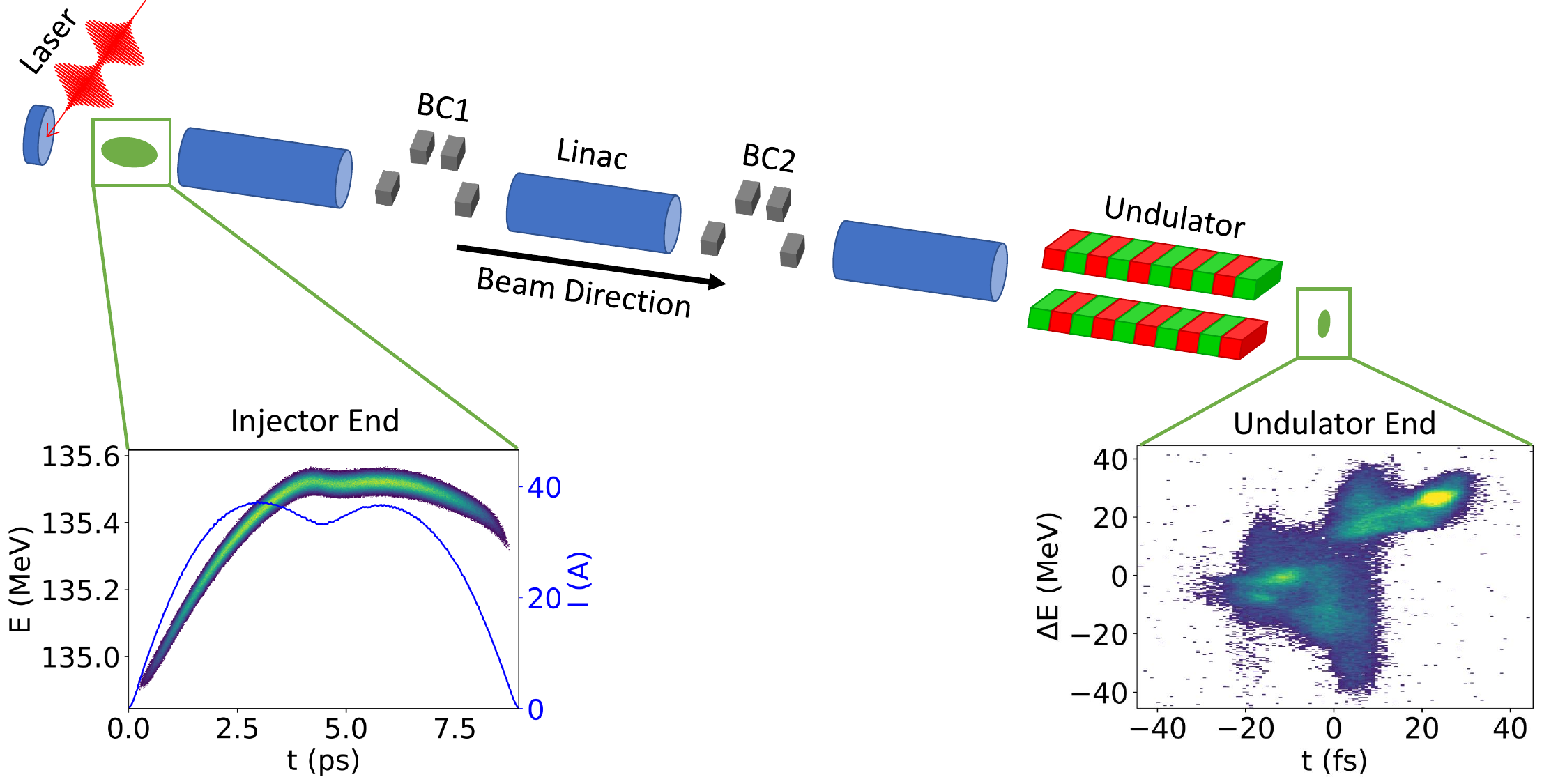}
\caption{\textbf{The accelerator schematic and electron bunch profiles}.
The top panel shows a representation of the LCLS linear accelerator. 
The bottom panels shows the simulated phase space and current profile of the electron bunch at the end of the photo-injector (left) and the measured phase space of the electron bunch after the undulator (right), respectively. 
The bunch head lies to the right.
}
\label{fig_s:Accelerator_Setup_and_Phase_Space}
\end{figure}

Figure~\ref{fig_s:Accelerator_Setup_and_Phase_Space} shows a schematic representation of the accelerator setup for our experiment.
A pulse stacker is used to generate a delayed replica of the photocathode laser. 
The two identical pulses are slightly delayed with respect to each other. 
The resulting laser temporal profile has a notch in the middle of the pulse and
generates an electron bunch with a modulated current profile. 
The current modulation is amplified by the microbunching instability as the bunch is accelerated and compressed in two magnetic chicanes (labeled BC1 and BC2).

The resulting electron bunch has a large current spike with roughly $10~$kA of peak current and $1.5~$fs duration. 
The spike a positive energy chirp (i.e. a positive correlation between longitudinal position and energy) on the order of 1\% of the total energy along the length of the spike. 
The chirp is generated by the longitudinal space-charge force generated by the spike and it is further enhanced using the coherent synchrotron radiation field in the XLEAP magnetic wiggler, placed before the FEL undulator.

Figure~\ref{fig_s:Accelerator_Setup_and_Phase_Space} shows the simulated longitudinal phase space at the end of the injector, as well as the measured longitudinal phase space at the end of the undulator. 
The high current spike cannot be fully resolved and it appears as a large vertical stripe.

\section{Properties of Two-Colour Pulses}

\subsection{Two-Colour Pulse Energies}
As discussed in the main text, a measurement of the pulse energy of both colours is desirable to employ these pulses in a pump/probe experiment. 
Here we describe the method used in this study to characterise the pulse energy of both pulses on a single-shot basis.

We measure the single-shot pulse energy of both pulses using the combined data from a gas monitor detector~(GMD) (which uses the photoionisation of rare gas atoms to infer the pulse energy~\cite{tiedtke2014absolute,heimann2018laser}), a variable line spaced~(VLS) grating based spectrometer~\cite{obaid2018lcls, hettrick1988resolving,chuang2017modular} and the co-axial velocity map imaging (c-VMI) spectrometer used in the streaking measurement.

The energy recorded by the GMD, $E_{\mathrm{GMD}}$, contains contributions from both $\omega$ and $2\omega$ pulses,
\begin{equation}
   E_{\mathrm{GMD}} = E_{\omega} +E_{2\omega}\frac{\sigma_{\omega,\mathrm{Kr}}}{2\sigma_{2\omega,\mathrm{Kr}}}, \label{eq_s:GMD_measurments} 
\end{equation}
where $E_{\omega}$ is the single-shot $\omega$ pulse energy, $E_{2\omega}$ is the single-shot $2\omega$ pulse energies, $\sigma_{\omega, \mathrm{Kr}} = 3.3~$Mb and $\sigma_{2\omega, \mathrm{Kr}}=0.8~$Mb are the absorption cross sections for the $\omega$ pulses ($370~$eV) and $2\omega$ pulses ($740~$eV), respectively, in the atomic krypton target used in the GMD. 

\begin{figure}[!ht]
\includegraphics[width=\linewidth]{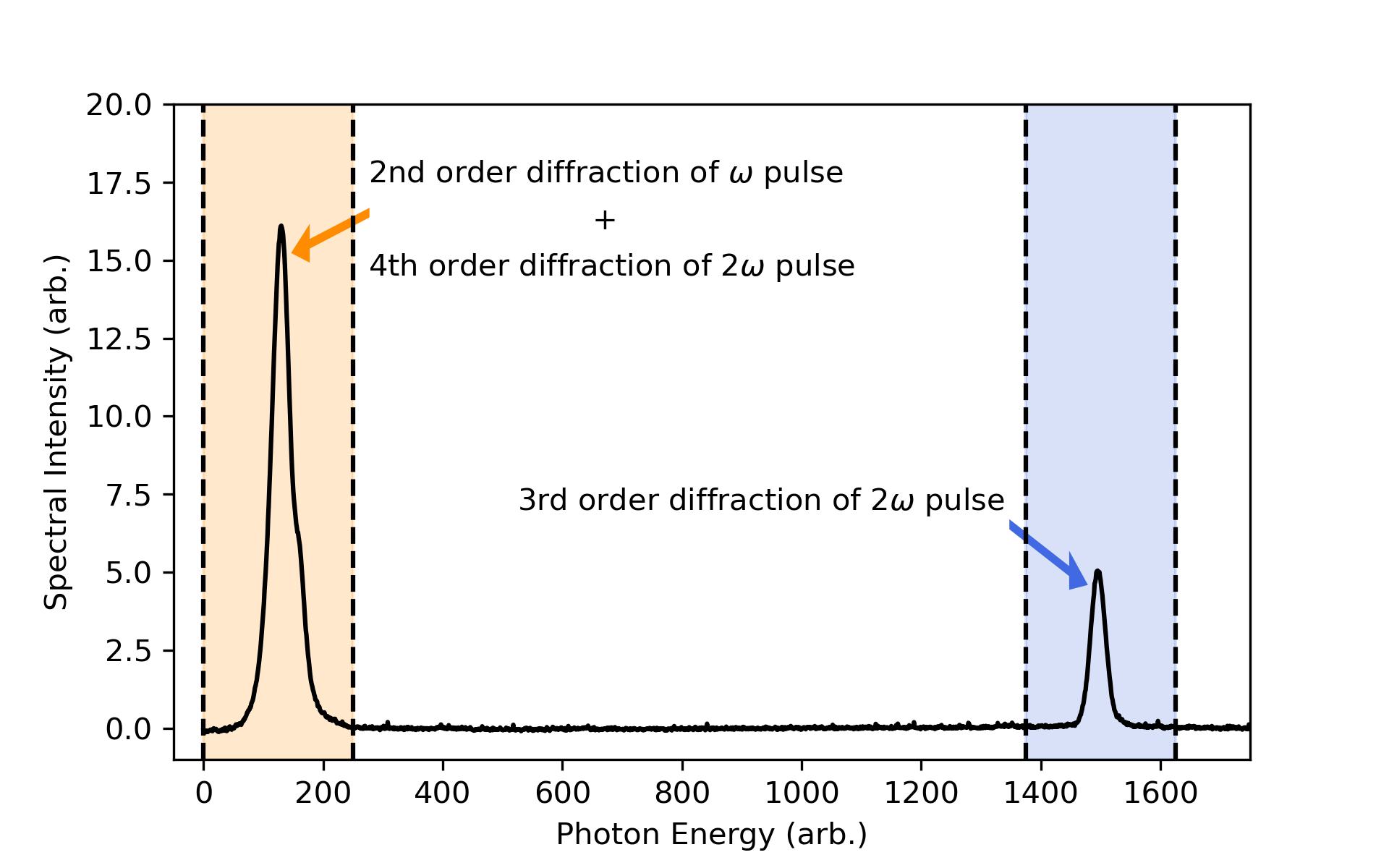}
\caption{
\textbf{Average spectrum measured on VLS}. 
The spectral peak on the left contains the 2nd order diffraction of $\omega$ pulses and the 4th order diffraction of $2\omega$ pulses. The spectral peak on the right is mainly the 3rd order diffraction of $2\omega$ pulses.}
\label{fig_s:average_VLS}
\end{figure}

Similarly, the VLS spectrometer contains information from both pulses because the CCD detector chip is large enough to image different orders of the $\omega$ and $2\omega$ pulses at the same time, as shown in Fig.~\ref{fig_s:average_VLS}.
The $\omega$ pulse spectrum is measured using the second-order diffraction, and it is polluted by the fourth-order diffraction of the $2\omega$ pulse. 
The $2\omega$ pulse is measured independently using the third-order diffraction, which does not overlap with any diffraction order of $\omega$ (see Fig. \ref{fig_s:average_VLS})

The c-VMI measurements~\cite{li2018co}, instead, contain independent information from each pulse but needs to be calibrated to the measured pulse energy.

 
We used the known cross sections for single-photon ionisation in CF4 to extract the average ratio of pulse energies of the $\omega$ and $2\omega$ pulses from the c-VMI data:
\begin{equation}
    \frac{E_{\omega}}{E_{2\omega}} = \frac{N_{e,\omega}}{N_{e,2\omega}} \frac{\sigma_F}{2\sigma_C},
\end{equation}
where $N_{e,\omega}$ and $N_{e,2\omega}$ are the electron yield in carbon and fluorine $K$-shell photoemission features, respectively, in the c-VMI data. 
$\sigma_C = 0.61~$Mb and $\sigma_F = 1.5~$Mb are carbon 1s and fluorine 1s partial cross sections, respectively, in CF$_4$~\cite{truesdale1984core, hitchcock1994bibliography,toffoli2006photoionization}. 
To obtain an absolute calibration, we identify a set of shots with negligible $2\omega$ contribution and use it to cross-calibrate the GMD and the $E_{\omega}$ photoelectron yield in the c-VMI (i.e. $N_{e, \omega}$ in the $\sim68$~eV carbon K-shell photoemission feature).
These two measurements provide an absolute calibration of the average pulse energy of the two colours, which can be used to calibrate the spectral energy density in the spectrometer on a single-shot basis. 
The calibration of the single-shot $2\omega$ pulse energy $E_{2\omega}$, is given by:
\begin{equation}
E_{2\omega} = \alpha_{2\omega} I_{2\omega}, \label{eq_s:pulse_energy_calibration_3}
\end{equation}
where $\alpha_{2\omega} = (1.77\pm 0.35)\times 10^{-2}~\mu$J and $I_{2\omega}$ is the integrated spectrum of the third-order diffraction of $2\omega$ pulses in the transparent blue area in Fig.~\ref{fig_s:average_VLS}.
The single-shot $\omega$ pulse energy $E_{\omega}$ is calculated by removing the contribution of $E_{2\omega}$ from $E_{\mathrm{GMD}}$ in Eq.~\ref{eq_s:GMD_measurments}.

\begin{figure}[!ht]
\includegraphics[width=\linewidth]{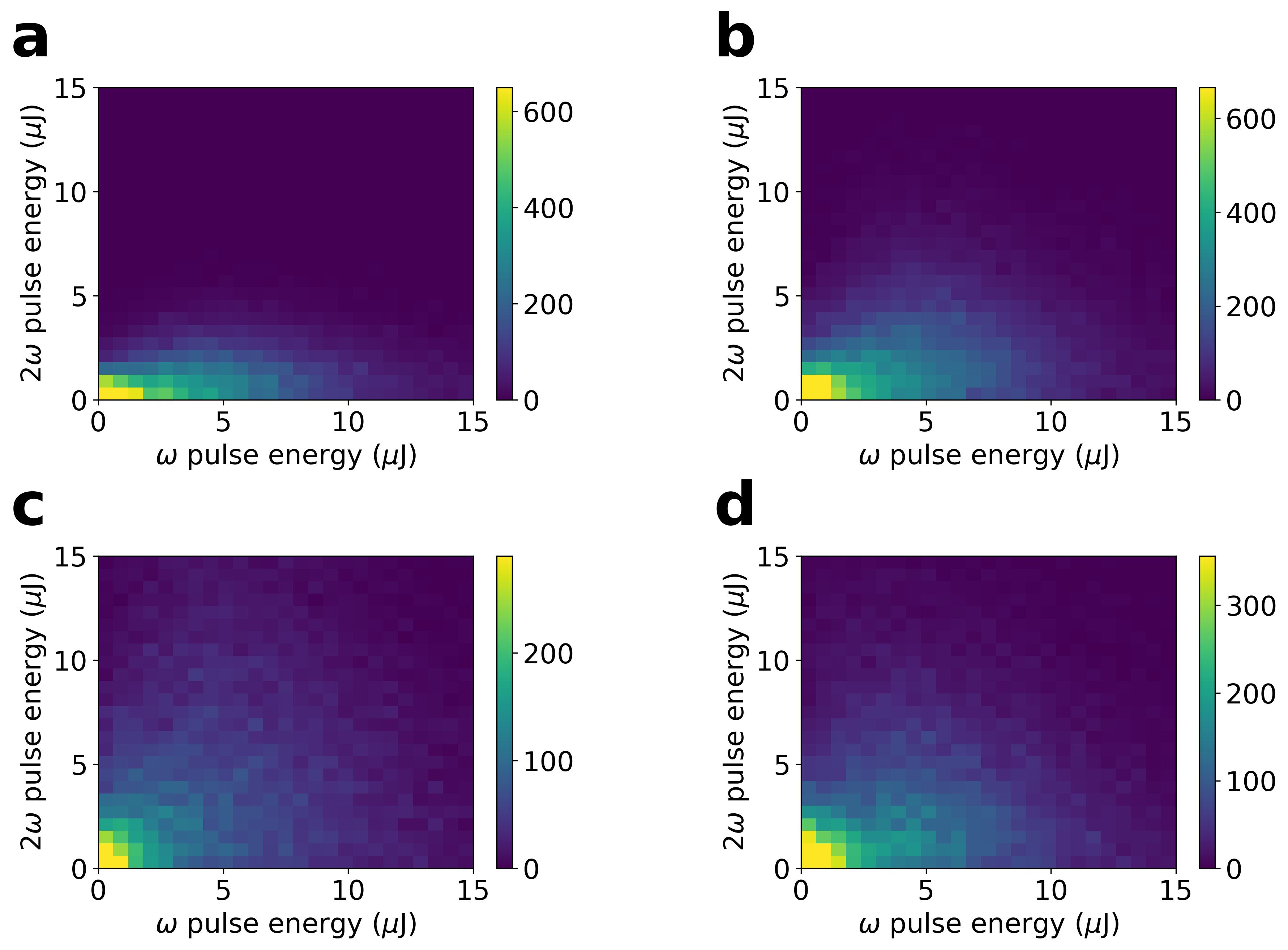}
\caption{
\textbf{Distributions of $\omega$ and $2\omega$ pulse energies}. 
\textbf{a-e}, Two-colour pulse energy distributions with different undulator beamline configurations for $2\omega$ pulses. 
\textbf{a}, Two-colour pulse energy distributions with no chicane delay and 1 undulator module for $2\omega$ pulses. 
\textbf{b}, Two-colour pulse energy distributions with no chicane delay and 2 undulator modules for $2\omega$ pulses. 
\textbf{c}, Two-colour pulse energy distributions with no chicane delay and 3 undulator modules for $2\omega$ pulses.
\textbf{d}, Two-colour pulse energy distributions with 1~fs nominal chicane delay and 4 undulator modules for $2\omega$ pulses.
Datasets shown in all panels from (\textbf{a}) to (\textbf{d}) have the same BC2 current~($2700~$A).
}
\label{fig_s:distribution_of_w_2w_pulse_energies}
\end{figure}


Fig.~\ref{fig_s:distribution_of_w_2w_pulse_energies} shows the distribution of single-shot measurements of $\omega$ and $2\omega$ pulse energies with different undulator beamline configurations. 
Unlike the conventional split-undulator mode~\cite{lutman2013experimental}, the intensities of $\omega$ and $2\omega$ pulses are not completely anti-correlated in the $\omega/2\omega$ mode. 
Taking Fig.~\ref{fig_s:distribution_of_w_2w_pulse_energies}~(b) as an example: The correlation coefficient between $\omega$ and $2\omega$ pulse energies is 0.15 on all shots in Fig.~\ref{fig_s:distribution_of_w_2w_pulse_energies}~(b).
When $\omega$ pulse energies are smaller than $5~\mu$J, $2\omega$ pulse energies have a positive correlation efficient ($0.28$) with $\omega$ pulse energies, since $2\omega$ pulses were built from microbunchings produced by $\omega$ pulses. 
The $\omega$ and $2\omega$ pulse energies gradually become anti-correlated (correlation efficient is $-0.08$) when $\omega$ pulse energies are larger than $5~\mu$J. 
This can be interpreted as follows: When the $\omega$ pulses are close to saturation, the large energy spread produced in the first undulator is detrimental to the amplification of $2\omega$ pulse and suppresses the harmonic microbunching. 
On the contrary, for small values of the $\omega$ pulse energy, the effect of the induced energy spread is negligible, and the two pulse energies are correlated because the harmonic bunching factor increases with the intensity of the first pulse.

When the magnetic chicane was tuned to a nominal delay of $1~$ fs, the correlation coefficient between $\omega$ and $2\omega$ pulse energies on all shots is significantly reduced to $0.04$, as shown in Fig.~\ref{fig_s:distribution_of_w_2w_pulse_energies}~(d). 
The nominal $1~$fs chicane delay partially suppressed the microbunching produced by the $\omega$ pulse in the electron beam and reduced the correlation coefficient between $\omega/2\omega$ pulse energies. 
However, the nominal $1~$fs chicane delay was not long enough to completely wash out the microbunching. 
On shots with $E_{\omega} < 5~\mu$J shown in Fig.~\ref{fig_s:distribution_of_w_2w_pulse_energies}~(d), the correlation coefficient between two pulse energies is $0.17$, which is still much higher than the typical negative correlation coefficient in a conventional split-undulator mode without any harmonic amplification.

\subsection{Two-Colour Pulse Bandwidths}

\begin{figure}[!ht]
\includegraphics[width=\linewidth]{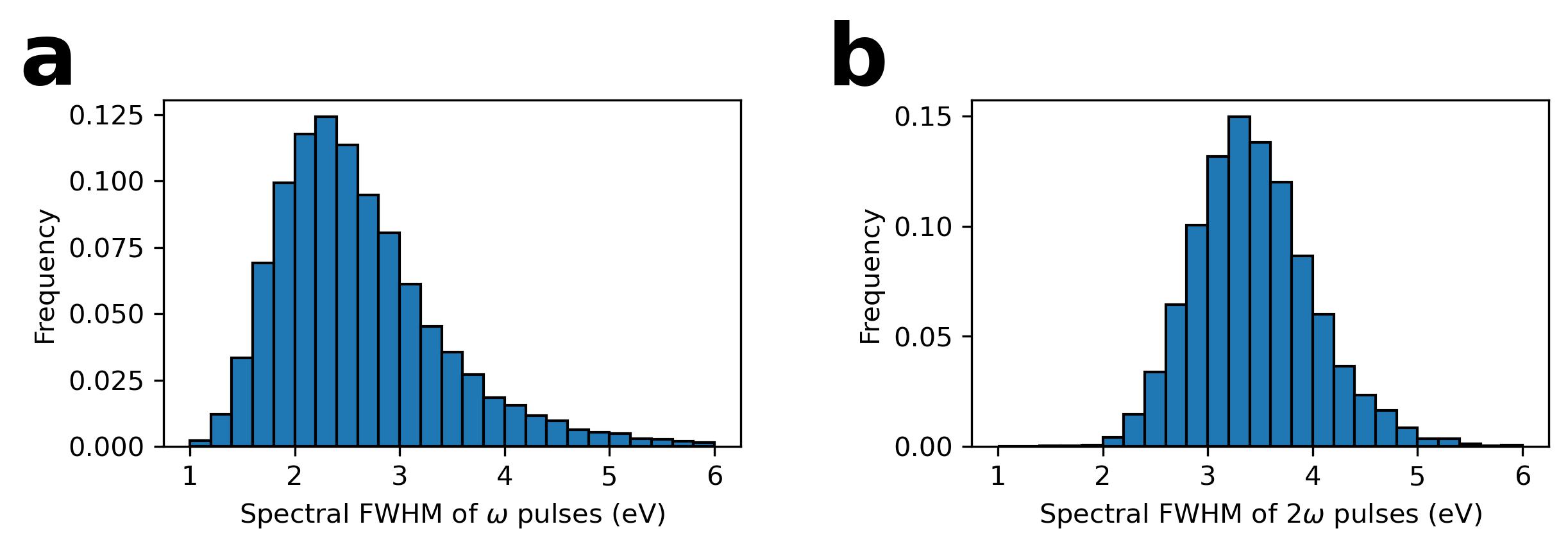}
\caption{
\textbf{Distribution of FWHM bandwidths}. 
Panels (\textbf{a}) and (\textbf{b}) show the distribution of full width at half maximum (FWHM) bandwidths of $\omega$ and $2\omega$ pulses, respectively.
}
\label{fig_s:distribution_of_w_2w_FWHM_bandwidths}
\end{figure}

\begin{figure}[!ht]
\centering
\includegraphics[width=0.8\linewidth]{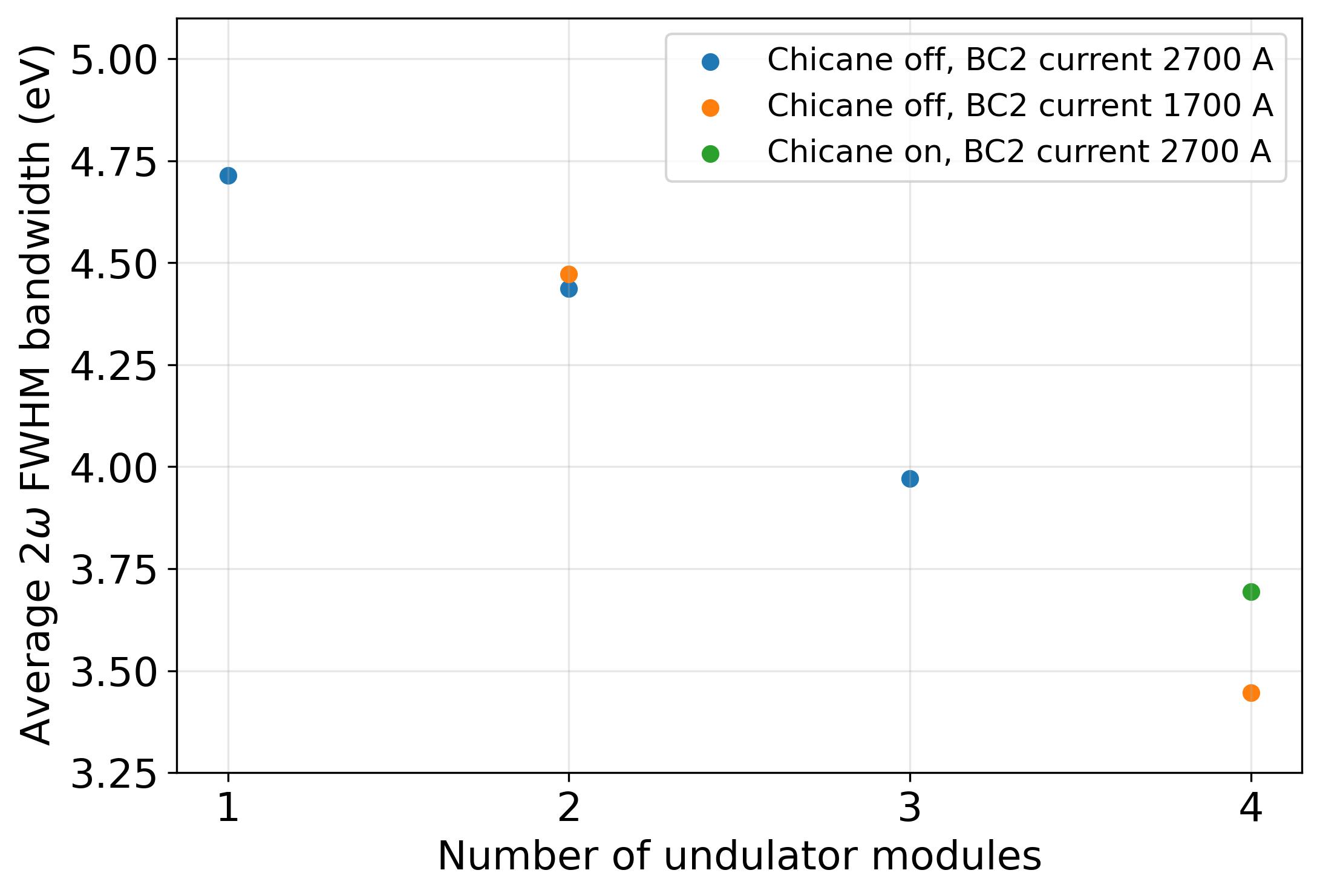}
\caption{
The average full width at half maximum (FWHM) bandwidths of $2\omega$ pulses at different $2\omega$ undulator beamline configurations and BC2 current profiles.}
\label{fig_s:2w_FWHM_bandwidth_vs_Nu}
\end{figure}

The single-shot bandwidth of the two colours is measured using the second-order diffraction of the $\omega$ pulses and the third-order diffraction of the $2\omega$ pulses.
Figure~\ref{fig_s:distribution_of_w_2w_FWHM_bandwidths} shows the histograms of the single-shot bandwidths of both colours. 
To minimise the effect of pollution from the high-order diffraction of the $2\omega$ pulse, we measure the $\omega$ bandwidth in a configuration where the $2\omega$ signal is negligible (i.e., chicane on and one undulator module in the second stage, with $E_{2\omega} < 0.2~\mu$J). The histogram of the $2\omega$ pulse corresponds to the configuration with 4 undulators and no chicane.
We report the average bandwidth from all undulator beamline and BC2 current configurations in Fig. \ref{fig_s:2w_FWHM_bandwidth_vs_Nu}. 
To improve the signal-to-noise ratio of the single-shot measurement we limit our analysis to shots with  pulse energies in the top 30\%  for datasets with more than one undulator module in the second stage, and the top 10\% for the one-undulator dataset. 
We believe that these bandwidths are representative of the entire datasets because the $2\omega$ pulse energies have negligible correlation with the $2\omega$ FWHM spectral bandwidths (the correlation coefficient is -0.018).

We process raw single-shot spectra by first applying a Gaussian filter with $\sigma=10$ pixels (0.72~eV for $\omega$ spectra, 1.74~eV for $2\omega$ spectra).
We extracted the full width at half-maximum (FWHM) of the two spectral peaks from these processed spectra. 
In the last step, the FWHM of the Gaussian filter is subtracted in quadrature from the FWHM bandwidths of the smoothed VLS measurements to obtain the FWHM bandwidths of the $\omega$ and $2\omega$ pulses of the raw data without smoothing. 
These experimental measurements of spectral widths are used to guide our start-to-end FEL simulations.


The average FWHM bandwidth in Fig.~\ref{fig_s:2w_FWHM_bandwidth_vs_Nu} decreases with the number of undulator modules for the $2\omega$ pulses. 
This trend is reproduced by the numerical simulations reported in Sec.~\ref{sec_s:Benchmark_Simulation} and is related to an increase of the pulse duration as a function of undulator length. 
The BC2 current is observed to have little effect on the $2\omega$ bandwidths. 
The BC2 current for each beamline configuration is listed in Table~\ref{tab_s:Configuration_of_each_run}.

\begin{table}[h]
\centering
\begin{tabular}{cccccc ccccc } 
 \hline
 \hline
 Chicane & Undulator Modules for $2\omega$ & BC2 Current\\ 
 \hline
 Off & 1 & $2700$~A\\
 Off & 2 & $2700$~A (Dataset A)\\
 Off & 2 & $1700$~A (Dataset B)\\
 Off & 3 & $2700$~A\\
 Off & 4 & $1700$~A\\
 On (Nominal $1~$fs delay) & 4 & $2700$~A\\
 \hline
 \hline
\end{tabular}
\caption{\label{tab_s:Configuration_of_each_run}
The $2\omega$ beamline configuration and the BC2 current of each dataset.}
\end{table}

\section{Time-Delay Analysis}

We use two independent methods to analyse the time-delay $\Delta \tau = \Delta \phi/2\pi\times T_{L}$ between $\omega/2\omega$ pulse pairs in each beamline configuration. 
Here $\Delta \phi$ is the difference between the streaking directions of two photoemission features and $T_{L}$ is the time period of the IR streaking laser. 
The first method exploits the correlation map of streaking-induced signal fluctuations in two photoemission features and does not rely on any single-shot streaking angle determination.
The second method (cosine fitting method) relies on single-shot sorting of the streaking angle of the inner photoline.

\subsection{Correlation Analysis}\label{sec_s:correlation_analysis}

In the first method, the average time delay of each beamline configuration is analysed by correlation analysis. 
The delay jitter is also estimated from the correlation analysis. 
The characteristic shape of the correlation coefficient map between the streaking-induced signal fluctuations in two photoemission features depends on the average time delay between two pulses. 
The amplitude of the correlation map is dependent on the delay jitter between the two pulses.
We extract the delay and the delay jitter by fitting the experimentally measured correlation maps to numerical simulations performed within the Strong Field Approximation (SFA).
In the correlation analysis, it is not required to determine the arrival time of XFEL pulses at each shot, which makes the delay fitting in the correlation analysis independent of systematic errors from any algorithm estimating the single-shot streaking direction.

Two assumptions are made in the correlation analysis:
\begin{enumerate}
    \item The momentum shift (or streaking amplitude) $\Delta p$ is relatively small compared to the widths of the photoemission feature.
    \item The streaking direction $\phi$ is independent of the IR laser intensities and the FEL pulse properties.
\end{enumerate}

The semi-classical picture states that the 2D streaked photoelectron momentum spectrum $M(p_x, p_y)$ can be approximated as the 2D unstreaked photoelectron momentum spectrum $M_0(p_x, p_y)$ ionised by the same XFEL pulse and shifted by $\Delta \vec{p} = (\Delta p \cos\phi, \Delta p \sin\phi)$:
\begin{equation}
M(p_x+\Delta p\cos\phi, p_y+\Delta p\sin\phi) = M_0(p_x, p_y), \label{eq_s:small_streaking_Cartesian_1}
\end{equation}
where $\Delta \vec{p} = e\vec{A}(t_0)$ at the time of ionisation $t_0$ in the semi-classical model. 
When the amplitude of the momentum shift $\Delta p$ is much smaller than the width of the photoemission feature, the streaking-induced change in the c-VMI signal can be approximated with the following linear expansion:
\begin{equation}
M(p_x, p_y)-M_0(p_x, p_y)
 \simeq -\frac{\partial M_0}{\partial p_x} \Delta p\cos\phi- \frac{\partial M_0}{\partial p_y} \Delta p \sin\phi, \label{eq_s:small_streaking_Cartesian_2}
\end{equation}
In polar coordinates, Eq.~\ref{eq_s:small_streaking_Cartesian_2} can be rewritten as
\begin{equation}
\begin{split}
M(\theta, p_r) - M_0(\theta, p_r)
 \simeq  -\frac{\partial M_0}{\partial p_r} \Delta p\cos(\phi-\theta) - \frac{1}{p_r} \frac{\partial M_0}{\partial \theta} \Delta p \sin(\phi-\theta), \label{eq_s:correlation_analysis_model_1}
\end{split}
\end{equation}
where $(p_x, p_y) = (p_r \cos\theta, p_r\sin \theta)$. 
We define the two linear terms on the right-hand side as the ``streaked component" of the photoelectron spectrum.

\begin{figure}[!ht]
\centering
\includegraphics[width=0.8\linewidth]{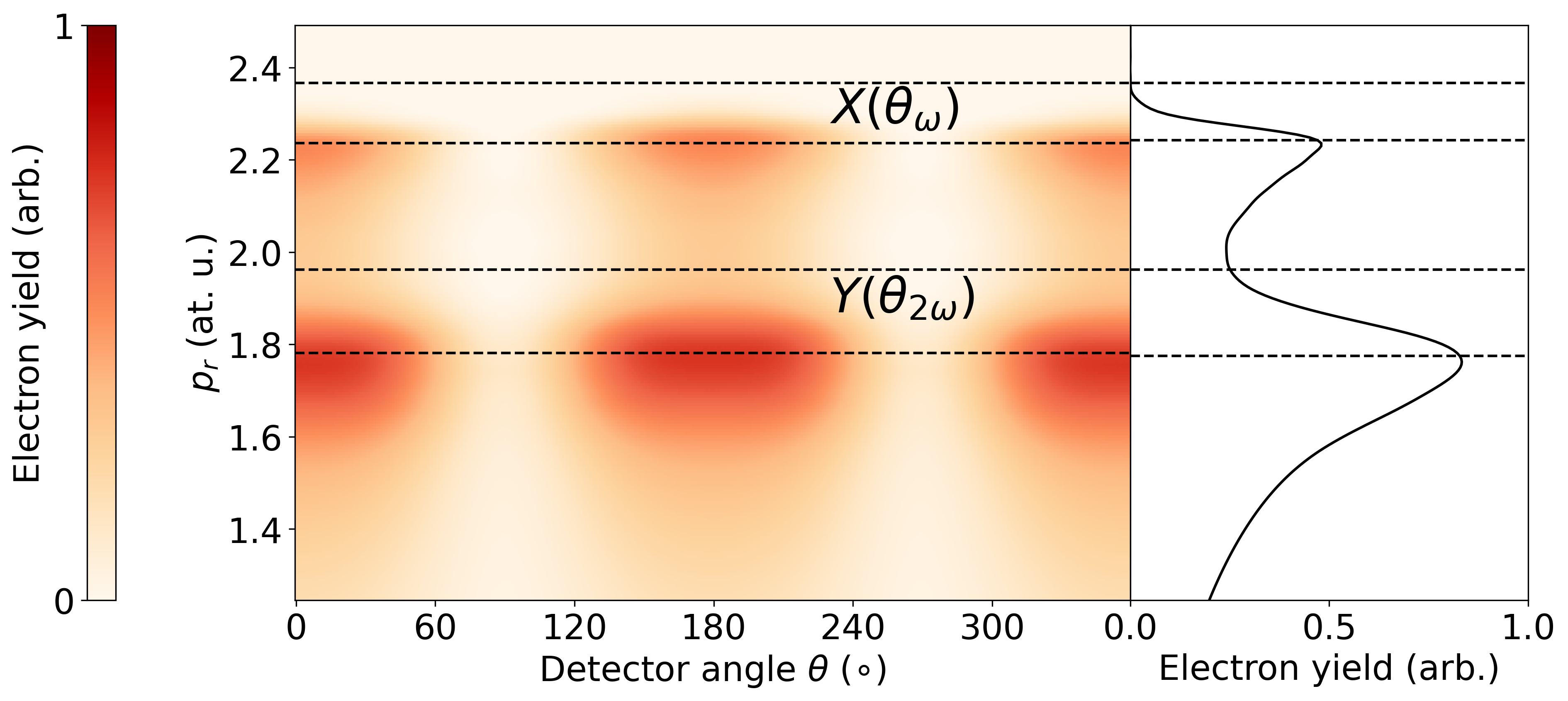}
\caption{
\textbf{The Region of Interest (ROI) in our delay analysis.}
The left panel shows the data from Fig.~1~\textbf{(b)} of the main text in polar coordinate. 
The 1D trace on the right-hand-side shows the electron yield integrated over all detector angles. 
Two pairs of dashed lines label inner and upper bounds of collecting electron yields in 2 photoemission features for the delay analysis
}
\label{fig_s:ROI_for_Delay_Analysis}
\end{figure}

Equation~\ref{eq_s:correlation_analysis_model_1} shows that the streaking-induced signal change in the photoelectron momentum spectrum is related to the gradient field of unstreaked momentum distribution. 
To maximise the streaking-induced signal change in our delay analysis, we selected the regions of interest (ROIs) corresponding to the high-energy flank of the photoemission features (i.e. the region above the radial maximum of the electron yield), as shown in Fig.~\ref{fig_s:ROI_for_Delay_Analysis}. 

In a simplified model, we neglect the angular gradient term in Eq.~\ref{eq_s:correlation_analysis_model_1} and study the contribution of only the radial gradient field to the covariance maps. 
By integrating Eq.~\ref{eq_s:correlation_analysis_model_1} in ROI, the streaked integrated signal $I(\theta)$ at the detector angle $\theta$ can be modelled as 
\begin{equation}
\begin{split}
I(\theta)
&=\int \mathrm{d}p_r~p_r M(\theta, p_r) \\
&= \int \mathrm{d}p_r~p_r M_0(\theta, p_r) - \int \mathrm{d}p_r~p_r \frac{\partial M_0}{\partial p_r} \Delta p \cos(\phi-\theta) \\
& = I_0(\theta)+\Delta p I_1(\theta)\cos(\phi-\theta), \label{eq_s:correlation_analysis_model_2}
\end{split}
\end{equation}
where $I_0(\theta) = \int \mathrm{d}p_r~p_r M_0(\theta, p_r)$ is the unstreaked integrated signal and $I_1(\theta) = - \int \mathrm{d}p_r~p_r {\partial M_0}/{\partial p_r}$ represents the integrated radial gradient field.

We name the integrated signal defined in Eq.~\ref{eq_s:correlation_analysis_model_2} as $X$ for the carbon $K$-shell (outer) photoemission feature, and $Y$ for the fluorine $K$-shell (inner) photoemission feature. 
Using the same notation as in Eq.~\ref{eq_s:correlation_analysis_model_2}, the integrated carbon 1$s$ signals $X(\theta_{\omega})$ at the detector angle $\theta_{\omega}$ and the integrated fluorine 1$s$ signals $Y(\theta_{2\omega})$ at the detector angle $\theta_{2\omega}$ are the following:
\begin{equation}
\begin{split}
X(\theta_{\omega}) & = X_0(\theta_{\omega}) + \Delta p X_1(\theta_{\omega}) \cos(\phi_{\omega}-\theta_{\omega}),\\
Y(\theta_{2\omega}) & = Y_0(\theta_{2\omega}) + \Delta p Y_1(\theta_{2\omega}) \cos(\phi_{2\omega}-\theta_{2\omega}), \label{eq_s:correlation_analysis_model_3}
\end{split}
\end{equation}
where $\phi_{\omega}$ and $\phi_{2\omega}$ are the streaking directions of the carbon $K$-shell and the fluorine $K$-shell photoemission features, respectively. 
The same streaking amplitude $\Delta p$ is used in both lines in Eqs.~\ref{eq_s:correlation_analysis_model_3} since the amplitude of the vector potential of the IR streaking laser roughly remains constant within one laser period.

The goal of the correlation analysis is to use the joint streaking-induced signal changes in two photoemission features to analyse the average delay and the delay jitter between $\omega/2\omega$ pulses. We make use of the following covariance between the streaked portions of the photoelectron spectra:
\begin{equation}
\operatorname{Cov}[\Delta p X_1(\theta_{\omega}) \cos(\phi_{\omega}-\theta_{\omega}), \Delta p Y_1(\theta_{2\omega}) \cos(\phi_{2\omega}-\theta_{2\omega})], \label{eq_s:correlation_analysis_model_4}
\end{equation}
where the covariance between two variables is defined as:
\begin{equation}
\operatorname{Cov}[A, B] = \langle AB\rangle-\langle A \rangle \langle B \rangle,
\end{equation}
where $\langle \cdot \rangle$ denotes the statistical average.

Although the signal we are interested in is the covariance between the two different streaked photolines, the covariance between the unstreaked signals $\operatorname{Cov}[X_0(\theta_{\omega}), Y_0(\theta_{2\omega})]$ due to the stochastic properties of SASE XFEL is not negligible. 
In fact, this covariance between unstreaked signals could be much larger than the covariance between streaking-induced signal fluctuations. 
To better extract time delays from covariance and correlation analyses, it is desirable to separate the two contributions. 

To this end, we can exploit the inversion symmetry of unstreaked photoelectron momentum spectra measured in c-VMI to separate the signal fluctuations induced by streaking and from other sources of variability. 
The unstreaked photoelectron spectra have the following symmetry:
\begin{equation}
M_0(\theta, p_r) = M_0(\theta+\pi, p_r), \label{eq_s:symmetry_1}
\end{equation}
where we are assuming that non-uniformities in the gain of the detector are negligible.

When the IR streaking pulse temporally overlaps with the XFEL pulses, the vector potential $\vec{A}$ of the streaking laser shifts the electron momentum spectrum and breaks the symmetry in Eq.~\ref{eq_s:symmetry_1}. 
In our data analysis, we define the anti-symmetric part of the photoelectron momentum spectrum as:
\begin{equation}
\tilde{M}(\theta, p_r) = M(\theta, p_r) - M(\theta+\pi, p_r). \label{eq_s:symmetry_2}
\end{equation}
Since the unstreaked data have inversion symmetry, it follows that the anti-symmetric part of the momentum spectrum is given by the streaked component of the spectrum multiplied by two:
\begin{equation}
\tilde{M}(\theta, p_r) \simeq -2\Delta p\left(\frac{\partial M_0}{\partial p_r} \cos(\phi-\theta) + \frac{1}{p_r} \frac{\partial M_0}{\partial \phi} \sin(\phi-\theta)\right), \label{eq_s:symmetry_3}
\end{equation}
Similarly to Eq.~\ref{eq_s:correlation_analysis_model_3}, we ignore the angular gradient fields in our simplified model.
By defining $\tilde{X}(\theta_{\omega}) = X(\theta_{\omega})-X(\theta_{\omega}+\pi)$ and $\tilde{Y}(\theta_{2\omega}) = Y(\theta_{2\omega})-Y(\theta_{2\omega}+\pi)$, the leading terms in the anti-symmetric parts of two integrated signals are,
\begin{equation}
\begin{split}
\tilde{X}(\theta_{\omega}) &= 2\Delta p X_1(\theta_{\omega})\cos(\phi_{\omega}-\theta_{\omega}),\\
\tilde{Y}(\theta_{2\omega}) &= 2\Delta p Y_1(\theta_{2\omega}) \cos(\phi_{2\omega}-\theta{2\omega}),\label{eq_s:symmetry_4}
\end{split}
\end{equation}
which mainly contain the streaking-induced signal changes in two photoemission features.

The covariance of the two anti-symmetrised integrated signals is given by
\begin{equation}
\begin{split}
&\operatorname{Cov}[\tilde{X}(\theta_{\omega}), \tilde{Y}(\theta_{2\omega})] \\
& = 4\left\langle \Delta p^2 X_1(\theta_{\omega})Y_1(\theta_{2\omega})\cos(\phi_{\omega}-\theta_{\omega}) \cos(\phi_{2\omega}-\theta_{2\omega}) \right\rangle \\
& - 4\left\langle \Delta p X_1(\theta_{\omega})\cos(\phi_{\omega}-\theta_{\omega}) \right\rangle
\left\langle \Delta p Y_1(\theta_{2\omega})\cos(\phi_{2\omega}-\theta_{2\omega}) \right\rangle.\label{eq_s:symmetry_5}
\end{split}
\end{equation}
The distribution of the streaking direction $\phi$ is uniform, therefore each cosine term individually averages to zero: $\langle \cos(\phi-\theta)\rangle=0$. 
Hence, the second term in Eq. \ref{eq_s:symmetry_5} is zero. 
Furthermore, the variables $\Delta p$ and $\phi$ are statistically independent of each other. 
They are also independent of the fluctuations in the XFEL pulse properties. 
It follows that Eq.\ref{eq_s:symmetry_5} can be simplified as follows:
\begin{equation}
\operatorname{Cov}[\tilde{X}(\theta_{\omega}), \tilde{Y}(\theta_{2\omega})]  = 4\left\langle \Delta p^2  \right\rangle \left\langle X_1(\theta_{\omega})Y_1(\theta_{2\omega}) \right\rangle \left\langle \cos(\phi_{\omega}-\theta_{\omega}) \cos(\phi_{2\omega}-\theta_{2\omega}) \right\rangle. \label{eq_s:symmetry_6}
\end{equation}

Assuming that the time delay distribution is a Gaussian distribution with average delay $\Delta \tau = \Delta \phi/2\pi \times T_L$ and a root-mean-square (rms) delay jitter $\Delta \tau_{\mathrm{jitter}} = \Delta \phi_{\mathrm{jitter}}/2\pi \times T_L$, the delay-sensitive factor in Eq.~\ref{eq_s:symmetry_6} can be calculated as
\begin{equation}
\begin{split}
& \left\langle \cos(\phi_{\omega}-\theta_{\omega}) \cos(\phi_{2\omega}-\theta_{2\omega})  \right\rangle  \\
& = \frac{1}{2}\left\langle\cos(\phi_{\omega}+\phi_{2\omega}-\theta_{\omega}-\theta_{2\omega})\right\rangle +
\frac{1}{2}\left\langle\cos\left( (\phi_{2\omega}-\phi_{\omega}) - (\theta_{2\omega}-\theta_{\omega})\right)\right\rangle\\
& = \frac{1}{2}\left\langle\cos\left( (\phi_{2\omega}-\phi_{\omega}) - (\theta_{2\omega}-\theta_{\omega})\right)\right\rangle \\
& = \frac{1}{2} \int \mathrm{d}u ~ 
\cos\left( u - (\theta_{2\omega}-\theta_{\omega})\right)\times\frac{1}{\sqrt{2\pi} \Delta \phi_{\mathrm{jitter}}}\exp\left( -\frac{\left( u - \Delta \phi\right)^2}{2\Delta \phi^2_{\mathrm{jitter}}}\right) \\
& =\frac{1}{2} \cos\left( (\theta_{2\omega}- \theta_{\omega})- \Delta \phi \right) \exp\left( -\frac{\Delta \phi^2_{\mathrm{jitter}}}{2}\right). \label{eq_s:correlation_analysis_model_6}
\end{split}
\end{equation}
The result in Eq.~\ref{eq_s:correlation_analysis_model_6} has intuitive interpretations: 
streaking-induced signal fluctuations in the carbon $K$-shell (outer) photoemission feature at $\theta$ have the strongest joint variability with streaking-induced signal fluctuations in the fluorine $K$-shell (outer) photoemission feature at $\theta+\Delta \phi$. 
Additionally, the correlation between streaking-induced signal fluctuations in two photoemission features will gradually be washed out when the delay jitter increases.

By inserting Eq.~\ref{eq_s:correlation_analysis_model_6} into Eq.~\ref{eq_s:symmetry_6}, the covariance between anti-symmetrised signal fluctuations in two photoemission features is
\begin{equation}
\operatorname{Cov}[\tilde{X}(\theta_{\omega}), \tilde{Y}(\theta_{2\omega})] \simeq
2 \left\langle \Delta p^2  \right\rangle 
\left\langle X_1(\theta_{\omega})Y_1(\theta_{2\omega}) \right\rangle
\cos\left( (\theta_{2\omega}- \theta_{\omega})- \Delta \phi \right) e^{ -\frac{\Delta \phi^2_{\mathrm{jitter}}}{2} }. \label{eq_s:symmetry_4}
\end{equation}
This covariance map is very sensitive to intensity fluctuations, which makes it difficult to compare with simulated data. 
To remove this effect, we replace the covariance used in Eq.~\ref{eq_s:symmetry_4} with the Pearson correlation between $\tilde{X}$ and $\tilde{Y}$:
\begin{equation}
\begin{split}
\rho_{\tilde{X}, \tilde{Y}} (\theta_{\omega}, \theta_{2\omega})
& = \frac{\operatorname{Cov}[\tilde{X}(\theta_{\omega}), \tilde{Y}(\theta_{2\omega})]}{\sqrt{\operatorname{Var}[\tilde{X}(\theta_{\omega})]\operatorname{Var}[\tilde{Y}(\theta_{2\omega})]}} \\
& \simeq \frac{\left\langle X_1(\theta_{\omega})Y_1(\theta_{2\omega}) \right\rangle}{\sqrt{\left\langle X^2_1(\theta_{\omega}) \right\rangle  
\left\langle Y^2_1(\theta_{2\omega}) \right\rangle}}
\cos\left( (\theta_{2\omega}- \theta_{\omega})- \Delta \phi \right) e^{ -\frac{\Delta \phi^2_{\mathrm{jitter}}}{2} } \label{eq_s:correlation_map_on_asymmetric_part}.
\end{split}
\end{equation}
The interpretation of this correlation map is similar to the covariance map, with the most correlated region appearing for $\theta_{2\omega}=\theta_{\omega}+\Delta \phi$.

\begin{figure}[!ht]
\includegraphics[width=\linewidth]{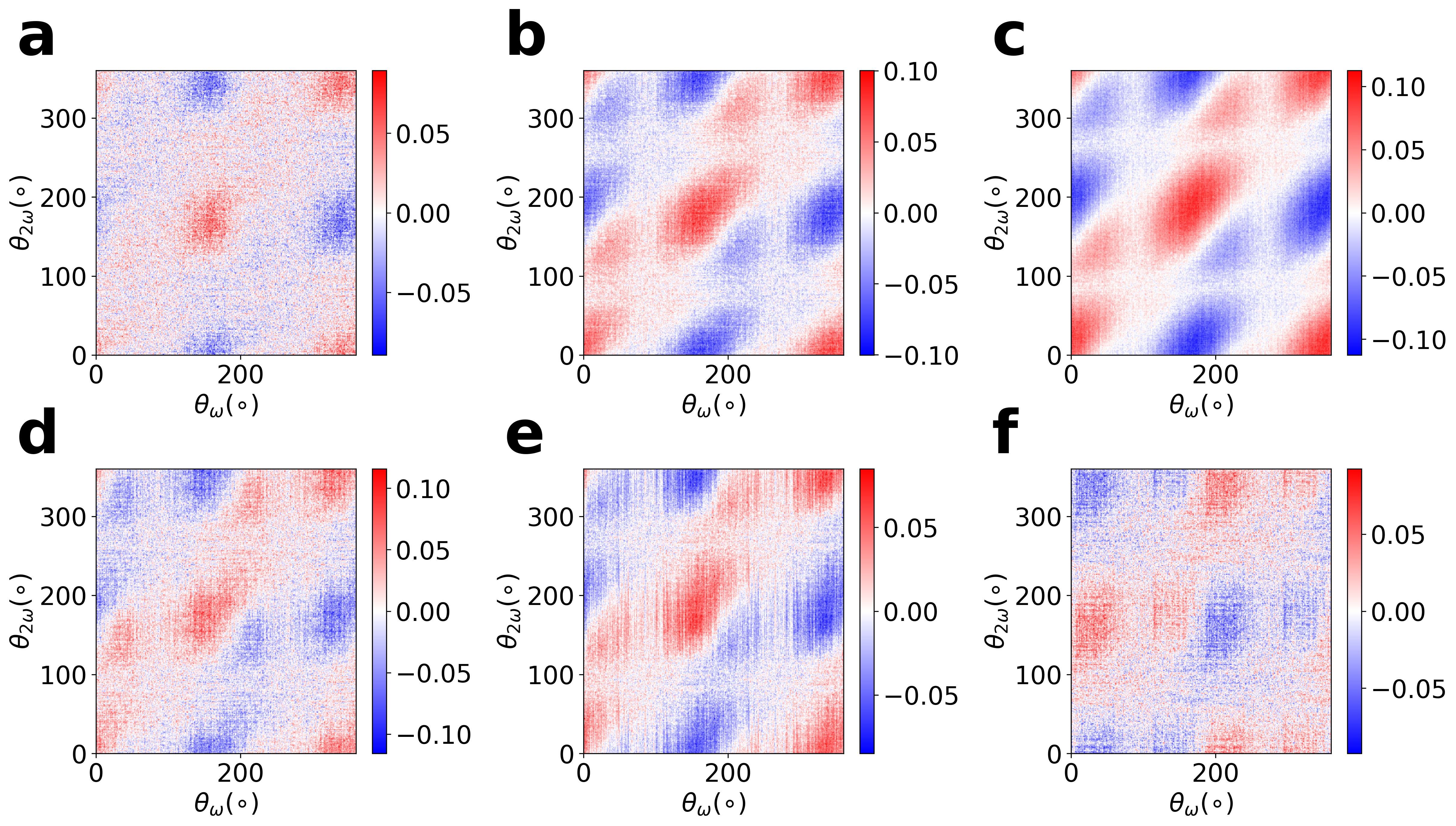}
\caption{\textbf{Correlation maps on data with different undulator beamline configurations}. 
\textbf{a-e}, Chicane turned off and 1 (\textbf{a}), 2 (\textbf{b}), 2 (\textbf{c}), 3 (\textbf{d}), and 4 (\textbf{e}) undulator modules for $2\omega$ pulse. Panels (\textbf{b}) and (\textbf{c}) are correlation maps calculated on data taken in datasets A and B respectively. 
\textbf{f}, Chicane turned on and 4 undulator modules for $2\omega$ pulses.
}
\label{fig_s:Correlation_Maps_of_All_Data}
\end{figure}

In constructing the experimental correlation map, we remove shots with low nominal GMD value ($<4~\mu$J) or with poor streaking amplitude.
Correlation maps calculated on the experimental data are shown in Fig.~\ref{fig_s:Correlation_Maps_of_All_Data}. 
A diagonal shift of highly correlated regions, which represents an average time delay, can be seen in each panel in Fig.~\ref{fig_s:Correlation_Maps_of_All_Data}. 
From panels (a) to (f) in Fig.~\ref{fig_s:Correlation_Maps_of_All_Data}, pixels with the best correlations shift more and more from the diagonal line, representing an increase in the average time delay when changing the undulator beamline configurations for $2\omega$ pulses. 
The correlation vanishes at $90^\circ$ and $270^\circ$ in both $\theta_{\omega}$ and $\theta_{2\omega}$ since there are few photoelectrons in nodes of dipolar photoemission features. 

\subsubsection{Analysis of Average Delays}

Average delays are fitted by comparing correlation maps shown in Fig.~\ref{fig_s:Correlation_Maps_of_All_Data} to simulated correlation maps.
We have performed SFA simulations to calculate photoelectron momentum spectra ionised by attosecond 2-colour XFEL pulses in the presence of a $1.3~\mu$m circularly polarised IR laser~\cite{li2018characterizing}. 
Correlation maps are calculated on these 2-colour photoelectron momentum spectra. 
Hyperparameters including XFEL pulse length distributions, pulse energy distributions, and photon energy distributions have been scanned to better benchmark correlation maps calculated on experimental data. 
In our simulations, we assume that the pulse duration is much shorter than the streaking laser wavelength (as is the case in the experiment).
In this impulsive limit, the parameter scans show that the shape of the correlation maps depends mainly on two parameters: the average delay and the ratio between the average $\omega/2\omega$ pulse energies. 
As predicted in Eq.~\ref{eq_s:correlation_map_on_asymmetric_part}, the delay jitter changes the amplitude but not the shape of correlation maps.
The simulations confirm that the average delay results in a diagonal shift in the correlation maps, as predicted by Eq.~\ref{eq_s:correlation_map_on_asymmetric_part} and observed in the experimental maps in Fig.~\ref{fig_s:Correlation_Maps_of_All_Data}. 

The $\omega/2\omega$ pulse energy ratio can also change the shape of the correlation maps due to the working principle of c-VMI. 
This is somewhat counter-intuitive: Ideally, the Pearson correlation defined in Eq.~\ref{eq_s:correlation_map_on_asymmetric_part} is independent of any linear scaling of $\tilde{X}$ and $\tilde{Y}$, suggesting that correlations should not be affected by the ratio of the $\omega/2\omega$ pulse energies. 
However, in a velocity map imaging spectrometer, the 3D Newtonian sphere of photoelectrons is projected into a 2D momentum spectrum, resulting in some high-energy carbon 1s electrons overlapping with low-energy electrons on fluorine 1s photoemission features. 
In other words, $X(\phi)$ contains only carbon 1s photoelectrons, but $Y(\phi)$ contains both fluorine $K$-shell photoelectrons and the inside signal of the carbon $K$-shell photoemission feature.
When $\omega$ pulse energies $E_{\omega}$ are much larger than probe pulse energies $E_{2\omega}$, the correlation in Eq.~\ref{eq_s:correlation_map_on_asymmetric_part} no longer characterises the joint variability of the streaking-induced signal of carbon $K$-shell and fluorine $K$-shell photoemission features. 
Instead, it only characterises the correlation between the inside and outside of the maximal electron yields in the carbon $K$-shell photoemission feature. 
On the other hand, when the $2\omega$ pulse energies $E_{2\omega}$ are strong enough compared to the $\omega$ pulse energies $E_{\omega}$, electron yields of the fluorine $K$-shell (inner) photoemission feature are strong enough and the projection of carbon $K$-shell electrons onto the fluorine $K$-shell photoemission feature is negligible. 
In this case, given an average delay $\Delta \tau$, the shape of the simulated correlation map no longer depends on the $\omega/2\omega$ pulse energy ratio and is even insensitive to the shapes of the distributions of $\omega$ and $2\omega$ pulse energies. 

In our SFA simulations, the FWHM durations of the $\omega$ and $2\omega$ pulses, both sampled with Gaussian distributions, are set to $600 \pm 150~$as and $400 \pm 100~$as, respectively. 
RMS photon energy jitters of both pulses were set to $2~$eV. 
The pulse energies of the $\omega$ and $2\omega$ pulses are sampled with gamma distributions and scaled to match the average $\omega/2\omega$ pulse energies calibrated from experimental data.
The RMS delay jitter is set to $400~$as in simulations to analyse the average delays, although the average delay is insensitive to this parameter. 
Average delays between $\omega/2\omega$ pulse pairs were scanned from $0~$as to $1400~$as with $10~$as per step. 
For each average delay, 1800 pairs of $\omega/2\omega$ pulses were used to simulate 2D photoelectron momentum spectra with random XFEL arrival time with respect to the IR laser. 
Given a simulated correlation map $\rho^{\mathrm{sim}}(\theta_{\omega}, \theta_{2\omega})$, the likelihood between the experimental correlation map $\rho^{\mathrm{exp}}(\theta_{\omega}, \theta_{2\omega})$ and the simulated correlation map $\rho^{\mathrm{sim}}(\theta_{\omega}, \theta_{2\omega})$ is calculated as a normalised inner product,
\begin{equation}
L(\rho^{\mathrm{exp}}, \rho^{\mathrm{sim}}) =
\frac{\left\langle\rho^{\mathrm{exp}}, \rho^{\mathrm{sim}}\right\rangle}{\sqrt{\left\langle\rho^{\mathrm{exp}}, \rho^{\mathrm{exp}}\right\rangle \left\langle\rho^{\mathrm{sim}}, \rho^{\mathrm{sim}}\right\rangle}}, \label{eq_s:inner_product_for_delay_fitting_1}
\end{equation}
where the inner product is defined as
\begin{equation}
\left\langle A, B\right\rangle = \sum^{360^\circ}_{\theta_{\omega}=1^\circ}\sum^{360^\circ}_{\theta_{2\omega}=1^\circ} A\left( \theta_{\omega}, \theta_{2\omega}\right) B \left( \theta_{\omega}, \theta_{2\omega}\right). \label{eq_s:inner_product_for_delay_fitting_2}
\end{equation}
For each experimental correlation map, a 1-D trace $L(\Delta \tau)$ as a function of simulated average delay $\Delta \tau$ is calculated between the experimental correlation map and the simulated correlation maps with the scan of $\Delta \tau$. 
The average delay is determined by maximising $L(\Delta \tau)$ after smoothing.

\begin{figure}[!ht]
\centering
\includegraphics[width=0.8\linewidth]{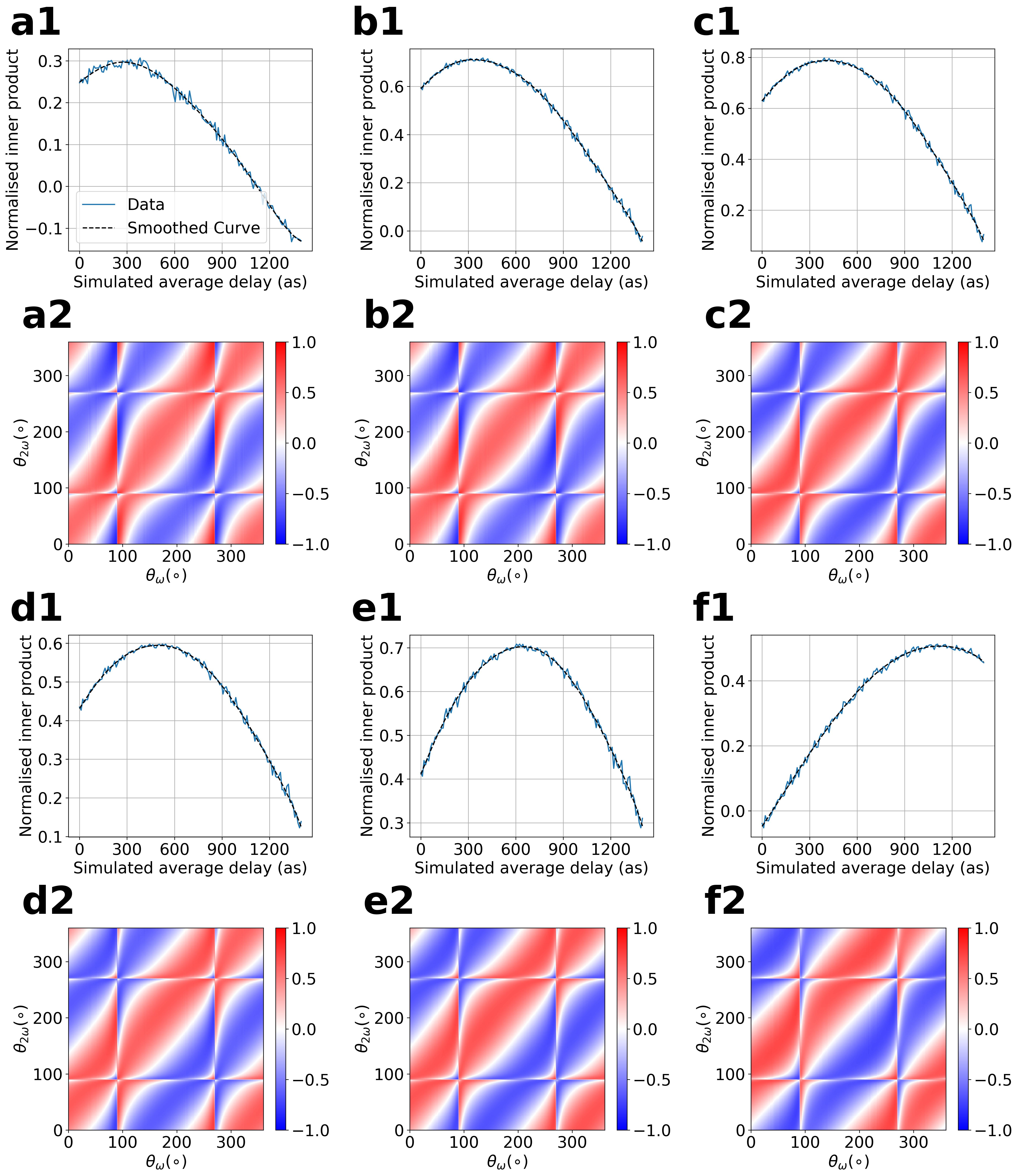}
\caption{\textbf{Normalised inner products and simulated correlation maps with the best delay fittings}. 
Panels with annotations ended with $\textbf{1}$ show the normalised inner products between the experimental and simulated correlations as a function of the simulated average delay in all undulator beamline configurations.
Panels with annotations ended with $\textbf{2}$ show the optimal correlation maps with the maximised normalised inner product in all undulator beamline configurations.  
\textbf{a-e}, Chicane turned off and 1 (\textbf{a}), 2 (\textbf{b}), 2 (\textbf{c}), 3 (\textbf{d}), and 4 (\textbf{e}) undulator modules for $2\omega$ pulse. Panels (\textbf{b}) and (\textbf{c}) represent normalised inner product between simulations and data taken in datasets A and B respectively. 
\textbf{f}, Chicane turned on and 4 undulator modules for $2\omega$ pulses. 
}
\label{fig_s:1D_normalised_Inner_Product_and_Best_Matched_Correlation_Maps}
\end{figure}

The normalised inner product of the simulated and experimental maps as a function of the average delay is shown in Fig.~\ref{fig_s:1D_normalised_Inner_Product_and_Best_Matched_Correlation_Maps} for different experimental configurations. 
From panels (a1) to (f1) in Fig.~\ref{fig_s:1D_normalised_Inner_Product_and_Best_Matched_Correlation_Maps}, the average delay, shown as the maximum in the normalised inner product, becomes larger. 
Panels (a2) to (f2) in Fig.~\ref{fig_s:1D_normalised_Inner_Product_and_Best_Matched_Correlation_Maps} show the simulated maps that best match the data. 

The uncertainty of the delay fitting is estimated by bootstrapping~\cite{efron1992bootstrap}. 
In each beamline configuration, we randomly resample the original dataset with replacements. 
We repeat this procedure 200 times and generate 200 resampled experimental correlation maps. 
We calculate the normalised inner product $L$ between the resampled correlation maps and the simulated maps based on Eq.~\ref{eq_s:inner_product_for_delay_fitting_1}. 
The 1-D normalised inner product $L(\Delta \tau)$ as a function of simulated average delay $\Delta \tau$ is smoothed by a 7th-order polynomial $L^{\mathrm{fit}}(\Delta \tau)$, the maximum of which is used as the fitted average delay for each experimental correlation map generated by bootstrapping. 
Finally, the uncertainty of the average delay is defined as three times the standard deviation of the distribution of the 200 delays obtained from the resampled images. 

\subsubsection{Analysis of the Pump/Probe Delay Jitter}

\begin{figure}[!ht]
\includegraphics[width=\linewidth]{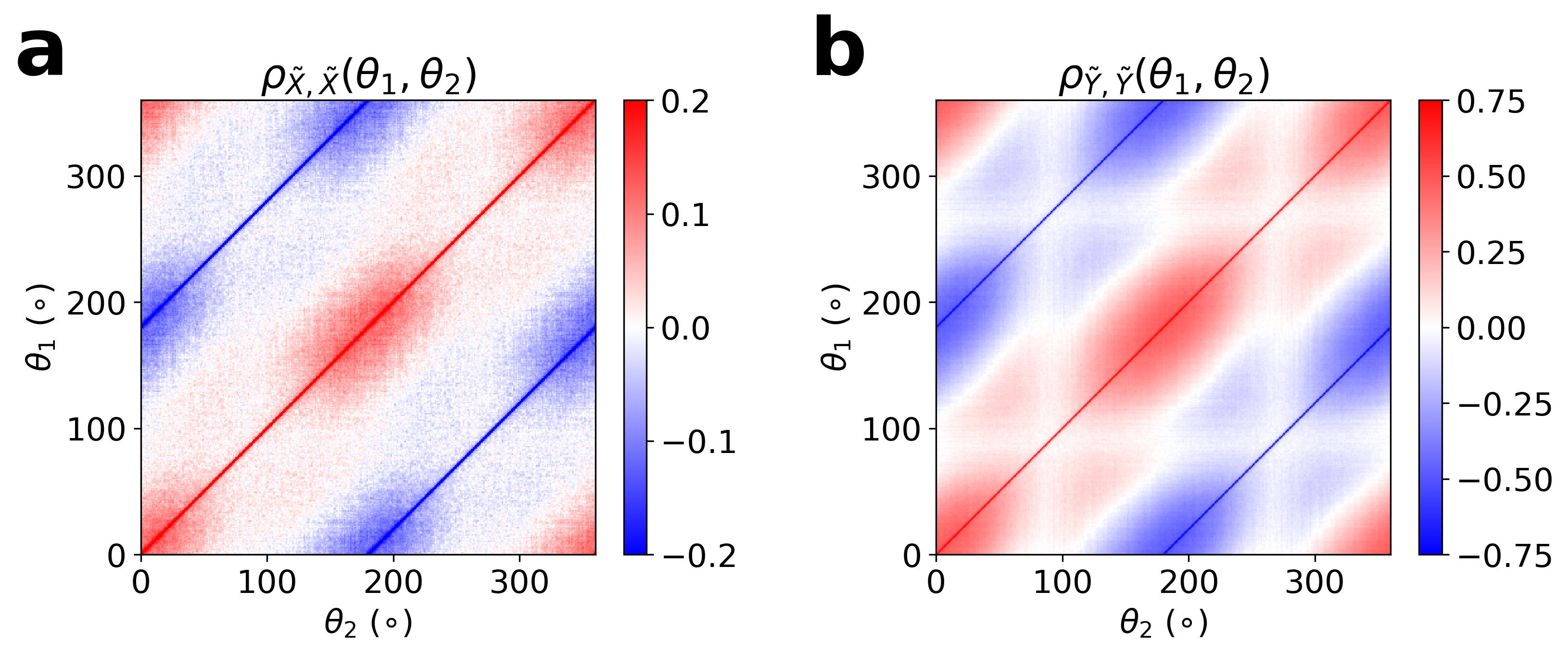}
\caption{\textbf{Autocorrelation maps of two photoemission features}.
Panel (\textbf{a}) and (\textbf{b}) show autocorrelation maps of $\tilde{X}(\theta_{\omega})$ and $\tilde{Y}(\theta_{2\omega})$, respectively. 
Diagonal lines in both maps are 1 since pixels on diagonal lines represent correlations of the same signals with themselves. 
The upper and lower limits of colourbars in both maps are adjusted to better represent the values of pixels in the rest of both maps.
}
\label{fig_s:Autocorrelation_of_w_2w_Photolines}
\end{figure}

We use the amplitudes of the correlation maps to estimate the jitter $\Delta \tau_{\mathrm{jitter}}$ of the time delay. 
As shown in Eq.~\ref{eq_s:correlation_map_on_asymmetric_part}, the delay jitter $\Delta \tau_{\mathrm{jitter}}$ decreases the magnitude of the correlation map. 
However, in our simulations, we have found that the delay jitter is not the only parameter that reduces the amplitude of the correlation maps. 
For example, this parameter can be reduced by counting noise in the detector. 
Furthermore, given the same level of counting noise, the strength of the streaking amplitudes can affect the magnitude of the simulated correlation maps. 

To construct a variable that depends mainly on the delay jitter $\Delta \tau_{\mathrm{jitter}}$, we  compute autocorrelation maps of anti-symmetric parts of the two photoemission features. 
Similarly to Eq.~\ref{eq_s:correlation_map_on_asymmetric_part}, autocorrelation maps of $\tilde{X}$ and $\tilde{Y}$ are defined as
\begin{equation}
\begin{split}
\rho_{\tilde{X}, \tilde{X}} (\theta_1, \theta_2)
& = \frac{\operatorname{Cov}[\tilde{X}(\theta_1), \tilde{X}(\theta_2)]}{\sqrt{\operatorname{Var}[\tilde{X}(\theta_1)]\operatorname{Var}[\tilde{X}(\theta_2)]}} \\
& \simeq \frac{\left\langle X_1(\theta_1)X_1(\theta_2) \right\rangle}{\sqrt{\left\langle X^2_1(\theta_1) \right\rangle  
\left\langle X^2_1(\theta_2) \right\rangle}}
\cos(\theta_2- \theta_1), \label{eq_s:autocorrelation_map_on_asymmetric_part_X}
\end{split}
\end{equation}
and
\begin{equation}
\begin{split}
\rho_{\tilde{Y}, \tilde{Y}} (\theta_1, \theta_2)
& = \frac{\operatorname{Cov}[\tilde{Y}(\theta_1), \tilde{Y}(\theta_2)]}{\sqrt{\operatorname{Var}[\tilde{Y}(\theta_1)]\operatorname{Var}[\tilde{Y}(\theta_2)]}} \\
& \simeq \frac{\left\langle Y_1(\theta_1)Y_1(\theta_2) \right\rangle}{\sqrt{\left\langle Y^2_1(\theta_1) \right\rangle  
\left\langle Y^2_1(\theta_2) \right\rangle}}
\cos(\theta_2- \theta_1), \label{eq_s:autocorrelation_map_on_asymmetric_part_Y}
\end{split}
\end{equation}
Since there is no delay between the arrival time of a pulse and itself, the autocorrelation maps of $\tilde{X}$ and $\tilde{Y}$ are independent of the delay time and delay jitter. 
Autocorrelation maps of two photoemission features calculated on the experimental data in dataset B are shown in Fig.~\ref{fig_s:Autocorrelation_of_w_2w_Photolines}. 
 
Simulations show that the amplitudes of the autocorrelation maps are also affected by the counting noise and the streaking amplitude,  except for pixels along the lines $\theta_1 = \theta_2$ and $\theta_1 = \theta_2+\pi$. 
As a consequence, we can construct a quantity that depends mainly on delay jitter $\Delta \tau_{\mathrm{jitter}}$ by defining a normalised trace from the correlation map between $\tilde{X}$ and $\tilde{Y}$, and two autocorrelation maps of $\tilde{X}$ and $\tilde{Y}$ themselves. 
We take the anti-diagonal trace $\theta_1 = 2\pi-\theta_2$ from each map,
\begin{equation}
\begin{split}
f_{\omega, 2\omega}(\psi) &= \rho_{\tilde{X}, \tilde{Y}} (\psi, 2\pi-\psi), \\
f_{\omega, \omega}(\psi) &= \rho_{\tilde{X}, \tilde{X}} (\psi, 2\pi-\psi), \\
f_{2\omega, 2\omega}(\psi) &= \rho_{\tilde{Y}, \tilde{Y}} (\psi, 2\pi-\psi). \\
\end{split}\label{eq_s:three_traces_for_jitter_analysis}
\end{equation}
We use a 5-th order polynomial to fit and smooth each 1-D trace in Eqs.~\ref{eq_s:three_traces_for_jitter_analysis} for $\psi \in [135^{\circ}, 225^{\circ}]$. 
The sharp peaks at $\psi = \pi$ in autocorrelation maps are ignored. 
A normalised correlation amplitude $g$ is defined from the 3 fitted 1-D traces,
\begin{equation}
g = \frac{\mathrm{max}~f^{\mathrm{fit}}_{\omega, 2\omega}(\psi)}{ \sqrt{\left(\mathrm{max}~f^{\mathrm{fit}}_{\omega, \omega}(\psi) \right) \left(\mathrm{max}~f^{\mathrm{fit}}_{2\omega, 2\omega}(\psi)\right)}}.
\end{equation}

The value of $g$ is insensitive to counting noise and streaking amplitudes because both affect all three correlation maps in the same way. Our numerical simulations confirm that the value of $g$ strongly depends on the delay jitter used in the simulations. 
This variable can now be used as a metric to estimate the delay jitter.

\begin{figure}[!ht]
\includegraphics[width=\linewidth]{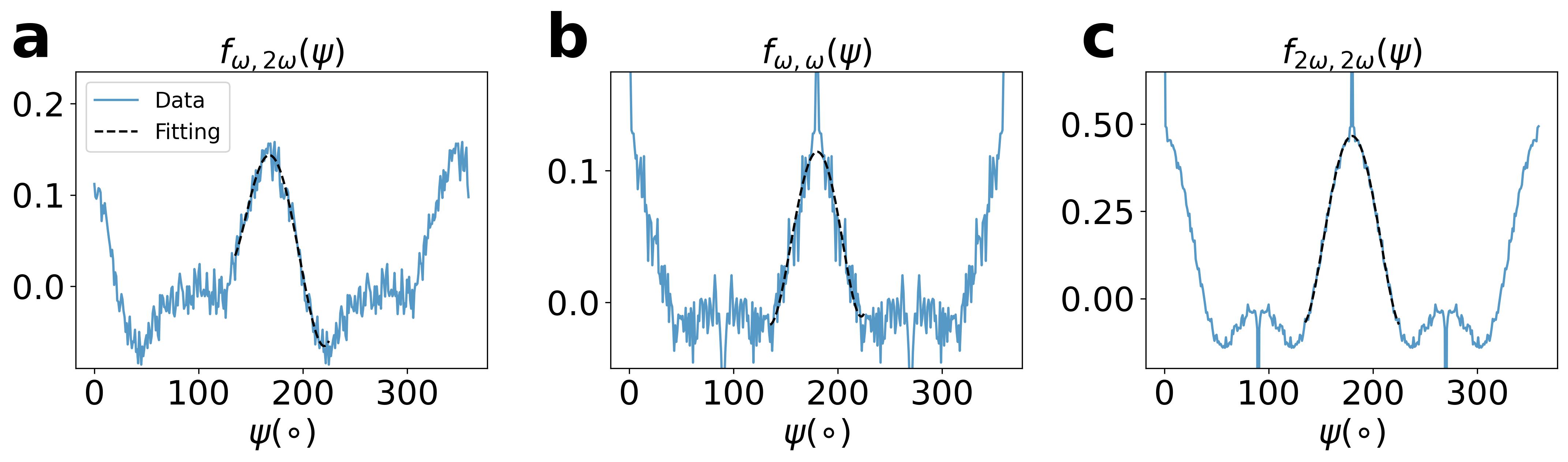}
\caption{\textbf{Anti-diagonal traces of three correlation maps}.
Panel (\textbf{a}) shows the antidiagonal trace of the correlation map shown in Fig.~\ref{fig_s:Correlation_Maps_of_All_Data}(c). 
Panels (\textbf{b}) and (\textbf{c}) shows anti-diagonal trace of autocorrelation maps in Figs.~\ref{fig_s:Autocorrelation_of_w_2w_Photolines}(a) and (b), respectively. 
The black dashed line in each panel represents the polynomial fitting of the anti-diagonal trace between $135^\circ$ and $25^\circ$. 
Sharp peaks $f_{\omega, \omega}(180^\circ)=1$ and $f_{2\omega, 2\omega}(180^\circ)=1$ are neglected in the polynomial fitting in panels (\textbf{b}) and (\textbf{c}).
}
\label{fig_s:Antidiagonal_Traces_of_3_Corr_Maps}
\end{figure}

Three antidiagonal traces taken from experimental correlation maps are shown in Fig.~\ref{fig_s:Antidiagonal_Traces_of_3_Corr_Maps}. 
The peak of $f_{\omega, 2\omega}(\psi)$ shown in Fig.~\ref{fig_s:Antidiagonal_Traces_of_3_Corr_Maps}(a) deviates from $\psi = 180^\circ$, in agreement with our analysis of the average delay ($434\pm49~$as in the cosine fitting analysis, $402\pm28~$as in the correlation analysis). 
All three traces between $135^\circ$ and $225^\circ$ are fitted by 5-th order polynomials for smoothing, shown as black dashed lines in all panels in Fig.~\ref{fig_s:Antidiagonal_Traces_of_3_Corr_Maps}.

\begin{figure}[!ht]
\includegraphics[width=\linewidth]{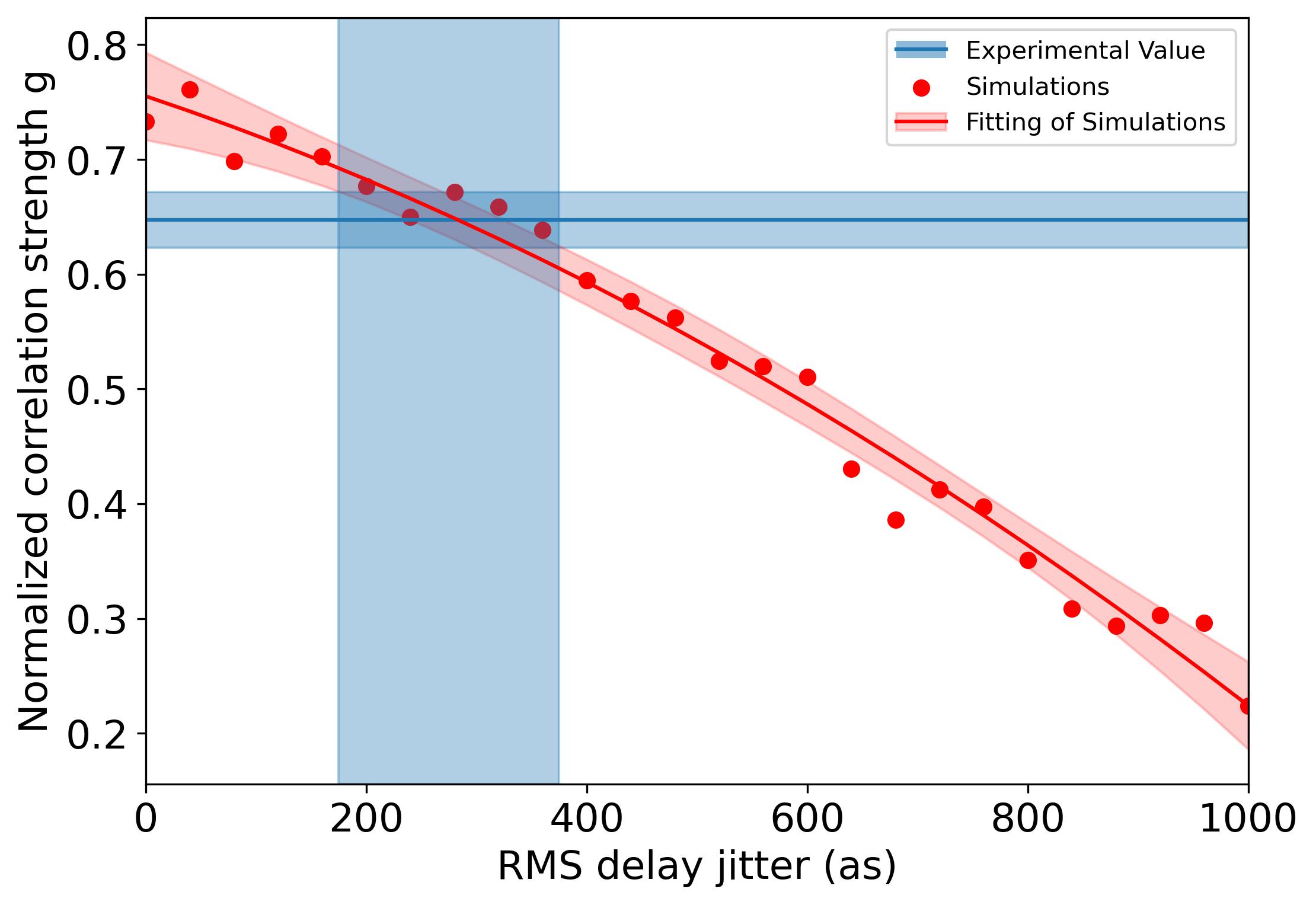}
\caption{\textbf{Fitting of rms delay jitter}.
Red dots represent the value of $g$ for each rms delay jitter $\Delta \tau_{\mathrm{jitter}}$ in simulations. 
The solid red line is the polynomial fitting of $g(\Delta \tau_{\mathrm{jitter}})$, and the transparent red area shows the uncertainty of fitting, $\pm 3\sigma_g$. 
The transparent horizontal blue line represents the value of $g = 0.64\pm 0.02$ calculated in experimental data. }
\label{fig_s:Fitting_of_Delay_Jitter}
\end{figure}

The comparison between the values of $g$ calculated in the experimental data and the numerical simulations is shown in Fig.~\ref{fig_s:Fitting_of_Delay_Jitter}. 
Matching the experimental value of $g = 0.64\pm 0.02$ to simulations yields an rms delay jitter of $270\pm 100~$as. 
This value can be interpreted as the upper bound of rms delay jitter in experimental data. 
Detailed investigation of the values of $g$ calculated in the experimental data shows that the value of $g$ still has a weak dependence on the streaking amplitude: the value of $g$ increases from $0.57\pm 0.02$ in shots that include all the streaking amplitudes to $0.64\pm 0.02$ in shots with the highest 30$\%$ streaking amplitudes. 
Our interpretation is that the correlation and covariance maps of the anti-symmetric parts of c-VMI images still contain fluctuations from other sources of jitter, because the detector response is not perfectly symmetric and linear.

\newpage
\subsection{Cosine Fitting Analysis on Sorted Data}\label{sec_s:cosine_fitting_method}

In the cosine fitting method, the momentum shift of the low-energy (fluorine $K$-shell) photoemission feature is used as a single-shot diagnostic of the streaking phase $\phi_{2\omega}$ of the $2\omega$ pulses. 
The images are grouped into bins corresponding to different streaking directions $\phi_{2\omega}$ of the fluorine $K$-shell photoemission feature. 
An average streaked 2D momentum spectrum $\overline{M}(p_x, p_y; \phi_{2\omega})$ is calculated by averaging all images in each bin of $\phi_{2\omega}$. 
Following the same model described in Eq.~\ref{eq_s:correlation_analysis_model_1}, for the low-energy (fluorine $1$s) photoemission feature, the average sorted streaked image $\overline{M}(p_x, p_y;\phi_{2\omega})$ can be approximated as,
\begin{equation}
\overline{M}(\theta, p_r;\phi_{2\omega}) - \overline{M}_0(\theta, p_r)
\simeq -\frac{\partial \overline{M_0}}{\partial p_r} \overline{\Delta p}\cos(\phi_{2\omega}-\theta) - \frac{1}{p_r}\frac{\partial \overline{M_0}}{\partial \theta} \overline{\Delta p} \sin(\phi_{2\omega}-\theta). \label{eq_s:cosine_fitting_model_2}
\end{equation}
where $\overline{M_0}(p_x, p_y)$ is the average image without streaking and $\overline{\Delta p}$ is the average momentum shift amplitude (or streaking amplitude). 
Following the second assumption made in Sec.~\ref{sec_s:correlation_analysis}, the average unstreaked image $\overline{M}$, the average gradient field $\nabla \overline{M}$, and the average streaking amplitude $\overline{\Delta p}$ are the same in all bins of $\phi_{2\omega}$. 
Equation~\ref{eq_s:cosine_fitting_model_2} can be further simplified if we wisely choose the ROI to be the region slightly above the radial maximum of the photoelectron yield, the same as in the correlation analysis (shown in Fig.~\ref{fig_s:ROI_for_Delay_Analysis}). 
In this ROI, the gradient field $\nabla \overline{M_0}$ is dominated by the radial gradient $\partial \overline{M_0}/\partial p_r$, which further simplifies Eq.~\ref{eq_s:cosine_fitting_model_2} to:
\begin{equation}
\overline{M}(\theta, p_r;\phi_{2\omega}) - \overline{M}_0(\theta, p_r) \simeq \overline{\Delta p} \left|\frac{\partial \overline{M_0}}{\partial p_r} \right| \cos\left(\phi_{2\omega}-\theta \right). \label{eq_s:cosine_fitting_model_3}
\end{equation}
Equation~\ref{eq_s:cosine_fitting_model_3} proves that the differential signal in the ROI of the fluorine $K$-shell photoemission feature has a cosine dependence on the streaking direction $\phi_{2\omega}$, with an absolute phase offset given by the detector angle $\theta$. 
For the carbon $K$-shell photoemission feature, the streaking angle $\phi$ in Eq.~\ref{eq_s:cosine_fitting_model_3} is replaced by $\phi_{2\omega}- \Delta\phi$, because the average delay $\Delta \tau = \Delta\phi/2\pi\times T_L$ results in an angular difference $\Delta \phi$ in the average streaking directions of two photoemission features. 
This explains why the two photoemission features will have a cosine dependence on $\phi_{2\omega}$ with an overall phase offset $\Delta\phi$ given by the average delay.

To fit the average delay $\Delta \tau = \Delta \phi/2\pi \times T_L$ between $\omega/2\omega$ pulses, we do cosine fittings with respect to $\phi_{2\omega}$ in 36 boxes above each photoemission feature in the polar coordinate.
Each box spans $5^\circ$ of the detector angles $\theta$. 
For each photoemission feature, there are 18 boxes between $-45^\circ \leq \theta \leq 45^\circ$, and another 18 boxes between $135^\circ \leq \theta \leq 225^\circ$. 
We neglected regions with little photoelectron signals in dipolar photoemission features to avoid numerical instabilities in cosine fittings and to ensure that the radial gradient dominates the gradient field. 
In each pair of boxes at the same detector angle $\theta$, an average phase shift $\Delta \phi(\theta)$ is calculated from two cosine fittings with respect to $\phi_{2\omega}$, one in the carbon $K$-shell photoemission feature and another one in the fluorine $K$-shell photoemission feature. 
A global phase shift $\Delta \phi\pm \delta_{\Delta \phi}$ is averaged on the 36 phase shifts calculated in the 36 pairs of boxes. 
The uncertainty $\delta_{\Delta \phi}$ is defined as the standard deviation of the phase shifts of cosine fitting calculated in 36 pairs of boxes. 
\begin{equation}
\begin{split}
\Delta \phi & = \frac{\sum_{\theta} \Delta \phi(\theta) }{36}, \\
\delta_{\Delta \phi} & = \sqrt{\frac{\sum_{\theta} \left( \Delta \phi(\theta)-\Delta \phi\right)^2}{36}}. \label{eq_s:average_phase_shift_1}
\end{split}
\end{equation}
The average delay analysed in the cosine fitting method is $(\Delta \phi\pm \delta_{\Delta \phi})/2\pi \times T_L$. 

\subsection{Comparison of Two Delay Analysis Methods}
\begin{figure}[!ht]
\includegraphics[width=\linewidth]{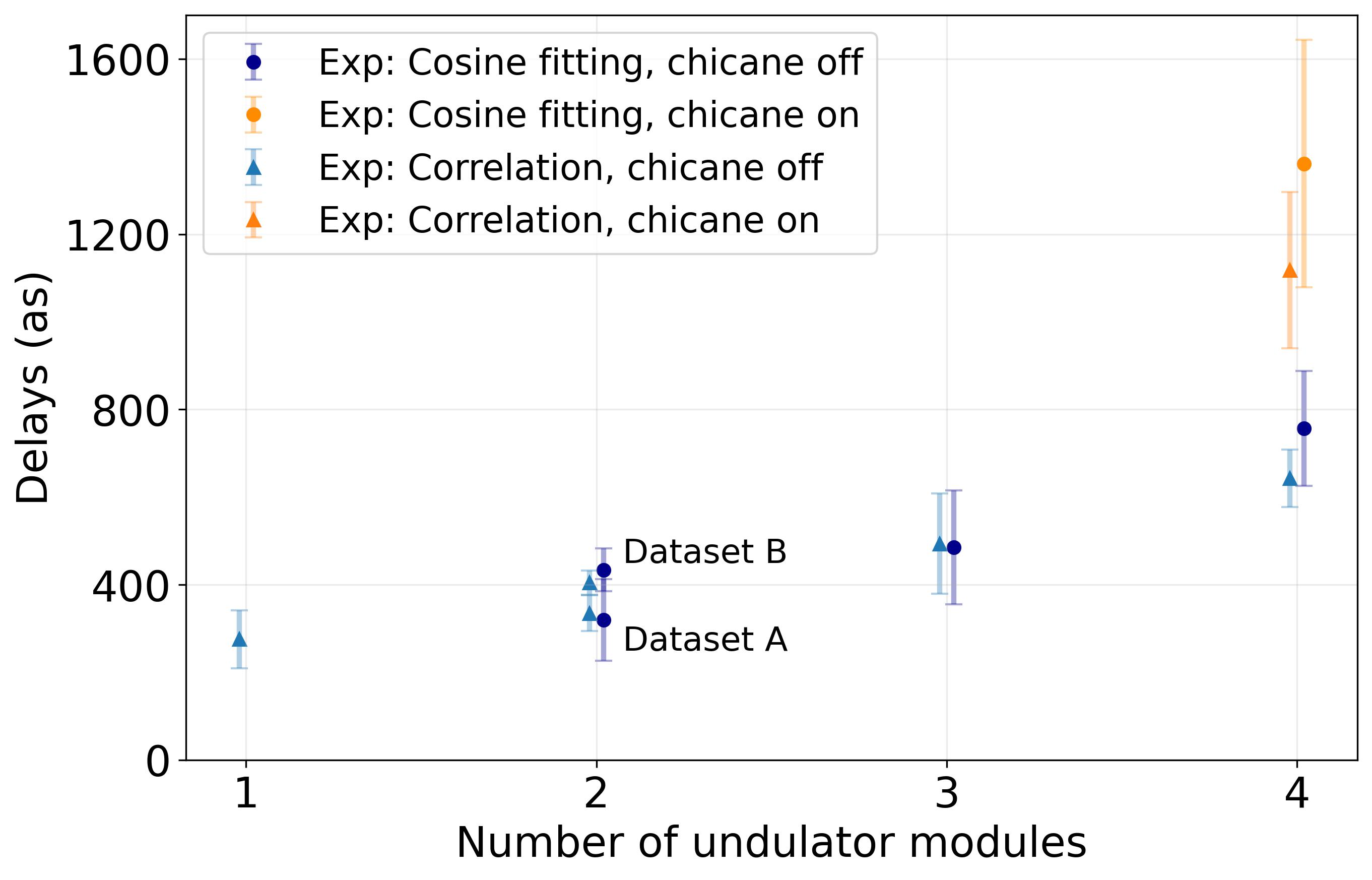}
\caption{Comparison between average delays fitted in two methods.}
\label{fig_s:Comparison_of_Delay_Analyses}
\end{figure}

Figure~\ref{fig_s:Comparison_of_Delay_Analyses} shows the comparison between two methods of analyzing average delays. 
The minimal delay with chicane off and 1 undulator module only for $2\omega$ pulse is not analysed in the cosine fitting analysis, because the photoelectron yield in the fluorine $K$-shell (inner) photoemission features ionised by $2\omega$ pulses was too low for robust single-shot determinations of the XFEL arrival time. 
The delay fittings of the two methods agree in all other beamline configurations. 
Figure~3 panel~(a) in the main text shows the average time-delays from the correlation analyses.

The first differential map, corresponding to an average delay $0.4~$fs, in Fig.~2 in the main text represents the dataset B.
By comparing datasets A and B shown in Fig.~\ref{fig_s:Comparison_of_Delay_Analyses}, our data analysis shows that changing the BC2 current has a negligible effect on the average delay between $\omega/2\omega$ pulses. 
The dataset B is not shown in Fig.~3(a) in the main text.

\section{XFEL Simulations of $\omega/2\omega$ Modes}
Our streaking diagnostic provides a high-resolution measurement of the time-delay between two pulses. However, streaking measurements are time-consuming and difficult to perform in parallel with other experiments.
To exploit the results of this paper in pump/probe experiments, it is therefore important to understand how these results scale with different beam properties and photon energies. 
To this end, we perform start-to-end simulations of our two-colour setup. 
In the first sub-section, we benchmark our streaking results and how they are affected by changes in the gain length. In the second sub-section, we study the scaling of the two-colour delay with the XFEL wavelength.

\subsection{Simulations of $365~$eV/$730~$eV Modes}

\subsubsection{Benchmark Simulation with Delay Chicane Off}\label{sec_s:Benchmark_Simulation}

\begin{figure}[!ht]
\includegraphics[width=\linewidth]{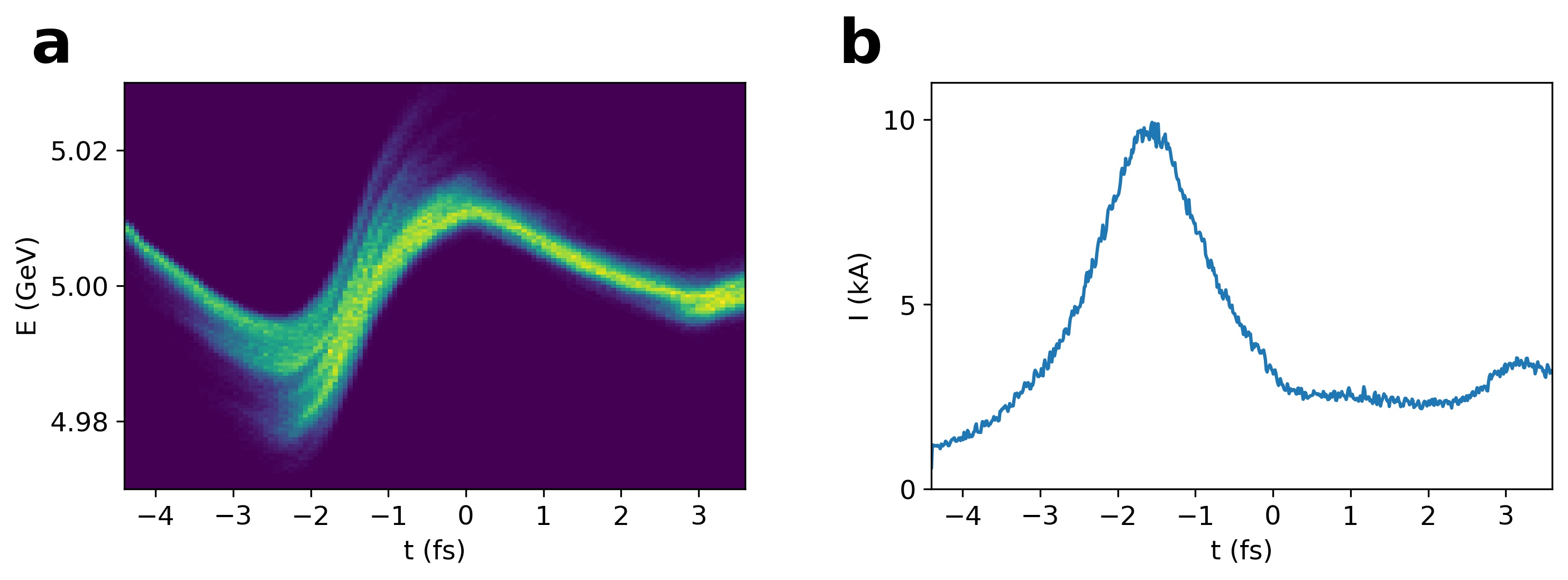}
\caption{\textbf{The ESASE current spike used in XFEL simulations}.
\textbf{a-b}, The longitudinal phase space distribution (\textbf{a}) and the current profile (\textbf{b}) of the e-beam used in XFEL simulations. 
The x-axis is the longitudinal bunch coordinate in the units of femtoseconds.
}
\label{fig_s:phase_space_and_current}
\end{figure}
Figure~\ref{fig_s:phase_space_and_current} shows the details of the ESASE current spike used in the benchmark simulation that reproduces the angular streaking measurements of time delays. 
Particle tracking simulations are performed with Elegant~\cite{borland2000elegant} to track the dynamics of electron bunch. 
The peak-to-peak energy modulation in the simulated ESASE current spike is $\sim 35$~MeV. 
The FWHM of the simulated ESASE current spike is $\sim 2~$fs.

We use the fully 3D code Genesis 1.3 version 4~\cite{reiche1999genesis, reiche2014update} to perform time-dependent XFEL simulations of $\omega/2\omega$ modes on the ESASE current spike shown in Fig.~\ref{fig_s:phase_space_and_current}. 
A two-stage XFEL simulation is performed to simulate the $\omega/2\omega$ modes. 
In the first stage simulation, the ESASE current spike shown in Fig.~\ref{fig_s:phase_space_and_current} is used to simulate the radiation of $\omega$ pulses starting from shot noise. 
The strongly bunched electron beam, stored as the slice-wise particle distribution, is dumped at the end of the first-stage simulation. 
The slice-wise pre-bunched electron beam is then loaded into the second-stage simulation to calculate the radiation of $2\omega$ pulses from the strong microbunching. 

\begin{figure}[!ht]
\begin{center}
\includegraphics[width=0.75\linewidth]{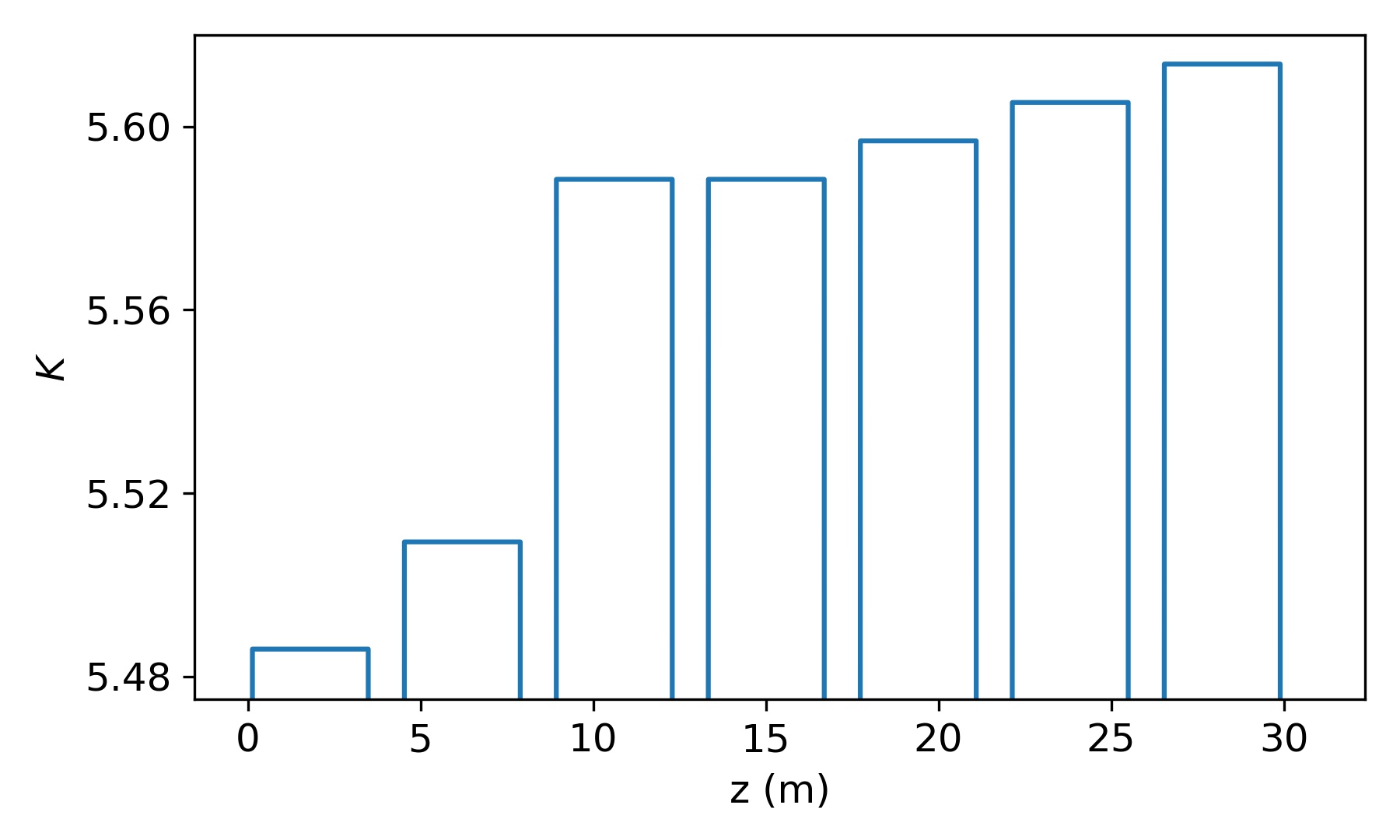}
\caption{The undulator taper used in XFEL simulations to generate $365~$eV ($\omega$) attosecond XFEL pulses with the electron beam shown in Fig.~\ref{fig_s:phase_space_and_current}.
 }
\label{fig_s:1st_Stage_Taper_for_Simulations}
\end{center}
\end{figure}

\begin{figure}[!ht]
\includegraphics[width=\linewidth]{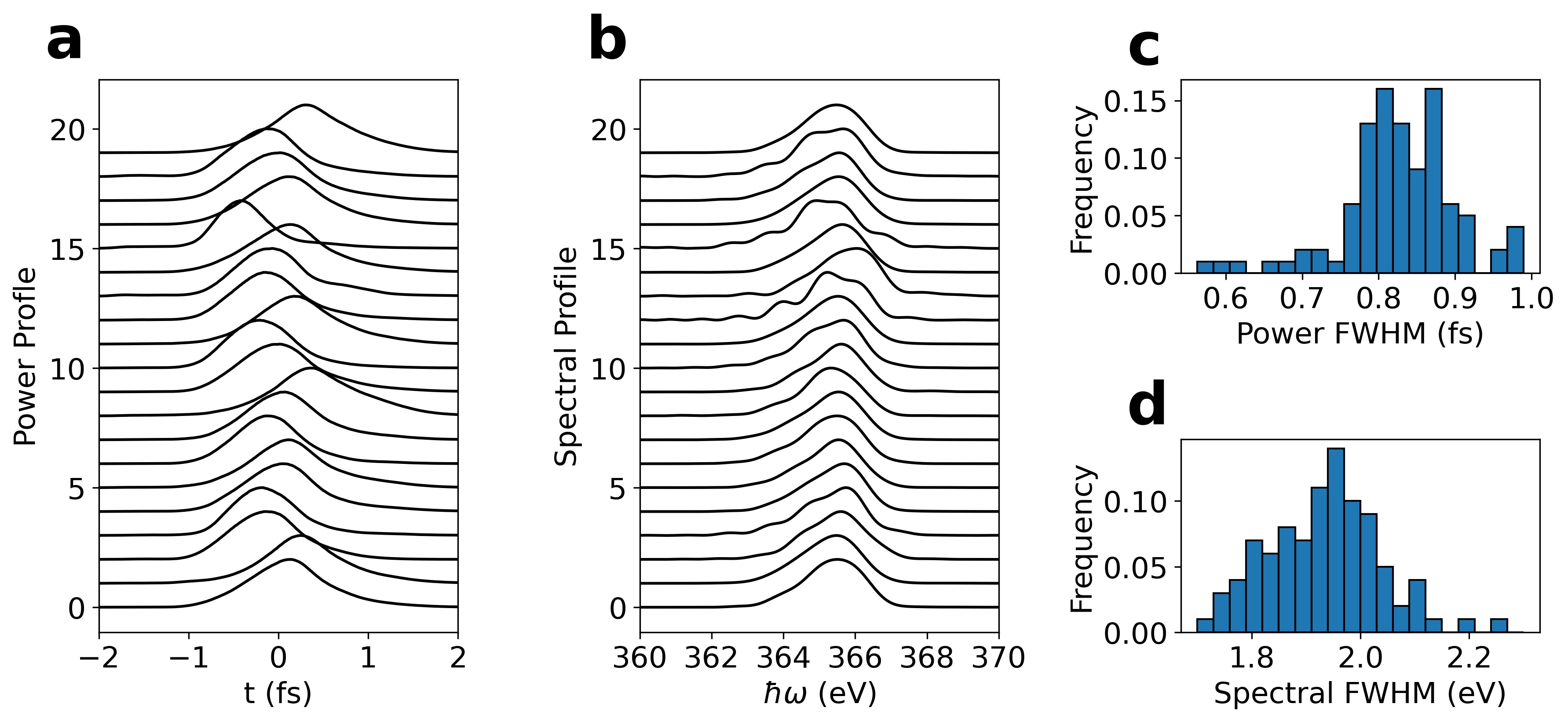}
\caption{\textbf{Simulated $\omega$ pulses}.
The XFEL simulations of $\omega$ pulses are done on the electron bunch shown in Fig.~\ref{fig_s:phase_space_and_current} and the undulator taper shown in Fig.~\ref{fig_s:1st_Stage_Taper_for_Simulations}.
\textbf{a-b}, Typical 20 shots of simulated $365~$eV ($\omega$) pulses in the temporal domain (\textbf{a}) and the spectral domain (\textbf{b}). 
\textbf{c-d}, Distributions of pulse lengths (\textbf{c}) and spectral widths (\textbf{d}) of 100 shots of XFEL simulations.} 
\label{fig_s:Statistics_of_Omega_Pulses}
\end{figure}

The undulator taper for $365~$eV ($\omega$) pulses in our first-stage simulations is shown in Fig.~\ref{fig_s:1st_Stage_Taper_for_Simulations}, which has the same shape as the taper used in the experiment (shown in the Methods section of the main text), except that the last two undulator modules are removed. 
We used only seven undulator modules in the first stage to avoid saturation of $\omega$ pulses in our simulations. 
The properties of simulated $\omega$ pulses are shown in Fig.~\ref{fig_s:Statistics_of_Omega_Pulses}. 
Fig.~\ref{fig_s:Statistics_of_Omega_Pulses}~(a) and (b) show 20 shots of $\omega$ pulses in the temporal domain and in the spectral domain, respectively. 
By comparing $\omega$ pulses in Fig.~\ref{fig_s:Statistics_of_Omega_Pulses}~(a) and the electron beam current profile in Fig.~\ref{fig_s:phase_space_and_current}~(b), the simulations show that the peaks of the $\omega$ pulses remain close to the falling edge of the ESASE current spike towards the bunch head, in agreement with the temporal gain-narrowing model in Ref.~\cite{baxevanis2018time}. 
Fig.~\ref{fig_s:Statistics_of_Omega_Pulses}~(c) and (d) show distributions of FWHM pulse lengths and spectral widths of 100 shots of XFEL simulations of $\omega$ pulses initialised with different random seeds. 
The FWHM spectral width of simulated $\omega$ pulses is $1.9\pm 0.2~$eV, which is $\sim30\%$ narrower than the experimental measurements. 
The small difference in FWHM bandwidths is probably due to a discrepancy in the slice energy spread caused by the use of a number of macroparticles smaller than the physical number of particles. 

Simulations of $2\omega$ pulses are done using slice-wise pre-bunched electron beams dumped at the end of the first-stage simulations.  
The value of $K$ in the second stage simulation is optimized to match the simulated time-delays with the angular streaking measurements of time-delays when the magnetic chicane is turned off. 
The simulated time delay $\Delta \tau$ is calculated as the difference between the centres of mass of $\omega$ and $2\omega$ pulses. 
The slippages of the $\omega$ and $2\omega$ pulses must be correctly calculated to correctly calculate the centres of mass of $\omega$ and $2\omega$ pulses as a function of $z$ in the second-stage undulator beamline. 
The 3D code Genesis 1.3 version 4 automatically processes the slippage of $2\omega$ pulses within bunch coordinate in the second stage. 
The slippage of $\omega$ pulses is calculated in the post-processing stage assuming propagation at the speed of light.

\begin{figure}[!ht]
\includegraphics[width=\linewidth]{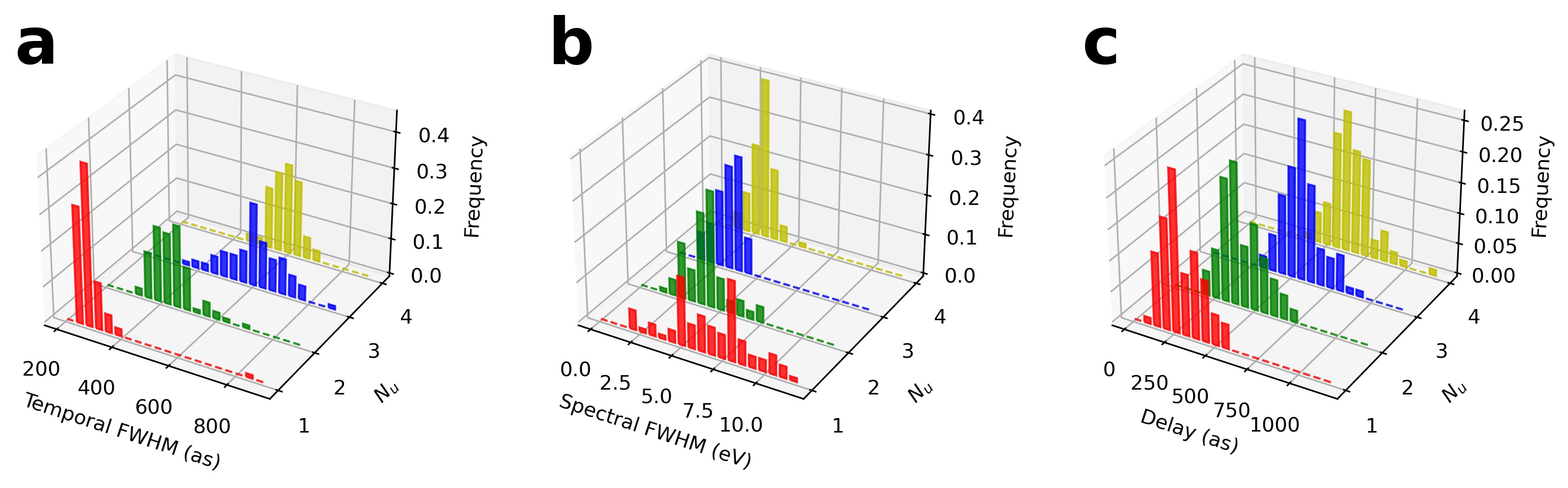}
\caption{\textbf{Properties of $2\omega$ pulses in XFEL simulations}. 
\textbf{a}, The distribution of FWHM pulse lengths of $2\omega$ pulses at the end of each undulator modules. 
\textbf{b}, The distribution of FWHM spectral widths of $2\omega$ pulses at the end of each undulator modules. 
\textbf{c}, The distribution of the time-delays between $\omega$ and $2\omega$ pulses at the end of each undulator modules. 
The time-delays are calculated as delays between centers of mass of $\omega$ and $2\omega$ pulses.}
\label{fig_s:Properties_of_2Omega_Pulses_in_Simulations}
\end{figure}

Figure~\ref{fig_s:Properties_of_2Omega_Pulses_in_Simulations} shows the properties of the simulated $2\omega$ pulses vs. the number of undulator modules $N_u$ in the second stage, without chicane delay. 
Figure~\ref{fig_s:Properties_of_2Omega_Pulses_in_Simulations} panel (a) shows the distributions of FWHM $2\omega$ pulse lengths at the end of each undulator module. 
The FWHM $2\omega$ pulse length is $300\pm90~$as, $440\pm70~$as, $610\pm100~$as, $590\pm50~$as at the end of the first, second, third and fourth undulator modules, respectively. 
In the first undulator module, short $2\omega$ pulses with peak power $\sim 1~$ GW are generated from coherent radiation from strong and nonlinear microbunching.  
Our interpretation of the simulated $2\omega$ pulse lengths is that the 1-GW-level pulse energy of $2\omega$ pulses, the ``seed", is not strong enough to directly start the superradiant effect.

Figure~\ref{fig_s:Properties_of_2Omega_Pulses_in_Simulations} panel (b) shows the distributions of FWHM $2\omega$ spectral widths at the end of each undulator module. 
The spectral width of the FWHM $2\omega$ gradually decreases from $6.9\pm2.3~$eV at the end of the first undulator module to $3.1\pm0.7~$eV at the end of the fourth undulator module, in agreement with the pulse length broadening in the temporal domain. 
Figure~\ref{fig_s:Properties_of_2Omega_Pulses_in_Simulations} panel (c) shows the distributions of time delays at the end of each undulator module. 
A quasi-linear change in average delays versus the number of undulator modules can be observed.

\subsubsection{Simulations on Electron Beams With Emittance Scans}

We performed simulations of $365~$eV/$730~$eV pulse pairs on electron bunches with scaled emittances to study how the electron beam quality can affect time delays when the delay chicane is turned off. 
A smaller (larger) emittance means that the transverse phase-space distribution of the electron beam is more compact (dilute), indicating a better (worse) electron beam quality. 
The projected normalised emittances of the ESASE current spike used in the benchmark simulations in Sec.~\ref{sec_s:Benchmark_Simulation} are $\epsilon_{n, x} = 1.17~\mu$m and $\epsilon_{n, y} = 0.45~\mu$m.  
We scaled the emittance by multiplying the 4D transverse phase space $(x, x^{\prime}, y, y^{\prime})$ of the whole ESASE current spike with the same scaling factor:
\begin{equation}
    x \to r^{1/2}x,~x^{\prime}\to r^{1/2}x^{\prime}, y \to r^{1/2}y,~y^{\prime}\to r^{1/2}y^{\prime}, \label{eq_s:emittance_Scaling}
\end{equation}
where $r$ is the emittance scaling ratio.
The longitudinal phase space of the current spike is not changed. 
The undulator beamline lattice profile used in simulations on electron beams with emittance scans is the same as that in the benchmark simulation discussed in Sec.~\ref{sec_s:Benchmark_Simulation}.

\begin{figure}[!ht]
\includegraphics[width=\linewidth]{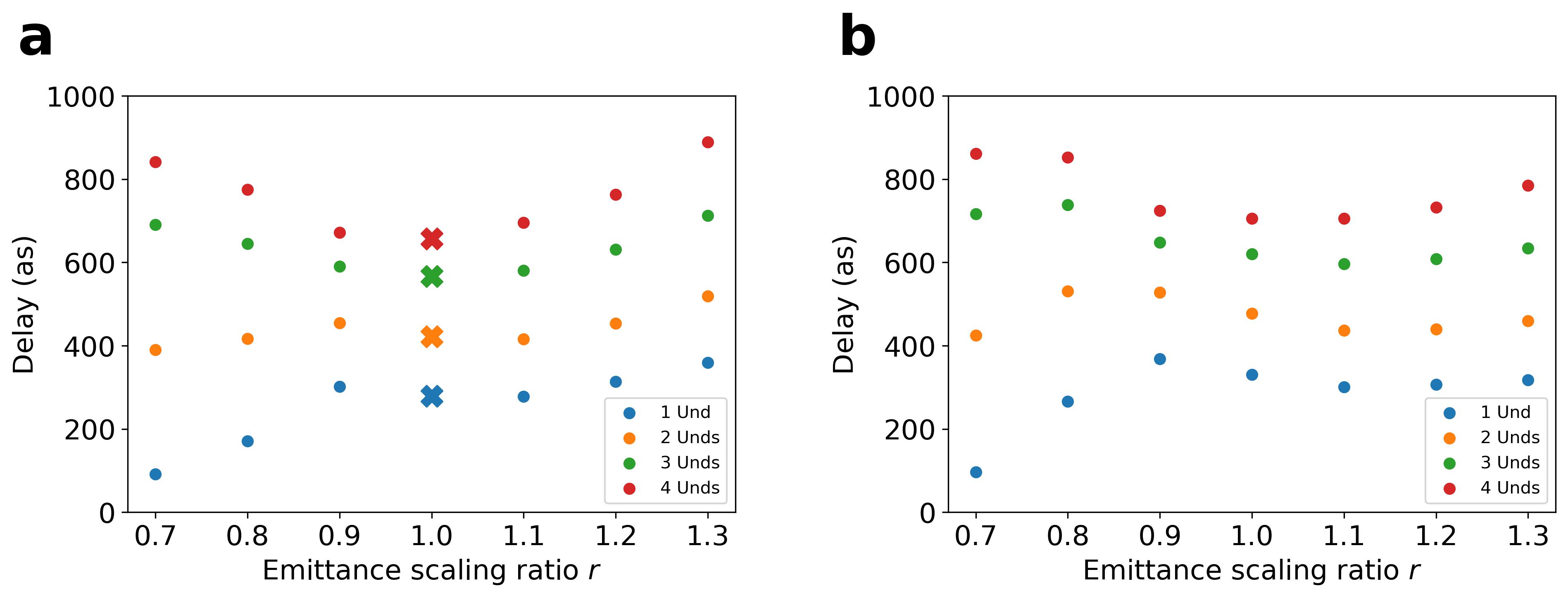}
\caption{\textbf{Time-delay simulations with electron beam emittance scans.}
XFEL simulations of $365~$eV/$730~$eV pulse pairs in the second stage are done on electron beams with scaled emittances. 
The plotted average delays are (\textbf{a}) the arithmetic mean of single-shot delays and (\textbf{b}) the delay between the means of two pulses averaged on all shots. 
Delays with $r=1$ plotted in panel (a) are simulations benchmarked with the experimental delay measurements. 
In both methods shown in panels (a) and (b), the change of average time-delays is within $40\%$ even if the electron beam emittance is increased by $30\%$ in both x and y. 
}
\label{fig_s:time_delay_vs_emittance_scan}
\end{figure}

Figure~\ref{fig_s:time_delay_vs_emittance_scan} shows the simulations of average time delays against the emittance scaling ratio, with the delay chicane turned off. 
We use two methods to calculate the mean of time delays averaged on all simulated shots. 
In the first method, the single-shot time delay was calculated as the difference between the centres of mass of two pulses in each shot of XFEL simulations. 
The mean time delay, shown in Fig.~\ref{fig_s:time_delay_vs_emittance_scan}(a), was calculated as the arithmetic mean of all single-shot delays. 
In the second method, we first calculated the means of the temporal profiles of two pulses averaged on all shots. 
Then, the average time delay shown in Fig.~\ref{fig_s:time_delay_vs_emittance_scan}(b) was calculated as the difference between the centres of mass of two average pulses. 
The average time delay calculated in the second method can be approximately regarded as the weighted arithmetic mean of the time delays averaged with respect to the pulse energies. 
In both methods, the minimal delays decrease to $\sim 100~$ when the electron beam emittance is reduced by $30\%$. 
Simulations show that the bunch tail radiates more $365~$eV photons if the electron beam emittance is $30\%$ smaller. 
More $365~$eV photons towards the bunch tail make the average time delay shorter. 
When increasing the electron beam emittance to $30\%$, the change in average time delays is less than $40\%$ in the first method or even $12\%$ in the second method. 
The discrepancy between average time delays defined in the two methods on degraded electron beams can be interpreted as follows: 
When the quality of the electron beam is significantly reduced, some $\omega/2\omega$ pulse pairs generated by the degraded electron beam have low pulse energies and noisy temporal shapes. 
These weak $\omega/2\omega$ pulse pairs could have time delays longer than $1~$ fs, which effectively makes the arithmetic mean of time delays longer. 
However, weak pulses have a negligible contribution to the mean of the temporal pulse profile, making the average delays defined in the second method robust to weak and noisy pulses. 
In conclusion, our simulation shows that the average delays between $\omega/2\omega$ pulse pairs are insensitive to degraded electron beam quality.

\subsubsection{Simulations with Delay Chicane Turned On}

We perform simulations of attosecond $365~$eV/$730~$eV pulse pairs with the delay chicane on to study how partially dispersed suppressed microbunching could affect time delays. 
The magnetic chicane used in our experiment to control the time-delay between $\omega/2\omega$ pulses is a 4-dipole magnetic chicane. 
The magnetic chicane was originally designed to perform the soft X-ray self-seeding experiments~\cite{feng2012system, ratner2015experimental}. 

In our FEL simulation with the delay chicane on, the simulations of $365~$eV ($\omega$) pulses are done in the same way as discussed in Sec.~\ref{sec_s:Benchmark_Simulation}. 
For a short chicane delay ($\leq 1~$fs), we directly set the chicane lattice in the second-stage beamline profile to produce an $1~$fs delay in Genesis 1.3 version 4. 

\begin{figure}[!ht]
\includegraphics[width=\linewidth]{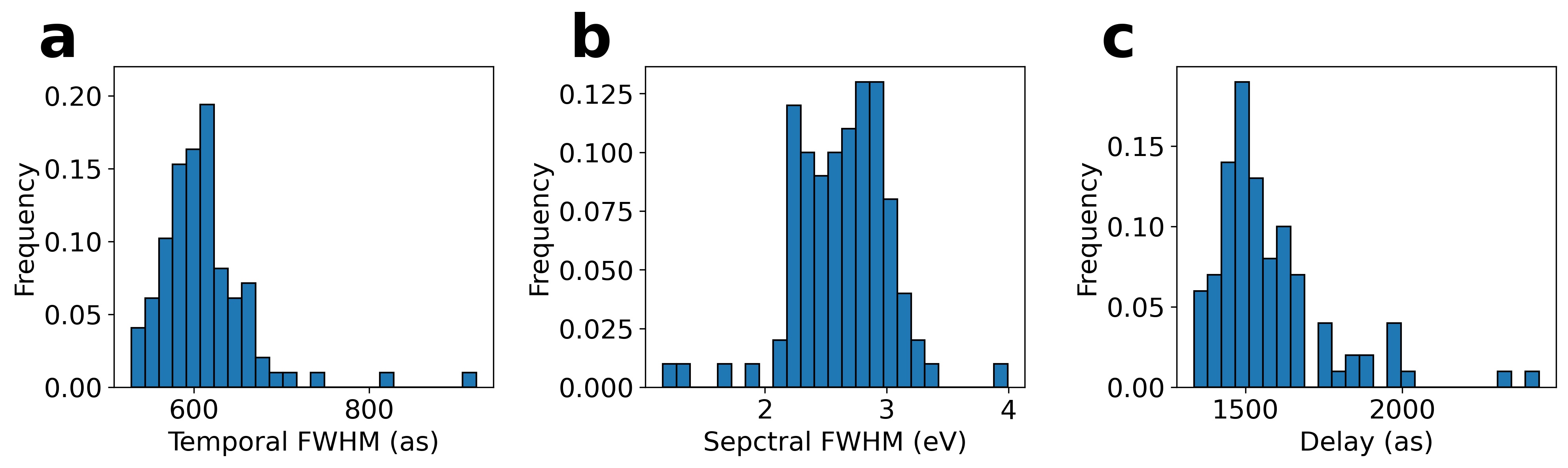}
\caption{\textbf{Simulated $730~$eV pulses with 1 fs chicane delay and 4 undulator modules for $730~$eV pulses}.
\textbf{a-c}, Distributions of simulated FWHM pulse lengths (\textbf{a}), FWHM spectral widths (\textbf{b}), and time-delays between pulse pairs (\textbf{c}). 
}
\label{fig_s:Properties_of_2Omega_Pulses_with_Chicane_On_in_Simulations}
\end{figure}

Figure~\ref{fig_s:Properties_of_2Omega_Pulses_in_Simulations} shows the simulations of $2\omega$ pulses from the pre-bunched electron beams with $1~$fs chicane delay and 4 undulator modules for $2\omega$ pulses. 
The FWHM pulse length and spectral width of simulated $2\omega$ pulses are $612\pm 54~$as and $2.6\pm 0.4~$eV. 
The time delay calculated as the difference in the centres of mass of two pulses is $1574\pm 187~$as, which is plotted as the ``Sim: $1~$fs chicane delay'' point in Fig.~3(a) of the main text. 
The discrepancy between the simulation and the angular streaking measurement of the ``Chicane On'' data point could possibly come from the inaccuracy of the actual delay produced by the magnetic chicane. 
The chicane was originally designed to produce delays from $100$s~fs to $1~$ps in the soft X-ray self-seeding experiments~\cite{feng2012system, ratner2015experimental}. 
The actual delay produced by the chicane could be a few hundred attoseconds different from the nominal value if the chicane is set to produce delays down to one femtosecond. 
More simulations show that the angular streaking measurement of the ``Chicane on" data point agrees with simulations performed with $0.6-0.75~$fs chicane delay.

\subsection{Simulations of $\omega/2\omega$ Modes at Different Photon Energies}\label{sec_s:Simulations_of_w_2w_Modes_at_Different_Photon_Energies}

We performed XFEL simulations of $\omega/2\omega$ modes at different XFEL photon energies to study the dependence of the time delays on XFEL wavelengths. 
The electron beam used in these simulations is the same beam used in the benchmark simulation discussed in Sec.~\ref{sec_s:Benchmark_Simulation}.

Similarly, as in the simulation of $365~$eV/$730~$eV pulse pair, a two-stage simulation is done at each pair of $\omega/2\omega$ photon energies: the first-stage simulation calculates the radiation of $\omega$ pulses starting from shot noise; the strongly bunched electron beam is dumped as a slice-wise particle distribution at the end of the first-stage simulation and reloaded into the second-stage simulation to calculate the radiation of $2\omega$ pulses from strong microbunchings. Since the FEL gain length decreases with the $\omega$ photon energy, different numbers of undulator modules are used for different photon energies in the first-stage simulations (ranging from 5 to 7) to prevent saturation of the first pulse.
In all first-stage simulations, the tapering of the undulator parameter $K$ starts from the fifth undulator module to match the energy chirp in the ESASE current spike. 
In all second-stage simulations, four undulator modules with the same $K$ are used for $2\omega$ pulses with the magnetic chicane off. 

\begin{figure}[!ht]
\includegraphics[width=\linewidth]{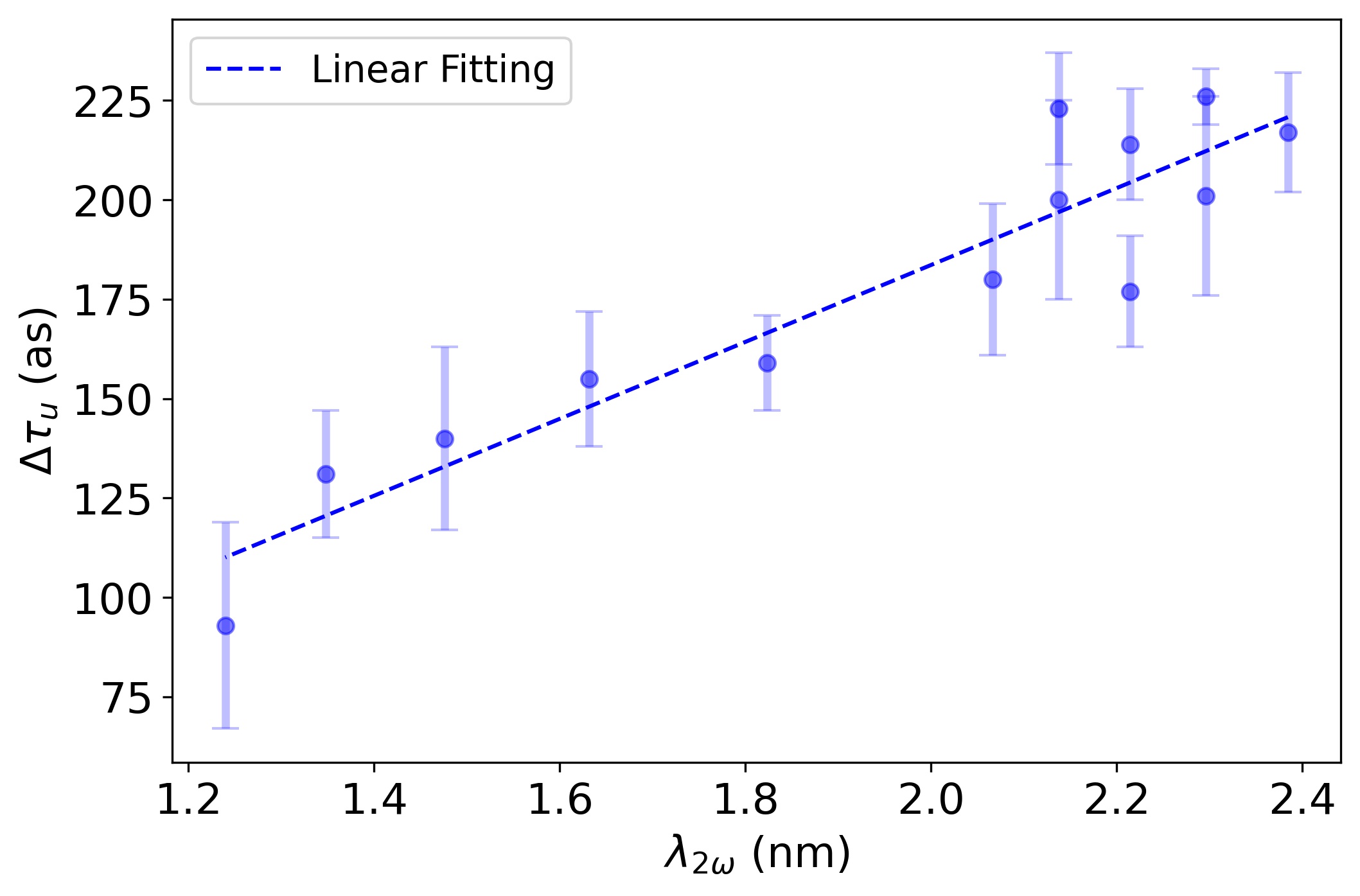}
\caption{\textbf{Delay change $\Delta \tau_u$ per undulator module vs wavelength $\lambda_{2\omega}$ of $2\omega$ pulse}.
Each dot represents the average delay change $\Delta \tau_u$ per undulator module averaged within $\pm 0.15\%$ of the optimal undulator parameter $K$ for $2\omega$ pulses at the corresponding $2\omega$ pulse wavelength. 
The error bar of each dot represents the standard deviation of $\Delta \tau_u$ calculated within $\pm 0.15\%$ of the optimal undulator parameter $K$. 
The optimal value of $K$ is determined by maximizing the average $2\omega$ pulse energy $E_{2\omega}$ in the undulator $K$ scan for $2\omega$ pulses. 
At $\lambda_{2\omega}=2.296~$nm ($540~$eV), $2.214~$nm ($560~$eV), and $2.1378~$nm ($580~$eV), two sets of FEL simulations of $\omega/2\omega$ modes are done with different undulator beamlines for $\omega$ pulses. 
}
\label{fig_s:delay_change_per_und_vs_wavelength}
\end{figure}

We have scanned $K$ in all second-stage simulations to study how time delays at different XFEL photon energies depend on undulator parameters. 
XFEL simulations show that the value of $\alpha$ is always close to the measured value of $0.7$ when the undulator parameter $K$ maximises the average $2\omega$ pulse energy $E_{2\omega}$. 
It follows that the change of delay $\Delta \tau_u$ per undulator module linearly scales with the $2\omega$ wavelength $\lambda_{2\omega}$ at the optimal $K$, as shown in Fig.~\ref{fig_s:delay_change_per_und_vs_wavelength}.
A linear fit between the delay change $\Delta \tau_u$ per undulator module and the $2\omega$ wavelength $\lambda_u$ shows the following:
\begin{equation}
    \Delta \tau_u [\mathrm{as}] = (91\pm 11) \lambda_{2\omega} [\mathrm{nm}] + (-10\pm22)[\mathrm{as}].
\end{equation}
When the chicane is turned off and the optimal $K$ is used for the $2\omega$ pulses, the average time-delay between $\omega/2\omega$ pulse pairs at the end of the $N_u$-th undulator module is approximately,
\begin{equation}
    \Delta \tau = t_0 + N_u\times\Delta \tau_u(\lambda_{2\omega}). \label{eq_s:estimation_of_total_delay}
\end{equation}
The offset term $t_0$ depends on the detailed electron bunch phase space and undulator beamline configuration. 
For example, for the electron bunch used in the benchmark simulation discussed in Sec.~\ref{sec_s:Benchmark_Simulation}, the value of $t_0$ will change from $\sim 100~$as to $\sim 300~$as by decreasing the $\omega$ pulse photon energy from $500~$eV to $260~$eV.

\section{Pump-probe experiment analysis}

\subsection{Estimation of Pump Fraction}\label{sec:supp:PumpFrac}
\begin{figure}[!ht]
\includegraphics[width=\linewidth]{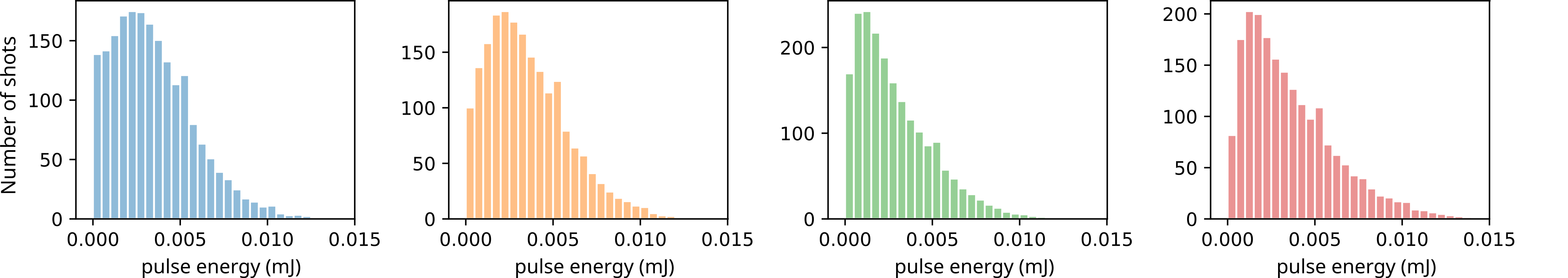}
\caption{ Histograms of pump pulse energy for each delay. The delay is increasing from left to right. 
}
\label{fig_s:Isat}
\end{figure}

Figure~\ref{fig_s:Isat} shows the pump energy distribution measured for each of the four delays. 
The average pulse energy across all delays is 1.8 $\mu$J.
Assuming a beamline transmission of $\sim60\%$~\cite{walter2022time}, and a beam diameter of 1.5~$\mu$m~(full width at half maximum) at the interaction point, we can estimate the pump fluence to be 6.1$\times 10^{-1}$~$\mu$J/$\mu$m$^2$. 
We estimate the saturation fluence to be 0.9$\times 10^{-1}$~$\mu$J/$\mu$m$^2$, based on the carbon $1s$ cross section at 300~eV~\cite{yeh_atomic_1985}. 
Then our pump fluence is $\sim~7$ times the saturation fluence and we assume that all the molecules in the focal volume are core ionised at least once, and thus we do not expect to see signal from ground-state~(un-pumped) molecules. 

\subsection{Covariance analysis plot}
\begin{figure}[!ht]
\centering
\includegraphics[width=0.75\linewidth]{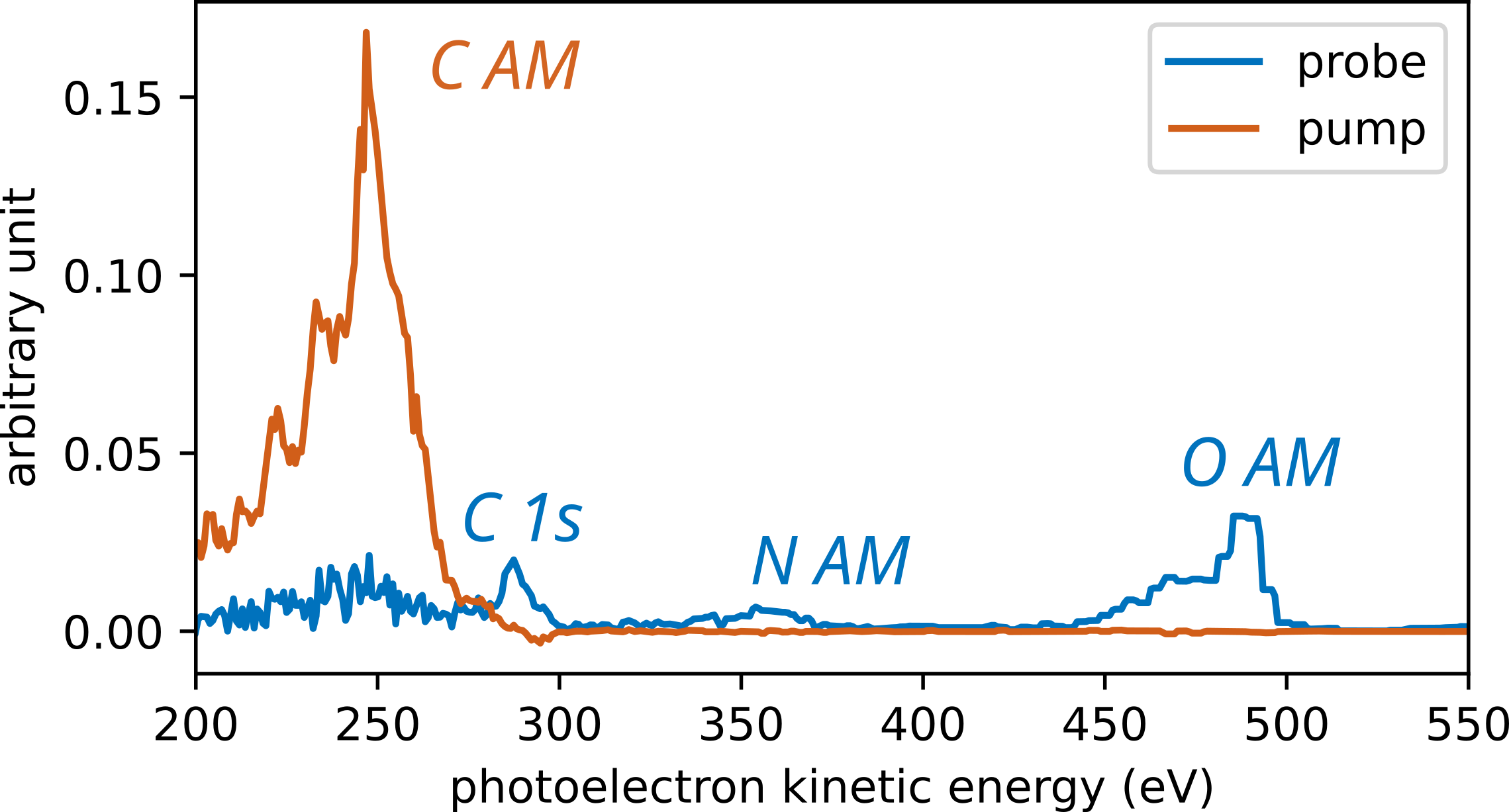}
\caption{ Intensity covariance. Blue shows the covariance of the photoelectron spectrum with the probe pulse energy. Orange is for photoelectron spectrum covariance with the pump intensity. AM refers to the Auger-Meitner features associated with carbon, nitrogen and oxygen sites. C 1s is the dispersing feature of interest. 
}
\label{fig_s:intensity_covariance}
\end{figure}

This section focuses on the covariance analysis used to process the raw experimental data. 
Covariance analysis is used to identify the different features in the measured electron spectra. 
Figure~\ref{fig_s:intensity_covariance} shows the covariance calculated between the electron spectra and the pulse intensity recorded on a shot-to-shot basis for both pump and probe. 
The orange curve shows the contributions from the pump pulse and the blue shows features depending on the probe pulse. 
We can identify the three Auger-Meitner emission features expected when core ionizing aminophenol molecules:~on a carbon site (C AM) due to ionization by the pump, and on the nitrogen (N-AM) and oxygen (O AM) sites for the probe pulse. 
This plot shows that the additional feature just below $300$~eV kinetic energy depends exclusively on the probe intensity and results from carbon $K$-shell~(C 1s) ionization by the probe. 
This rules out the possibility that the dispersive feature could be valence electrons produced by the pump.

\begin{figure}
\includegraphics[width=\linewidth]{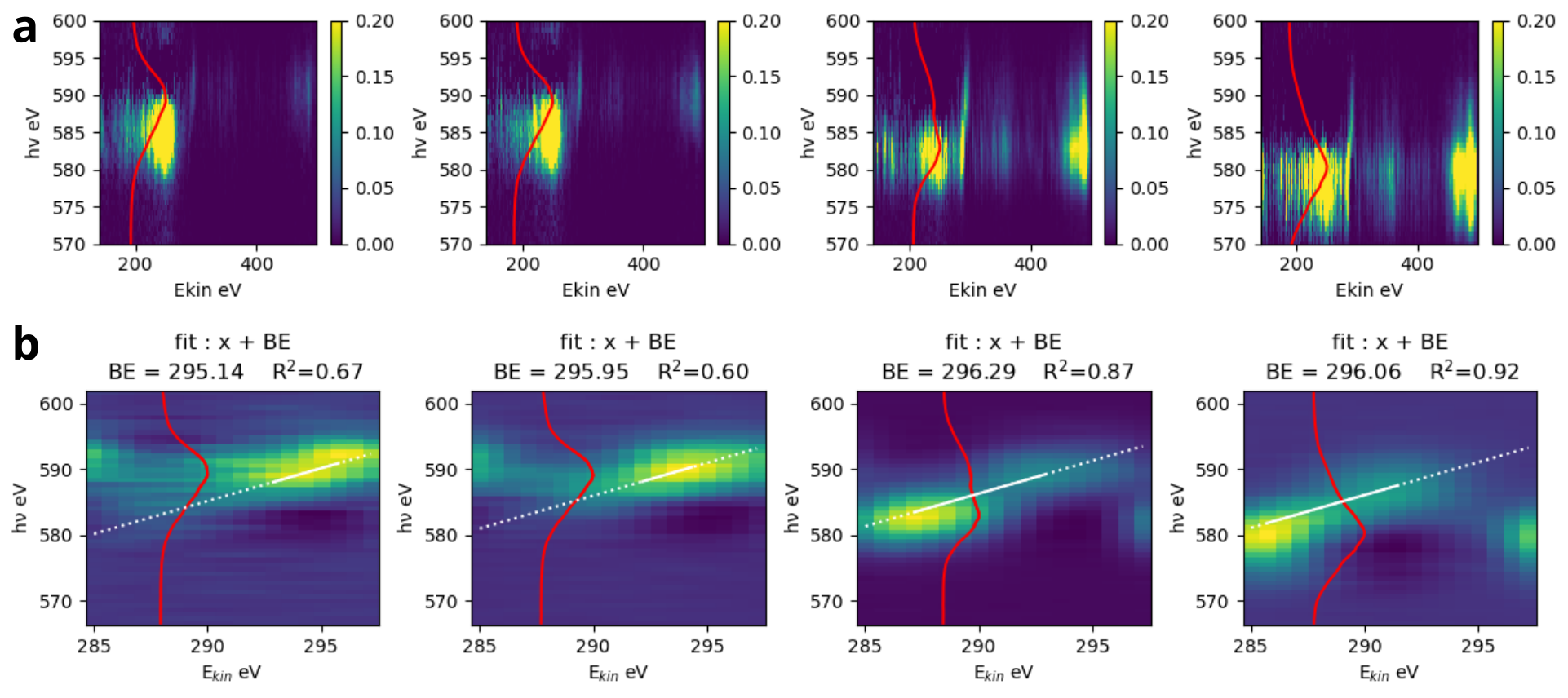}
\caption{ \textbf{a}, Covariance maps of photon spectra with full photoelectron spectra for each delay. The red curve shows the averaged photon spectra for each delay. 
\textbf{b}, Zoom in the ROI, i.e. around the C1s feature. Each pixel is the average of 5 consecutive bins of photoelectron kinetic energy(x-axis) centered around the original bin. 
The white line shows the fit across the photoline (dispersive photoemission feature) of the dispersion law stating $h\nu$ = $E_{kin}$ + BE where BE is the free parameter that we call binding energy in the main paper. 
}
\label{fig_s:covariance_analysis}
\end{figure}

Figure~\ref{fig_s:covariance_analysis} displays the analysis procedure of the photoelectron signal in order to retrieve the binding energy evolution with time delay. This procedure is detailed in the Method section of the paper.
 
\subsection{Modelling PCI : Post Collision Interaction effects} 
\label{Subsection:SI_PCI}

We model the interaction between the outgoing electrons according to the model of Russek and Mehlhorn~\cite{russek_mehlhorn_1986}. 
When a fast electron~($\sim300$~eV) produced by the probe pulse passes the slow electrons~($\sim1-10$~eV) ionised by the pump pulse, the electrostatic screening that the fast electron experiences increases and thus the energy of the fast electron increases.
If the fast electron overtakes the slow electrons at a radius $\rho$, the energy of the fast electron increases by $\rho^{-1}$, and as a consequence of energy conservation, the slow electron from the pump decreases by this same amount.

To determine $\rho$, we calculate the time, $\tau$, when the electrons from the probe pulse overtake the slow electrons from the pump: 
\begin{equation}
     \tau = \int_{R_0}^{\rho} \frac{dr}{v_1(r)}\, =\,\int_{R_0}^{\rho} \frac{dr}{v_2(r)} + \Delta t,
     \label{eq_s:PCI}
\end{equation}
where $R_0$ is the radii at which the electrons originate, 
$v_{1}=\sqrt{2(E_1+1/r)}$ is the velocity of the electrons ionised by the pump pulse~($E_1=1-10$~eV), $v_2(t)=\sqrt{2(E_2+2/r)}$ is the velocity of the electrons ionised by the probe pulse~($E_2=300$~eV), and $\Delta t$ is the time delay between the pump and the probe pulses.
For this calculation we have assumed $R_0=1$~a.u., but the results are rather insensitive to the choice of $R_0$.

\begin{figure}[h!]
\includegraphics[width=\linewidth]{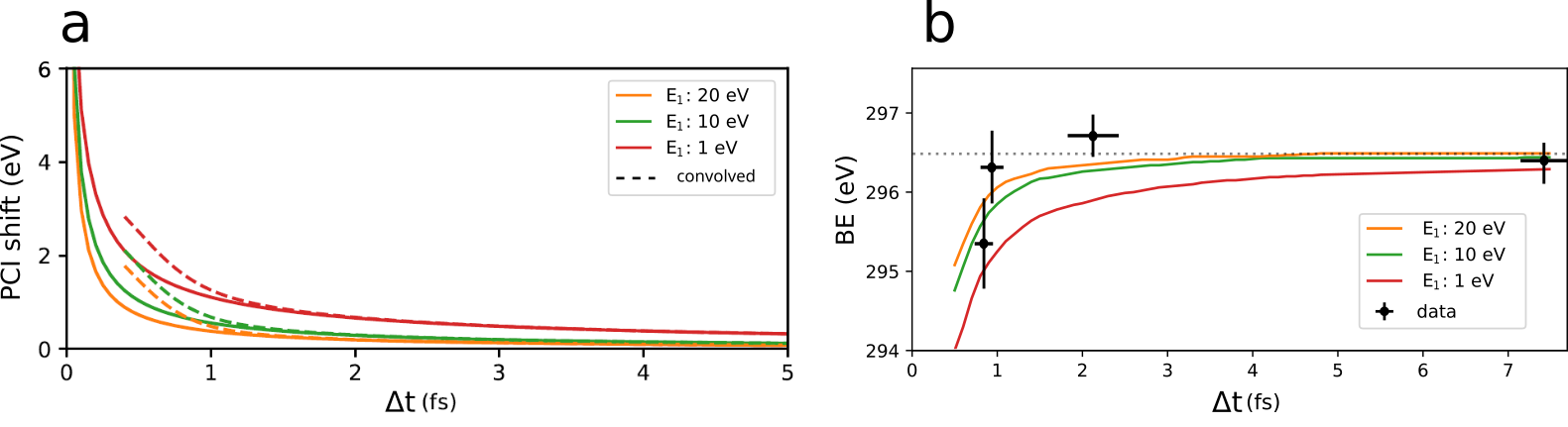}
\caption{\textbf{a}, Energy shift due to post-collision interaction of the two photoelectrons as a function of delay time. Convolution with finite pulse duration. 
\textbf{b}, Implementation of PCI effect in the original simulation (without coherence effects). 
Plain curve shows exclusively the PCI effect with slow electron at 10 eV. Black crosses are our data points. 
The gray dotted line indicates the fixed position of calculated BE$_{DCH}$.
}
\label{fig_s:PCI}
\end{figure}

Figure~\ref{fig_s:PCI} panel a shows the PCI shift of the photoelectron produced by the probe pulse, evaluated numerically as a function of pump-probe delay.
There is a strong dependence on the kinetic energy of the electron ionised by the pump~($E_1$). 
The PCI shift can be as big as a few eV at early delays ($\Delta t <$ 1~fs) but vanishes quickly within the first few fs.
The calculated PCI shift needs to be convolved with the finite duration~($\sim500$~as) of the pump and probe pulses, which is shown as the dashed lines in Fig.~\ref{fig_s:PCI}-\textbf{a}. 
This convolution, increases the PCI shift for short delays~($0.5-1$~fs).

Figure~\ref{fig_s:PCI} panel b, compares the measured change in binding energy as a function of the delay between the pump and probe pulses, along with the results of the PCI model applied to the binding energy of a carbon $K$-shell electron in a core-ionised state, BE$\sim296.5$~eV~\cite{zhaunerchyk_disentangling_2015}. 
This is the energy required to ionise an isolated molecule with a single core hole~(SCH) to create a two-site double core hole~(tsDCH), which is shown as a dashed line in Fig.~\ref{fig_s:PCI}(b). 
The PCI calculation is done for several values of the photoelectron kinetic energy ionised by the pump pulse~($E_1$). 

\subsection{Including the SCH lifetime }

Beyond $\sim2$~fs, other ultrafast processes become important to understand any experimental measurement.
The core-excited ion decays \textit{via} Auger-Meitner emission with a lifetime of $9.5-10.5$~fs, which induces an additional increase in the carbon $K$-shell binding energy~\cite{zhaunerchyk2015disentangling}.
As mentioned above, the carbon $K$-shell binding energy of a core-excited ion is $\sim296.5$~eV. 
The carbon $K$-shell binding energy of the dication states populated by AM decay is calculated to be in the range $~\sim301.5-304$~eV~\cite{zhaunerchyk2015disentangling}. 
To model the AM decay effect in the data, we  model the photoelectron kinetic energy spectrum as the sum of two contributions,~(1)~a feature representing subsequent core-level ionization of a core-ionised cation to create a double core hole~(DCH)~\cite{young2010femtosecond}, and~(2)~a feature corresponding to core-level ionization of a dication produced by the AM decay. 
This model neglects any contribution from ground state molecule, as discussed in section~\ref{sec:supp:PumpFrac} 
\begin{table}
\centering
\begin{tabular}{|c |c| c|} 
 \hline
  SCH & ts - DCH & PAP\\ 
  && (SCH in dication) \\[0.5ex] 
  \hline \hline
  290.56 & 296.53 & 302.24 \\
 $\pm$ 0.67 eV & $\pm$ 0.91 eV & $\pm$ 1.17 eV \\

 \hline
\end{tabular}
\caption{Calculated core binding energies from Ref.~\cite{zhaunerchyk2015disentangling}. Here the `$\pm$' gives the 
standard deviation of the calculated values for the 6 carbon atoms. 
It does not stand for the error in the calculation. 
PAP stands for the consecutive photoemission from the pump - Auger-Meitner emission - photoemission from the probe. }
\label{tab_s:zhaunerchyk}
\end{table}
%
Our model for the photoelectron kinetic enegy spectrum, $S(E)$, can we written as:
 \begin{equation}
S(E)=A_{{DCH}}(t) \times e^{-\frac{(E-BE_{{DCH}})^2}{2\sigma^2}} + A_{{PAP}}(t) \times e^{-\frac{(E-BE_{{PAP}})^2}{2\sigma^2}},
\end{equation}
 where $A_{DCH}(t)= e^{-t/\tau}$ is the intensity of the double-core hole feature produced by the probe~(i.e. ionization of a SCH), $A_{PAP}(t)= 1-e^{-t/\tau}$ is the intensity of the feature that corresponds to core-ionization of a dication molecule, and $\sigma$ is the width of each photoelectron feature, which we set to $\sigma = 3$~eV, based on the experimental data recorded at long pump/probe delays~($7$~fs). 
We have found that the binding energy shift predicted by the model is rather insensitive to the choice of $\sigma$.
    
\begin{figure}[h]
    \includegraphics[width=\textwidth]{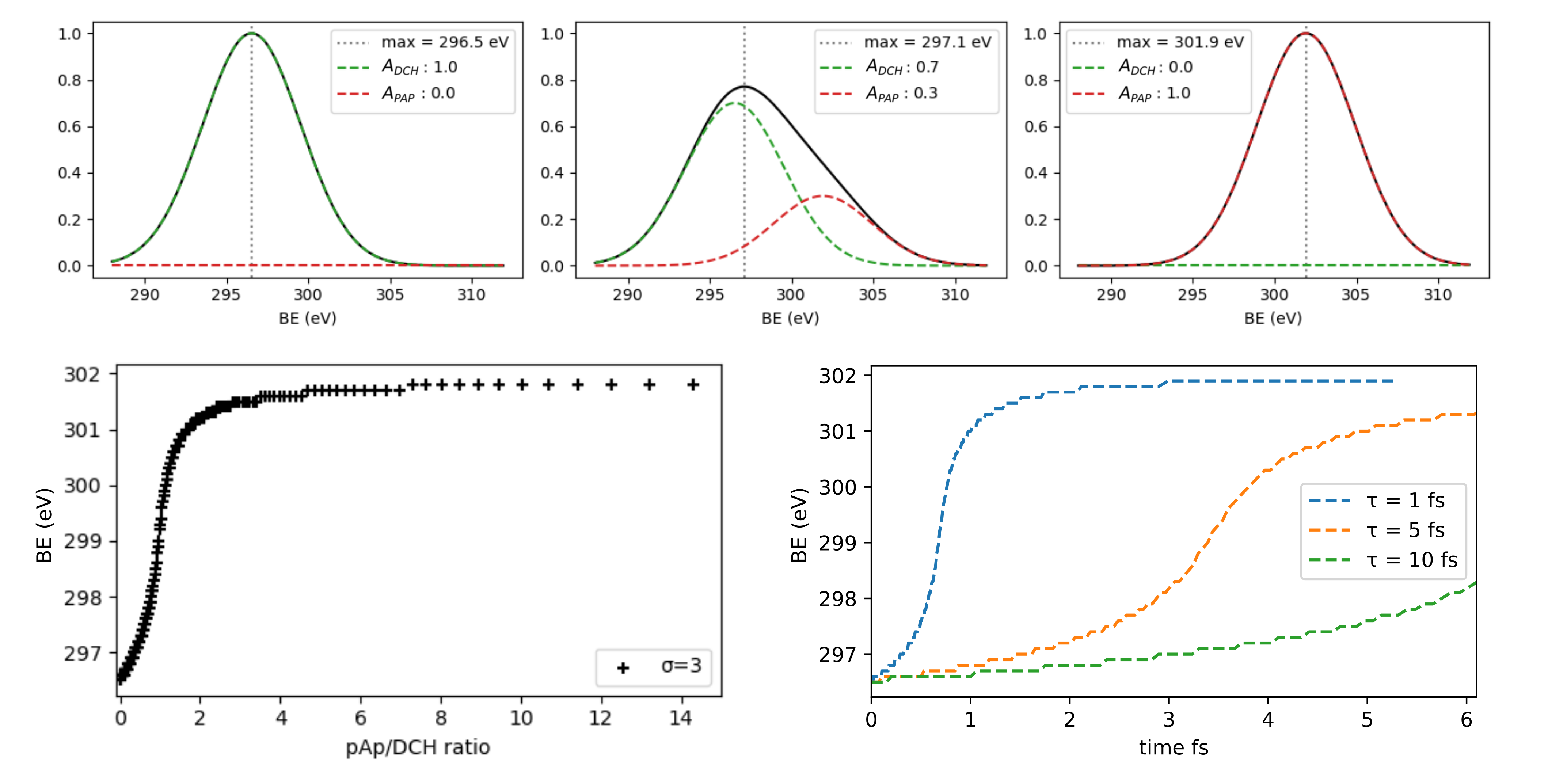 }
    \caption{\textit{top - }Simulated photoelectron peaks for different amplitude ratio DCH/PAP. The maxima of the sum of these two peaks are indicated in dotted line. The global width and global BE are estimated by a single Gaussian fitting over the feature at the latest delay.
    \textit{left - }Maximum position of the two Gaussian simulation in binding energy BE wrt. PAP/DCH amplitude ratio. \textit{right - }Peak position mapped from amplitude ratio to time for different $\tau$. }
    \label{fig_S:Gaussian_simulation}
\end{figure}

The top row of Fig.~\ref{fig_S:Gaussian_simulation} shows the photoelectron binding energy spectrum for different values of $A_{DCH}$ and $A_{PAP}$, demonstrating how the composite spectrum~(black line) shifts in binding energy as the ratio changes.
The peak of the photoelectron feature is shown in the lower left panel of Fig.~\ref{fig_S:Gaussian_simulation}.
%
%
From these equations, one can map the amplitude ratio to time via the decay time parameter  $\tau$ as follows : $t = \tau \times \textnormal{ln}(\frac{A_{PAP}}{A_{DCH}}+1)$ and get the time evolution of the peak position. 
The lower right panel of Fig.~\ref{fig_S:Gaussian_simulation}  shows the evolution of the BE as a function of time and for different values of $\tau$. 

\begin{figure}[h!]
    \centering
    \includegraphics[width=0.5\textwidth]{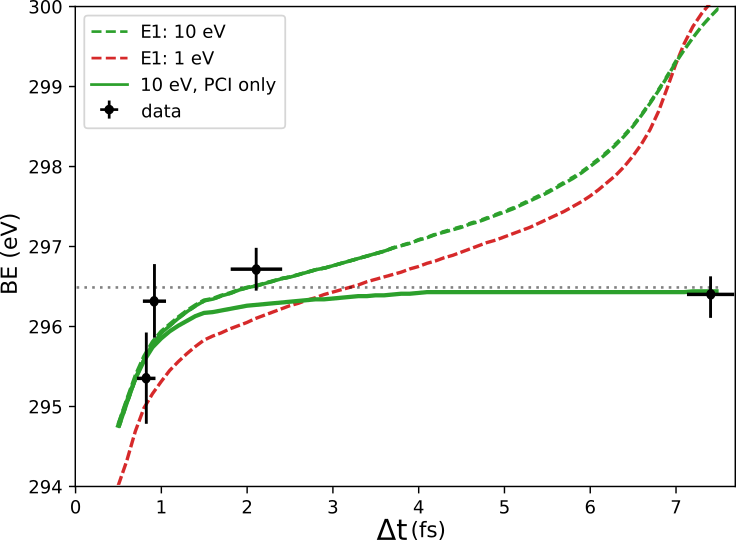 }
    \caption{Combination of PCI and SCH lifetime effects in the effective binding energy simulation. Plain curve shows exclusively the PCI effect for slow electron of 10 eV. Dashed curves shows the combination of the same PCI effect plus SCH lifetime of 10 fs. Black crosses are our data points for 1~eV and 10~eV slow electron. The gray dotted line indicates the fixed position of calculated BE$_{DCH}$. }
    \label{fig_S:PCI+SCH}
\end{figure}
%
The results of this model combined with the predicted PCI shift are shown in Figure~\ref{fig_S:PCI+SCH}.
Including the finite AM core-vacancy lifetime improves the agreement for the first three data points. 

\subsection{Behavior at longer delays and electronic coherence}
\label{Subsection:SI_coherences}
The simple model for PCI and AM-decay developed in the previous subsection does not explain the longer-term behavior of the molecule. 
This can be seen in the disagreement between the simulation and the experimental data point at the largest delay~($7$~fs).
A detailed understanding of the intricate dynamics excited by the pump would require significantly more data and is beyond the scope of this paper. 
In this subsection we discuss some of the possible explanations for the observed discrepancy at large delays.

Core ionisation and subsequent AM decay can trigger ultrafast nuclear motion, resulting in additional shifts of the core-level binding energy. 
For instance, dissociation of a proton~($H^+$) could happen very quickly and would lead to a significant shift of the binding energy to lower values, i.e., opposite to the binding energy shift induced by the AM decay.
Following the proton loss, the binding energy of the resulting cation~(NC$_6$H$_6$O$^+$) would be similar to the core-ionised aminophenol molecule~(dashed line in Figure~\ref{fig_S:PCI+SCH}). 

Coherent excitation by the pump pulse may also play a role in the dynamics of the molecule.
Given the large bandwidth of the pump pulse, it may be possible to simultaneously excite multiple core-hole states, resulting in a superposition of core hole states~\cite{inhester_detecting_2019,li2022attosecond}.
The temporal evolution of such a coherent superposition could then cause oscillations in the ionization amplitudes for the probe-photoelectron spectrum that manifest themselves in a temporal shift in the observed binding energy.
The fastest oscillation frequency that could be observed in such a coherent electron wave-packet signal is estimated by the largest energy difference between two core hole binding energies, that is $\Delta E=1.7$~eV~\cite{zhaunerchyk2015disentangling} and corresponds to an oscillation frequency of $h/\Delta E=2.4$~fs. 
This value highlights that such oscillations are definitely accessible with the here considered pump-probe delays.

One has to consider that electronic coherence effects can considerably be suppressed by several aspects:
\begin{enumerate}
    \item The geometrical fluctuations of the molecule in its vibrational ground-state and the above-mentioned nuclear dynamics typically lead to a rapid dephasing of the electronic wave-packet~\cite{vacher_electron_2017,arnold_electronic_2017}.
    In addition, electronic coherence is suppressed by interaction with the departing photoelectron~\cite{arnold_molecular_2020}.
    \item For the present experiment, we measure probe photoelectrons averaged across all emission directions and all molecular orientations with respect to the polarization of the x-rays. This implicit averaging leads to significant suppression of signals attributed to electronic coherence~\cite{inhester_detecting_2019}.
    \item For potential coherence effects in the PAP signal, one has to consider that the probe photoelectron signal is averaged over many possible AM channels, leading to further suppression of electronic coherence.
\end{enumerate}

Overall, the current experimental setup, particularly the accessibility of attosecond delay times in an X-ray-pump/X-ray-probe experiment, opens up new avenues to explore coherent electron dynamics.
We cannot rule out that such features are already visible in the current measurements, or that they are responsible for the discussed discrepancy at 7~fs. 
We expect that electronic coherence effects should definitely be visible when combining the here-developed experimental setup with coincident detection of ions and electrons, resolving the molecular-frame electron angular distribution.

\bibliographystyle{unsrt}
\bibliography{refs, referencesJPC}